\documentclass[12pt]{article}
\pdfoutput=1
\usepackage{putex}
\usepackage{graphicx}
\usepackage{epstopdf}
\usepackage{enumerate}
\usepackage{cite}
\usepackage{tensor}
\usepackage{breqn}
\usepackage{slashed}
\usepackage{hyperref}
\usepackage{float}
\usepackage[bottom]{footmisc}
\usepackage[utf8]{inputenc}
\usepackage{amsmath}
\usepackage[dvipsnames,svgnames,table]{xcolor}

\usepackage{tikz}
\usetikzlibrary{positioning}
\usetikzlibrary{decorations.pathmorphing}
\usetikzlibrary{decorations.markings, decorations.pathmorphing}
\tikzset{graviton/.style={
	decorate,
	decoration={snake, amplitude=1, segment length=5},
}}

\numberwithin{equation}{section}
\hypersetup{
	pdfstartview={FitH},    
	pdftitle={The gravitational S-matrix from the path integral: asymptotic symmetries and soft theorems},    
	pdfauthor={Isen, Kraus, Monten, Myers},     
	colorlinks=true,       
	linkcolor=blue,          
	citecolor=blue!59!black! ,        
	filecolor=magenta,      
	urlcolor=blue!75!black,           
	linktoc=all
}

\makeatletter
\g@addto@macro\bfseries{\boldmath}
\makeatother

\numberwithin{equation}{section}

\newcommand{\eq}[2]{\begin{align}\label{#1}#2\end{align}}

\newcommand {\be} {\begin {equation}}
\newcommand {\ee} {\end {equation}}

\newcommand{\p}{\partial}

\newcommand\rt{{\rightarrow}}
\def\eps{\epsilon}

\newcommand{\rf}[1]{(\ref{#1})}
\newcommand{\rff}[1]{\ref{#1}}

\newcommand{\Lb}{\overline{L}}

\newcommand{\psib}{\overline{\psi}}

\newcommand{\Oc}{{\cal O}}

\newcommand{\zb}{\bar{z}}

\newcommand{\phib}{\overline{\phi}}

\newcommand{\xh}{\hat{x}}

\newcommand{\qv}{\vec{q}}

\newcommand{\scrI}{{\cal I}}

\newcommand{\hb}{\overline{h}}
\newcommand{\mathscr}{\mathcal}

\newcommand{\Ic}{\mathcal{I}}

\newcommand{\veps}{\varepsilon}
\newcommand{\yh}{\hat{y}}
\newcommand{\yb}{\bar{y}}

\newcommand{\epsh}{\hat{\veps}}

\newcommand{\vepsh}{\hat{\veps}}
\newcommand{\Vb}{\overline{V}}

\newcommand{\htil}{\tilde{h}}

\newcommand{\dr}{d}

\newcommand{\Wc}{{\cal W}}

\begin{document}

	\institution{UCLA}{ \quad\quad\quad\quad\quad\quad\quad\ ~ \, $^{1}$Mani L. Bhaumik Institute for Theoretical Physics
		\cr Department of Physics \& Astronomy,\,University of California,\,Los Angeles,\,CA\,90095,\,USA}

    \institution{CERN}{ \quad\quad\quad\ ~ \, $^{2}$Theoretical Physics Department, CERN, 1211 Geneva 23, Switzerland}

	\institution{UvA}{~~\quad\quad\quad\quad\quad\quad \,
	$^{3}$Institute for Theoretical Physics, University of Amsterdam, \cr
	~~\quad\quad\quad\quad Science Park 904, Postbus 94485, 1090 GL Amsterdam, The Netherlands
}
	\title{\LARGE The gravitational S-matrix from the path integral:\\[-1cm] asymptotic symmetries and soft theorems \\
    \normalsize
	}
	
	\authors{Jack Isen$^{1}$, Per Kraus$^{1}$, Ruben Monten$^{2}$,  Richard M. Myers$^{3}$}
	
	\abstract{We extend a previously developed formulation of the S-matrix, based on a path integral with asymptotic boundary conditions, to include gravity.   The path integral defines a Carrollian boundary partition function whose invariance under asymptotic symmetries implies Ward identities obeyed by the associated  boundary correlators, which are simply related to standard S-matrix elements.  We develop this in the context of extended BMS transformations at tree level.  Modulo well-known subtleties associated with  poles in the superrotations and corner terms, this leads to an efficient derivation of the leading and subleading soft graviton theorems from BMS symmetry.   Our general arguments are verified by explicit diagrammatic computation of specific terms in the partition function, which are shown to satisfy the Ward identities.   We also show how, in our context, the subleading soft theorem is fixed by 
    Poincaré Ward identities together with the leading soft theorem.  
 }
	
	\date{}
    \preprint{CERN-TH-2026-044}
    
	\maketitle
	\setcounter{tocdepth}{2}
	\begingroup
	\hypersetup{linkcolor=black}
	\tableofcontents
	\endgroup

\section{Introduction}

Asymptotic symmetries, together with their connection to the S-matrix via soft theorems, provide an organizational framework for gauge theory and gravity. They may also offer clues toward a dual description of gravitational physics in asymptotically flat spacetime.   Given the success of the analogous program in anti-de Sitter (AdS) spacetimes, it is natural to develop a flat-space formalism that mirrors the one developed in AdS \cite{Gubser:1998bc, Witten:1998qj}. Concretely, we take as the fundamental object a path integral with specified asymptotic boundary conditions, which serves as a generating functional for boundary correlation functions, which are in turn related to standard S-matrix elements. Building on earlier work \cite{Arefeva:1974jv} (see also \cite{Balian:1976vq,Jevicki:1987ax}), such a formalism has been developed in \cite{Kim:2023qbl, Kraus:2024gso, Kraus:2025wgi} for the case of QED and Yang--Mills theory; the extension to gravity is treated here.

\begin{figure}[ht]
    \centering
    \includegraphics[scale=0.3]{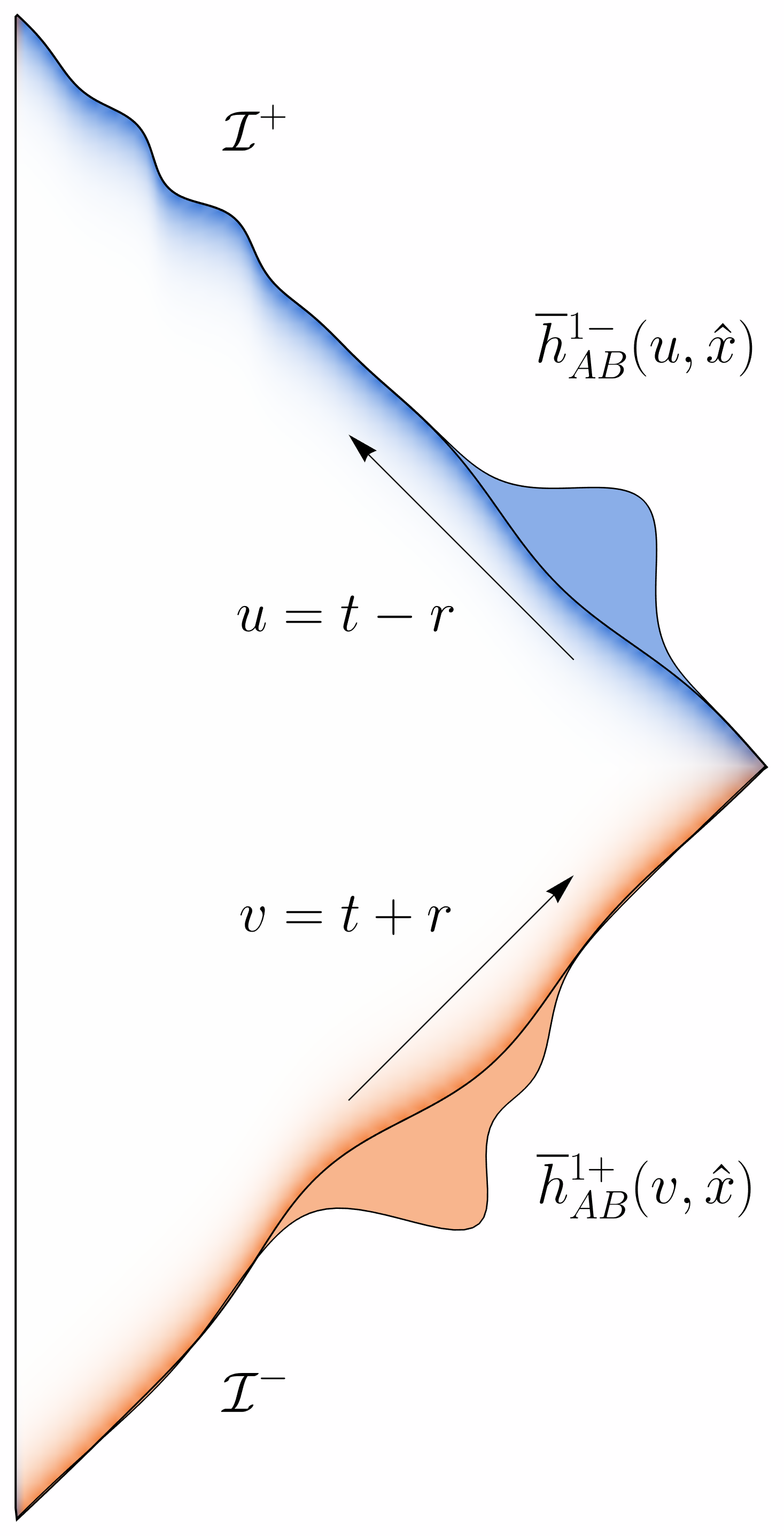}
    \caption{Minkowski Penrose diagram with asymptotic boundary conditions on the leading asymptotic field data; a pure negative (positive) frequency waveform is specified along $\scrI^+$($\scrI^-$).}
    \label{fig:triangle_with_waveform}
\end{figure}

The starting point is a specification of asymptotic boundary conditions on the metric and massless\footnote{There is no technical barrier to describing massive fields in this formalism, but since their boundary data lives at timelike infinity, instead of null infinity, the putative holographic structure is less clear. However, see e.g. \cite{Have:2024dff,Ammon:2025avo}. Hence we will restrict attention to massless fields in this work.} matter fields,  denoted as $\overline\Phi = (\overline{h}_{\mu\nu}, \phib)$ and depicted in figure \ref{fig:triangle_with_waveform}.  These are analogous to the boundary ``sources" in the AdS analog.\footnote{We recall that the ``source" moniker is based on how such data appear in the dual CFT, while from the bulk perspective these are just boundary conditions.  Here we use the term source in the same spirit.}
We denote the resulting bulk path integral --- equivalently, at tree level, the on-shell action --- as a functional of these sources by $Z[\hb_{\mu\nu},\phib]$. Schematically, this takes the form
\eq{INTRO1}{
    Z[\overline \Phi] = \int_{\overline\Phi}\!\mathcal{D}\Phi \, e^{iI[\Phi, \overline\Phi]}~.
}
This object serves as a generating functional for boundary $n$-point correlators $W_n$ (we suppress field indices for now) defined on null infinity,
\eq{INTRO2}{
    \ln Z[\overline\Phi] = \sum_n \frac{1}{n!}\int[\dr^3 y]W_n(y_1,\ldots,y_n)\overline\Phi(y_1)\cdots \overline\Phi(y_n)~.
}
Such correlators are often referred to as \textit{“Carrollian correlators”}, and provide objects to be computed by some hypothetical Carrollian CFT.\footnote{Carrollian theories have seen extensive development as candidate holographic duals to asymptotically flat spaces in recent years \cite{Duval:2014uoa,Duval:2014uva,Hartong:2015xda,Hartong:2015usd,Bagchi:2016bcd,Ciambelli:2018wre,Ciambelli:2019lap,Bagchi:2019clu,Donnay:2022aba,Bagchi:2022emh,Donnay:2022wvx,Bagchi:2023fbj,Nguyen:2023vfz,deBoer:2023fnj,Mason:2023mti,Alday:2024yyj,Cotler:2024xhb,Ciambelli:2025mex,Nguyen:2025sqk,Poulias:2025eck,Surubaru:2025fmg,Lipstein:2025jfj,Agrawal:2025bsy,Fiorucci:2025twa,Hartong:2025jpp,Cotler:2025npu}. See the recent reviews \cite{Bagchi:2025vri,Ciambelli:2025unn,Nguyen:2025zhg,Ruzziconi:2026bix} for a collection of perspectives. Celestial holography is another major candidate proposal, see the reviews \cite{Strominger:2017zoo,Raclariu:2021zjz,Pasterski:2021rjz,Pasterski:2021raf,McLoughlin:2022ljp,Donnay:2023mrd}. The map between these proposals was first worked out in \cite{Donnay:2022aba,Bagchi:2022emh,Donnay:2022wvx,Bagchi:2023cen}.} As such, we refer to \eqref{INTRO1} as the \textit{“Carrollian partition function”}. Importantly, after suitable differentiation and Fourier transformation, these correlators are identified with flat space S-matrix elements.\footnote{The map between momentum space amplitudes and Carrollian correlators was first proposed in \cite{Donnay:2022aba,Bagchi:2022emh,Donnay:2022wvx,Bagchi:2023cen}, and follows naturally from the functional \eqref{INTRO2} as shown in \cite{Kraus:2024gso,Kraus:2025wgi}.} One of the major goals of this work is to give a precise definition, in perturbation theory around Minkowski space,  of the schematic expressions \eqref{INTRO1} and \eqref{INTRO2} in the presence of dynamical gravity.\footnote{Since top-down examples of a holographic duality are scarce in asymptotically flat spacetimes, we work from the bottom-up, meaning we will give an entirely bulk account of \eqref{INTRO1}. However, see \cite{Costello:2022wso,Costello:2022jpg,Costello:2023hmi} for twistorial work from the top down.} 

One virtue of this approach is that it yields a clean way, technically and conceptually, of packaging asymptotic symmetries, in this case supertranslations and superrotations forming the (extended) BMS group \cite{Bondi:1962px,Sachs:1962zza,Barnich:2009se}, and their link to leading \cite{Weinberg:1965nx} and subleading \cite{Cachazo:2014fwa} soft graviton theorems. In particular, symmetries of the action imply symmetries of the partition function, and hence Ward identities for the boundary correlators/S-matrix elements appearing in \eqref{INTRO2}.\footnote{Here we will not consider possible subtleties in the transformation of the path integral measure.}  We distinguish two cases, depending on whether the symmetry transformation preserves Minkowski space (unbroken symmetry) or does not (broken symmetry).    Under an unbroken symmetry the field transformations are  linearly realized, $\delta_\lambda \overline\Phi \propto \overline\Phi$, so the variation  of \eqref{INTRO2} will not mix terms of different $n$, implying term-by-term Ward identities
\eq{INTRO3}{
    L_\lambda W_n = 0
}
for some operator $L_\lambda$. Poincar\'e transformations are of this type, in which case \rf{INTRO3} implies momentum conservation and Lorentz invariance of the S-matrix. 

We contrast this with the case of spontaneously broken gauge/diffeomorphism transformations, which act on the asymptotic gauge field/metric data with a shift, $\delta_\lambda \overline\Phi \propto \overline \Phi + C_\lambda$ for some field-independent $C_\lambda$. 
Since the action of such a transformation changes the number of field factors in each term of \eqref{INTRO2}, boundary correlators of different $n$ must mix such that $\delta_\lambda Z = 0$, resulting in a Ward identity with schematic form
\eq{INTRO4}{
    L'_\lambda W_{n+1} + L''_\lambda W_n = 0~.
}
Large gauge transformations, including supertranslations and superrotations, are typically of this type, which is why their Ward identities are able to reproduce soft theorems that relate amplitudes of different multiplicity.    In particular, applied to (extended) BMS transformations, \rf{INTRO4} becomes the (sub)leading soft graviton theorem.\footnote{ \label{subtle}There is a subtlety here: asymptotic symmetries act at strictly zero frequency, while soft theorems involve the limiting behavior as the frequency goes to zero.  Keeping track of this distinction can involve factors of $2$, as illustrated by the relation $\lim_{\omega \rt 0^+ } \int_{-\infty}^\infty \! dx  \frac{ e^{i\omega x} }{ x-i\eps} = 2 \int_{-\infty}^\infty  \frac{dx}{ x-i\eps}$, where the conditionally convergent integrals are defined by imposing the cutoff $x \in [-L,L]$ with $L$  taken to infinity at the end.  See \cite{Kraus:2025wgi} for a careful discussion.   An analogous subtlety arises in the covariant phase space approach \cite{He:2014laa}. In this paper we define such divergent integrals using the prescription \eqref{eq:cosReg} and find that it leads to consistent results.
}
 
Of course, the connection between asymptotic symmetries and soft theorems in gravity has already been developed in great detail over the past decade \cite{Barnich:2009se,Barnich:2010eb,Barnich:2010ojg,Barnich:2011mi,Strominger:2013jfa,He:2014laa,Kapec:2014opa,Campiglia:2014yka,Campiglia:2015yka,Campiglia:2015kxa,Distler:2018rwu,Campiglia:2019wxe,Campiglia:2020qvc,Campiglia:2021bap,Capone:2023roc,Agrawal:2023zea,Choi:2024ygx,Campiglia:2024uqq,Choi:2024ajz}. Most of this work has been presented in the language of covariant phase space, symplectic forms, Poisson brackets, and so on. Our motivation to instead approach this connection from the path integral perspective is partly driven by the clean packaging of  asymptotic symmetries and their associated Ward identities. Beyond this, one may also note that:
\begin{itemize}
    \item[1)] The perturbative quantization of gauge theories and gravity is typically more convenient in the path integral formulation. As such, the path integral may provide a similarly convenient framework for understanding the known loop corrections to soft theorems, e.g. \cite{Bern:2014oka,Sahoo:2018lxl, Ma:2023gir}, by making contact with general methods such as in \cite{Baulieu:2025itt}.
    
    \item[2)] Since \eqref{INTRO1} directly generates scattering amplitudes, we are forced to work with the boundary conditions relevant for scattering, which differ from those implied by the classical initial value problem adopted in much of the asymptotic symmetry literature \cite{Strominger:2017zoo}. Perturbatively, the distinction is between using the retarded/advanced versus the Feynman propagator, and the difference in asymptotics can be seen in very simple examples such as in the field produced by a moving point charge \cite{AtulBhatkar:2021txo, Kim:2023qbl}; see also \cite{Briceno:2025cdu}.
    
    \item[3)] The general ``compute the path integral as a functional of boundary conditions"  framework provides a simple route for embedding flat space physics within AdS/CFT, since as one takes the AdS radius to infinity the AdS boundary data turn into data on null infinity, e.g. \cite{Susskind:1998vk,Polchinski:1999ry,Giddings:1999jq,Gary:2009ae,Gary:2009mi,Penedones:2010ue,Fitzpatrick:2011jn,Fitzpatrick:2011ia,Hijano:2019qmi,Hijano:2020szl,Komatsu:2020sag,Li:2021snj,Duary:2022pyv,Bagchi:2023fbj,Campoleoni:2023fug,deGioia:2024yne,Marotta:2024sce,Alday:2024yyj}, yielding a relation between the flat space and AdS partition functions \cite{Kraus:2024gso}. 
\end{itemize}
 
 \vspace{.2cm} 
 \noindent
 {\bf  \Large  Overview}
  \vspace{.2cm} 
 
We now give a more detailed account of the contents of this work. To define the path integral \eqref{INTRO1}, we must define our asymptotic boundary conditions and the boundary terms included in the action. We address both points in section \ref{action}. In brief, since our basic observable is the S-matrix (an in-out quantity), we impose boundary conditions on both past and future null infinity by specifying the positive or negative frequency content of certain metric and matter field components. Having specified the boundary conditions, the action must be chosen to admit a compatible variational principle, that correctly generates the S-matrix. This principle serves to determine what boundary terms, which live on null infinity, should be added to the action.\footnote{\label{local finite counterterms}A similar perspective can be adopted in AdS \cite{Papadimitriou:2007sj}. Strictly, there is a remaining ambiguity whereby one can add to $I$ a functional of the sources alone. In AdS this is the ambiguity of adding local and finite counter terms, and in the present case can be fixed by demanding equivalence with LSZ \cite{Kim:2023qbl,Kraus:2025wgi}.} 

In determining the boundary terms, one must decide when to keep or discard terms that are themselves total derivatives on null infinity, i.e. so-called corner contributions.\footnote{It may be possible to fix the corner ambiguities by an on-shell action evaluation, such as in \cite{Upadhyay:2025ged}, but we leave this to future work.} Whether such terms contribute depends on assumptions regarding field falloffs near the boundaries of null infinity (large retarded times). Corner terms could also be cancelled by contributions from spatial infinity, like in the anomaly term of \cite{Compere:2011ve}. In this work we generally assume these terms can be discarded, either due to falloff conditions or terms at spatial infinity. However, this point merits further study pending a better understanding of the S-matrix boundary conditions at the edges of null infinity.

From the resulting action we identify the operators conjugate to variations of the boundary metric components, and give the precise map between the Carrollian boundary correlators and momentum space S-matrix elements. Since the boundary conditions fix half the frequency content of the metric, instead of the induced metric, it is not obvious that the conjugate operator has an interpretation as a boundary stress tensor. It would be interesting to connect to existing discussions of the holographic stress tensor, e.g. \cite{Kapec:2016jld,Kapec:2017gsg,Saha:2023hsl,Bagchi:2024gnn,Ruzziconi:2024kzo,Ciambelli:2025mex}.

In section \rff{asymp} we define the asymptotic symmetry transformations corresponding to supertranslations and superrotations, and then study invariance of the action under these transformations.    The story is straightforward for supertranslations.   For superrotations, we work in the extended BMS framework \cite{Barnich:2009se} in which infinitesimal  superrotations are  labelled by  meromorphic vector fields on the celestial sphere.  Here we run into three well-known issues: justifying the antipodal matching of the superrotation parameters, the (mild) violation of the asymptotic fall-off conditions, and the appearance of diverging integrals along null infinity.

The first of these relates to spatial infinity, where supertranslations are understood.
Indeed, previous work \cite{Mann:2005yr,Mann:2006bd, Mann:2008ay,Compere:2011ve} has shown that a good variational principle is possible for gravity modulo corner terms. 
Nevertheless, one might still worry that the conjugate operator to the Goldstone mode of supertranslations receives a contribution from spatial infinity, which would alter the Ward identity. One implication of \cite{Compere:2011ve} is that this does not occur so long as the supertranslation Goldstone mode obeys a de Sitter wave equation along the hyperbolic resolution of spatial infinity. It was shown in \cite{Capone:2022gme,Compere:2023qoa} that all relevant solutions to this wave equation, supplemented by a parity condition, obey antipodal matching of the supertranslation function between future and past null infinity.%
\footnote{A curious feature of \cite{Compere:2011ve} is the anomaly term, which is not invariant under supertranslations, and which may disrupt the off-shell invariance of the action. However, on-shell this term is a corner, so at the level of our remarks here, which may be viewed as tree level and ignoring corners, the anomaly term does not enter. Furthermore, the absence of loop corrections in the leading soft graviton theorem would seem to suggest that this term remains inactive even at loop order, yet this term remains to be understood in detail.} In section \ref{spatial} we give a brief review of how this process works. The answer to the analogous question of the superrotation mode appears to remain unknown in the literature, though see \cite{Fiorucci:2024ndw}. We leave this question to future work.

The second issue is that the extended BMS transformations fail to preserve the asymptotic boundary conditions at the locations of the poles.    To respect the asymptotics we are thus forced to define a modified transformation of the fields in which we simply delete these pole terms.      However, this maneuver creates an obstacle to demonstrating invariance of the action, since the modified transformation is not a bona fide diffeomorphism, leading to potential  noninvariance  at the poles. Nonetheless, if we assume that these non-invariant contributions cancel, we formally obtain a Ward identity. This is shown in section \ref{ward}, where we also demonstrate the unbroken Ward identities implied by Poincar\'e invariance. Understanding the cancellation of non-invariant contributions is likely tied to the treatment of corner terms discussed above.

Analysis of the superrotation Ward identities, undertaken in section \ref{Soft-thm}, leads to the third issue: the associated charge contains an  integral over null infinity which is divergent and must be defined. A similar issue appears in the Ward identity for supertranslations, where we encounter a conditionally convergent integral, whose proper treatment is  critical to finding the correct leading soft theorem after translating the Ward identity into S-matrix statements; see footnote \rff{subtle}. We give a prescription for these divergent and conditional integrals as follows: 
\begin{align}
\label{eq:cosReg}
    \int_{-\infty}^\infty du \, f(u) \equiv \lim_{\omega \to 0^+} \int_{-\infty}^\infty du \, \cos(\omega u) f(u)
    \ ,
\end{align}
which involved exchanging the order of the integral and the limit.
We find that this convention reproduces the correct tree level subleading soft theorem, but the basic principles for selecting this prescription remain unclear. However, we further show in section \ref{poincare} that the normalization is unambiguously fixed by the leading soft theorem and Poincar\'e invariance (together with a few mild assumptions), so any rule for defining the integral that respects   Poincar\'e invariance  should produce the same result.\footnote{This may be understood as a position space version of the argument appearing in \cite{Bern:2014vva}.} This explains why our manipulations of divergent quantities are more robust than they might appear.

In section \ref{diags} we compute explicit flat space Witten diagrams for processes that illustrate the leading and subleading soft graviton theorems. We use these examples to explicitly demonstrate invariance of the partition function under supertranslations and superrotations in the form \eqref{INTRO2}, providing a useful check of our abstract arguments.  We also clarify the relation between soft theorems for the standard correlators  versus  those for ``reduced correlators," which are defined by removing the overall momentum conserving delta function.

Mainly for purposes of comparison to existing literature on the subject, in section \ref{charges} we use Noether's theorem to derive the conserved charges associated with the asymptotic symmetries. These charges may also be  interpreted as fluxes, since they are expressed as integrals over null infinity and thus represent the flow through null infinity of a canonical charge defined at spatial infinity.   This is the analog of the familiar statement that in electromagnetism the electric charge may be expressed either as an integral over a codimension-2 surface at spatial infinity, or as an integral over  a codimension-1 Cauchy surface, the two expressions being equivalent due to the Gauss law constraint.

Finally, throughout this work we ignore the fact that the gravitational S-matrix is IR divergent beyond tree level. Accordingly, our results should be understood as tree level statements. Properly understanding loop effects would require identifying IR-safe observables or introducing an IR regulator, which can modify the soft properties of amplitudes, e.g. \cite{Bern:2014oka, Sahoo:2018lxl,Ma:2023gir}. While we leave this for future work, we expect the tools developed here to carry over once such a regulator is imposed, or once suitably IR-finite observables are identified within our framework.

\section{Gravitational action}
\label{action}

Our goal is to define a generating functional for the S-matrix in asymptotically flat spacetimes in terms of a path integral with boundary conditions, i.e. \eqref{INTRO1}. To do so, we will follow the previous analyses for scalar and gauge fields in \cite{Kim:2023qbl,Kraus:2024gso,Kraus:2025wgi}. A complete definition of \eqref{INTRO1}  must specify the action $I$, including any boundary terms, and the precise boundary conditions imposed on the fields.

In the interest of keeping our focus on the special features of gravity in the path integral framework, we will refer the reader to \cite{Kraus:2025wgi} for a detailed review of the basic procedure, accompanying subtleties, and proof of correspondence with the more standard LSZ procedure. An underlying assumption here is that the fields have a nice $1/r$ expansion near null infinity, the precise form of which we will extract from linearized solutions.  This assumption seems reasonable at tree level (and is bolstered by our explicit computations) but is problematic at loop level, where logs associated with IR divergences can appear.   For this reason, among others,  our results should be viewed as holding at tree level. The action to be used inside the path integral is then fixed by demanding a good variational principle, though see the comments in footnote \ref{local finite counterterms}.

Besides having a good variational principle, we would like an action that is invariant under asymptotic symmetry transformations corresponding to certain diffeomorphisms that act non-trivially at the boundary. Indeed, this invariance will encode the soft theorems. At the same time, in order to set up perturbation theory we need to add a gauge fixing term which, by definition, is not invariant under diffeomorphisms. The clash between these two features is avoided by adopting background field gauge, where the gauge fixing term depends explicitly on the choice of a background field. The virtue of this is that the full action, including the gauge fixing term, is invariant under a diffeomorphism that acts simultaneously on the background and dynamical fields. Since we choose this background field to encode the asymptotic boundary conditions, this makes it manifest that the bulk action is invariant under the asymptotic symmetries, leaving only boundary terms to be checked for invariance. We add that, since the partition function generates the S-matrix, it is gauge invariant and so does not depend on the specific choice of gauge fixing term. The virtue of background field gauge is that it makes the invariance under asymptotic symmetries manifest.

Due to the assumed falloffs, boundary terms vanish beyond quadratic order for fluctuations around Minkowski spacetime.   We therefore use the expansion of the action to quadratic order to fix the boundary terms.

A peculiarity of the Einstein--Hilbert action is that it contains terms of first order in fluctuations around Minkowski space.  These terms are total derivatives and so do not affect the equations of motion, but since total derivatives can contribute to the on-shell action we need to pay attention to them.  In particular, since there is no ``1-point S-matrix", we need to verify that there is no contribution to the on-shell action at first order in fluctuations. We check that this is indeed the case, modulo corners, once we take into account asymptotic falloffs and gauge conditions.

Finally, we will frequently encounter terms that are total derivatives on past and future null infinity.  We systematically neglect such terms, under the optimistic assumption that they either vanish due to falloffs on null infinity, or cancel between  $\Ic^+_-$ and $\Ic^-_+$.  Alternatively, it may be the case that additional boundary terms on     $\Ic^+_-$ and $\Ic^-_+$ are needed to cancel such contributions.  Presently, this is difficult to resolve given our incomplete understanding of falloffs along null infinity.  A constructive route to fixing these ambiguities would be to demand on-shell equality  between the action written in this section and the partition function computed in section \rff{diags} via flat space Witten diagrams. In particular, the Witten diagrams  are well established as yielding agreement with the S-matrix as computed by LSZ.

\subsection{Boundary conditions}

To motivate our boundary conditions, it is instructive to first consider the solutions of linearized gravity, and ask what asymptotics the free fields obey. Expanding around Minkowski space as 
\eq{BC1}{
    g_{\mu\nu} = \eta_{\mu\nu} + h_{\mu\nu}
}
it is useful to adopt the De Donder gauge, which in standard Cartesian coordinates takes the form
\eq{BC2}{
    \p^\mu h_{\mu \nu} - \frac{1}{2}\p_\nu h^\mu_{\ \mu} = 0~,
}
so the linearized Einstein equations take the form
\eq{BC3}{
    \p^\lambda\p_\lambda h_{\mu\nu} = 0
}
where indices have been raised using the inverse of $\eta_{\mu\nu}$.
This is nothing but the scalar wave equation in each component, and so the field equations admit the mode solution
\eq{aa}{ h_{\mu\nu}(x) = \sum_{\alpha =\pm} \int {d^3 q\over (2\pi)^3}   \big[ \eps^{\alpha *}_{\mu\nu}(\qv) a_\alpha(\qv)  e^{iq\cdot x} + \eps^{\alpha }_{\mu\nu}(\qv) a_\alpha^\dagger(\qv)  e^{-iq\cdot x} \big]}
where $q^0 = \omega_q = |\vec q|$, $\alpha = \pm$ denotes the helicity, and the polarization vectors are symmetric, transverse, and traceless,
\eq{BC4}{
    q^\mu \epsilon^\pm_{\mu\nu}(\vec q) = \eta^{\mu\nu}\epsilon_{\mu\nu}^\pm(\vec q) = 0,\ \ \ \epsilon_{\mu\nu}^\pm(\vec q) = \epsilon_{\nu\mu}^\pm(\vec q)~.
}

To analyze the asymptotics of the solutions \eqref{aa} in the vicinity of $\scrI^+$, it is useful to work in the retarded coordinates $(u, r, \hat x)$ defined by
\eq{BC5}{
    \vec x^2 = r^2, \ \ \ t = u + r,\ \ \ \hat x = \frac{\vec x}{r}~.
}
For many computations it is also useful to introduce the stereographic coordinates $(z, \zb)$ on the sphere, related to our other coordinates by
\eq{BC6}{
    \hat x(z, \zb) = \frac{1}{1 + z \zb}(z + \zb, -i(z - \zb), 1 - z \zb),\ \ \ z = \frac{x^1 + ix^2}{x^3 + r},\ \ \ \zb = \frac{x^1 - ix^2}{x^3 + r}~.
}
In these coordinates, the Minkowski metric becomes
\eq{BC7}{
    ds^2 = \eta_{\mu\nu}d x^\mu dx^\nu &= -\dr u^2 - 2\dr u \dr r + r^2 \gamma_{AB}\dr x^A \dr x^B,\cr
    \gamma_{AB}\dr x^A \dr x^B &= 2\gamma_{z\zb}\dr z \dr \zb,\ \ \ \gamma_{z \zb} = \frac{2}{(1 + z\zb)^2}~.
}
Throughout we will use indices $A, B,C,\ldots$ to denote coordinates on the sphere. It will also be useful to define the ``unit'' null vector pointing to a specified point on the celestial sphere by
\eq{jj1}{
    n^\mu(z, \zb) = (1, \hat x(z, \zb))~.
}

In these coordinates, a standard saddle point analysis \cite{Strominger:2017zoo} of \rf{aa}, subject to suitable regularity assumptions on the mode data, gives the asymptotic expansion,
\eq{BC8}{
    h_{\mu\nu} = {i\over 2(2\pi) ^2} {1\over r} \int_0^\infty\! d\omega \Big(- \eps^{\alpha *}_{\mu\nu}(\omega \xh) a_\alpha(\omega \xh) e^{-i\omega u} + \eps^{\alpha }_{\mu\nu}(\omega \xh) a^\dagger_\alpha(\omega \xh) e^{i\omega u} \Big)+ \mathcal{O}(r^{-2})
}
as $r\rt \infty$ at fixed $u$, where we stress that this relation is specific to  Cartesian coordinates.  

Our boundary condition on $\Ic^+$ will be to fix the physical helicity modes $a_\alpha^\dag$, i.e.~the negative frequencies, but importantly this is not equivalent to fixing the negative frequency part of the symmetric tensor $h_{\mu\nu}$ at order $1/r$, because the 10 components here are not independent. Said differently, there are only two physical degrees of freedom, parametrized here by the two helicity modes, and \eqref{BC8} instructs us how to embed this gauge invariant data into the 10 components of the boundary field.   It is convenient to go to retarded coordinates and think of the mode data as being packaged in the large $r$ expression for the symmetric traceless tensor $h_{AB}$.

To demonstrate how this works, we fix the polarization tensor
\eq{c4}{
    \epsilon_{\mu\nu}^\pm(\vec p) &= \epsilon_\mu^\pm(\hat p)\epsilon^\pm_\nu(\hat p)
}
where the polarization vectors $\epsilon^\pm_\mu$ have Cartesian components
\eq{jj3}{
    \epsilon^+_\mu(\hat p(z, \zb)) = \frac{1}{\sqrt{2}}(-\zb, 1, -i, -\zb),\ \ \ \epsilon^-_\mu(\hat p(z, \zb)) = \frac{1}{\sqrt{2}}(- z, 1, i, -z)~.
}
Note that we are labelling $\hat{p}$ by a point on the ``celestial momentum sphere" using the same expression as for $\hat{x}$ in \rf{BC6}.   We also note the relation
\eq{ddd}{ \epsilon_{\mu}^{\pm*}  (\vec p) = \epsilon_{\mu}^{\mp} ( \vec p  )~.  }

In retarded coordinates, the components of these  polarization  tensors are  position dependent, as follows from the definition   $\eps^\alpha_{\mu'} = {\p x^\mu\over \p x^{\mu'}} \eps^\alpha_{\mu}$. 
In the saddle point expression  \rf{BC8} the direction of the  momentum vector is locked to the position vector, $\hat{p} = \hat{x}$.  The locked polarization vectors then take the form\footnote{For a generic momentum $p(w, \overline w)$ at a point $\hat x(z, \zb)$ on the sphere, $\epsilon^+_\mu \dr x^\mu = \sqrt{2}\frac{\overline w - \zb}{1 + z \zb}\dr r - \frac{\overline w}{\sqrt{2}}\dr u + \sqrt{2}r\frac{\zb (\zb - \overline w)}{(1 + z\zb)^2}\dr z + \sqrt{2}r\frac{1 + \overline w z}{(1 + z\zb)^2}\dr \zb$ and similarly for $\epsilon^-$.}
\eq{an}{ &\eps^+_r (\xh) =0 ~,\quad   \eps^+_u( \xh) =-{1\over \sqrt{2}}\zb~,\quad  \eps^+_z(\xh) =0~,\quad  \eps^+_{\zb}( \xh) = {\sqrt{2}r\over 1+z\zb}~,\cr
&\eps^-_r ( \xh) =0 ~,\quad   \eps^-_u( \xh) =-{1\over \sqrt{2}}z~,\quad  \eps^-_{z}( \xh) = {\sqrt{2}r\over 1+z\zb} ~,\quad  \eps^-_{\zb}( \xh) =0~. }
Let us also clarify a potentially confusing point regarding the definition of $\eps^{\pm *}_{\mu\nu}$ in retarded coordinates.  The definition is 
\eq{an75}{  \eps^{\pm *}_r(\vec{p} )=  {\p x^\mu \over \p r}  \eps^{\pm *}_\mu(\vec{p})~,\quad 
 \eps^{\pm *}_u(\vec{p} )=  {\p x^\mu \over \p u}  \eps^{\pm *}_\mu(\vec{p})~,\quad
  \eps^{\pm *}_A(\vec{p} )=  {\p x^\mu \over \p x^A}  \eps^{\pm *}_\mu(\vec{p})~. }
In particular, complex conjugation does not act on the ${\p x^\mu \over \p x^{\mu'}}$ factor. 
From the relation  \rf{ddd} in Cartesian coordinates, it follows that 
\eq{an76}{  \eps^{\pm *}_r(\vec{p} )=  \eps^{\mp}_r(\vec{p})~,\quad
\eps^{\pm *}_u(\vec{p} )=  \eps^{\mp}_u(\vec{p} ) ~,\quad
\eps^{\pm *}_A( \vec{p} )=  \eps^{\mp}_A(\vec{p} )~.  }
It is also useful to strip off the factor of $r$ in the angular components by defining 
\eq{an77}{ \hat\varepsilon_A^\alpha(\hat x) = \frac{1}{r}\epsilon^\alpha_A(\hat x) ~,\quad \hat\varepsilon_A^{\alpha *} (\hat x) = \frac{1}{r}\epsilon^{\alpha *}_A(\hat x)   }
in terms of which we have the completeness relation 
\eq{BC10}{
    \hat\varepsilon_{AB}^\alpha \hat\varepsilon^{*AB}_\beta = \delta_\beta^\alpha~,
}
and we also note $\vepsh^{\pm  *}_A = \vepsh^{\mp}_A $. Another useful fact is 
\eq{aj91}{ \vepsh_{zz}^- = \vepsh_{\zb\zb}^+ = \sqrt{\gamma} = {2\over (1+z\zb)^2}~.}

From the explicit form of these polarization vectors, we have the large $r$ expression\footnote{Here and throughout, a superscript appearing as $h_{\mu\nu}^n$ denotes the coefficient of the $r^n$ term in the large $r$ expansion.}   
\eq{aj89}{ h_{AB}(r,u,\hat x)  = r h_{AB}^1(u,x^A) +  {\cal O}(r^0) }
with 
\eq{aj}{ h^1_{zz}(u,\hat x) & =  -{i\over 8\pi^2}  \vepsh^-_{zz} \int_0^\infty d\omega \big[ a_+(\omega \xh)e^{-i\omega u}  -a^\dagger_-(\omega \xh) e^{i\omega u} \big] \cr
 h^1_{\zb\zb}(u,\hat x) & =   -{i\over 8\pi^2}  \vepsh^+_{\zb\zb} \int_0^\infty d\omega \big[ a_-(\omega \xh)e^{-i\omega u}  -a^\dagger_+(\omega \xh) e^{i\omega u} \big] ~. }
The trace part vanishes at this order, $h^1_{z\zb} = 0$. 
This expression  makes it manifest that the physical mode data are packaged as 
$h^{1-}_{zz} \sim a_-^\dag$ and $h^{1-}_{\zb\zb} \sim a_+^\dag$.

With these free field expressions as motivation, we define our boundary conditions by assuming that the Cartesian components of the metric perturbation start at order $1/r$ near null infinity, which after converting to retarded coordinates implies the leading falloffs
\begin{align}
\label{g65}
&\begin{aligned}
h_{rr} &\sim r^{-2}~,\quad & h_{ur} &\sim r^{-2}~,\quad & h_{rz} &\sim h_{r\zb} &\sim  r^{-1}   \cr
h_{uu} &\sim r^{-1}~,\quad & h_{uz} &\sim h_{u\zb} \sim  r^0~,\quad & \cr
h_{zz} &\sim h_{\zb\zb} \sim r~,\quad & h_{z\zb} &\sim r^0~,\cr
\end{aligned} 
\end{align}
where we have additionally suppressed the trace of $h_{AB}$ and radial components by an additional order since the leading term constructed from the polarization \eqref{an} lacks these components --- of course these components are allowed to appear at subleading order, which is fluctuating and hence outside our control. Though we focus on the behavior near $\scrI^+$ here, we make analogous assumptions near $\scrI^-$.  Our boundary conditions on $\scrI^+$ therefore take the form 
\eq{g65zz}{ \lim_{x^\mu \rt \scrI^+} \frac{1}{r} h_{AB}(x^\mu) = \overline h^{1-}_{AB}(u, \hat x) + \text{(pos. freq.)}  }

Our working hypothesis is to assume the falloff conditions \eqref{g65} in the full nonlinear theory. In fact, the range of validity of this statement is not clear since there is the possibility of $\ln r$ enhancements due to interactions; nonetheless we will provisionally assume that such effects, if present, do not affect our  analysis.

For the matter fields --- in this paper we will focus on a massless scalar field --- we will similarly assume a $1/r$ type fall-off near null infinity,
\begin{align}
    \phi(r, u, \xh) &= \frac{1}{r} \phi_{-1}(u, \xh) + \ldots
\end{align}
with boundary conditions
\begin{align}
\label{eq:scalarData}
    \lim_{x^\mu \to \Ic^+} r \phi &= \overline \phi_{-1}^-(u, \xh) + \text{(pos. freq.)}
    \ ,
\end{align}
and vice versa on $\Ic^-$.

\subsection{Bulk action and gauge fixing}

Up to total derivatives (aka boundary terms) which are discussed below, the bulk action takes the form of the Einstein--Hilbert action, supplemented by a gauge fixing term and a corresponding ghost action, 
\eq{ACT1}{
    I_{\text{bulk}} = I_{\text{EH}} + I_{\text{gf}} + I_{\text{gh}}~.
}

As discussed  earlier, we work in background field gauge. To define this we introduce a  background metric $\overline g_{\mu\nu}$ and write
\eq{c14}{ g_{\mu\nu}  = \overline{g}_{\mu\nu} + \mathsf{h}_{\mu\nu}}
so that $\mathsf{h}_{\mu\nu}$ will be the dynamical field we integrate over inside the path integral. We take the background field to obey the asymptotic boundary conditions so\footnote{To avoid any confusion, this means that the field $h_{\mu\nu}$ discussed above is $\overline h_{\mu \nu} + \mathsf h_{\mu\nu}$.}
\eq{c15}{
    \overline g_{\mu\nu} = \eta_{\mu\nu} + \overline h_{\mu\nu}
}
where $\overline h_{\mu\nu}$ is any field configuration, for example a free field, obeying the boundary conditions \eqref{g65} on $\scrI^+$ with $h_{AB}^{1-} = \overline h_{AB}^{1-}$ held fixed, and with the corresponding conditions on $\scrI^-$. In particular, this implies that $\mathsf{h}_{AB}^{1-} = 0$ since $\overline h_{AB}$ already saturates the boundary condition.

Background field gauge is based on thinking of $ \mathsf{h}_{\mu\nu}$ as a spin-$2$ field propagating on the background field metric $\overline{g}_{\mu\nu}$.  The terms in the action are then
\eq{ACT2}{
    I_{\text{EH}} &= \frac{1}{16\pi G}\int\dr^4 x \sqrt{-g}R,\cr
    I_{\text{gf}} &= -\frac{1}{16\pi G}\frac{\xi}{2}\int\dr^4 x \sqrt{-\overline g}\overline g^{\mu\nu}C_\mu C_\nu,\ \ \ C_\lambda \equiv \overline g^{\mu\nu}\Big( \overline \nabla_\mu  \mathsf{h}_{\nu \lambda} - \frac{1}{2}\overline \nabla_\lambda \mathsf{h}_{\mu\nu} \Big),\cr
    I_{\text{gh}} &= \frac{1}{16\pi G}\int\!\dr^4 x \sqrt{-\overline{g}}\overline g^{\mu\nu} c^{*\lambda}\Big( \overline \nabla_\mu \delta_c\mathsf{h}_{\nu \lambda} - \frac{1}{2}\overline \nabla_\lambda \delta_c\mathsf{h} \Big)~.
}
Here the covariant derivative $\overline \nabla$ is the Christoffel connection of $\overline g_{\mu\nu}$, and $\delta_c \mathsf{h}_{\mu\nu}$ the action of the diffeomorphism, as defined below in \eqref{c16}, generated by the ghost vector field $c^\mu$.
The ghosts $c^\mu$ will play no role in our analysis; they are included  here for completeness but will henceforth be ignored.   We have also allowed for an arbitrary gauge fixing parameter $\xi$, though it will later be convenient to set $\xi=1$.

The virtue of this action is that, while it necessarily breaks gauge transformations acting solely on the dynamical field,
\eq{c16}{
    \delta_\xi \overline g_{\mu\nu} = 0,\ \ \
    \delta_\xi \mathsf{h}_{\mu\nu} = \nabla_\mu \xi_\nu + \nabla_\nu \xi_\mu~,
}
it is manifestly  invariant under the background gauge transformation
\eq{hd4}{ \delta_\xi \overline{g}_{\mu\nu} & =  \overline{\nabla}_\mu \xi_\nu + \overline{\nabla}_\nu \xi_\mu \cr
& =   \xi^\alpha \p_\alpha \overline{g}_{\mu\nu} +\overline{g}_{\mu\alpha} \p_\nu \xi^\alpha+ \overline{g}_{\alpha\nu}\p_\mu \xi^\alpha\cr
\delta_\xi \mathsf{h}_{\mu\nu} & = \xi^\alpha \p_\alpha \mathsf{h}_{\mu\nu} + \mathsf{h}_{\mu\alpha} \p_\nu \xi^\alpha+ \mathsf{h}_{\alpha\nu}\p_\mu \xi^\alpha }
which acts simultaneously on the background and dynamical  fields. These are just two alternative ways of breaking up the transformation of the full metric,
\eq{hd1}{
    \delta_\xi g_{\mu\nu} &= \nabla_\mu \xi_\nu + \nabla_\nu \xi_\mu\cr
    &= \xi^\alpha \p_\alpha g_{\mu\nu} + g_{\mu\alpha} \p_\nu \xi^\alpha + g_{\alpha\nu}\p_\mu \xi^\alpha
}
into parts acting on the background and the fluctuation.  

Perturbation theory is thus rendered well-defined. In the case that the background field transformation labelled by the vector field $\xi^\mu$ does not die off at infinity, it may act on our boundary conditions.   Invariance of the action under such large   background gauge transformations will  in later sections be used to deduce Ward identities.

\subsection{Boundary terms} 

The boundary terms accompanying the bulk action \eqref{ACT1} are determined by demanding a good variational principle.  Due to large $r$ falloffs and the general considerations discussed in \cite{Kraus:2025wgi}, insofar as our goal is to match with the usual S-matrix elements computed by LSZ, it is sufficient to restrict attention to the terms in the action that are  at most quadratic in fluctuations around Minkowski space.   To this end we quote the result for the expansion of the Einstein--Hilbert term, writing $g_{\mu\nu} = \eta_{\mu\nu} + h_{\mu\nu}$, and keeping all terms,
\eq{hc17}{ &\sqrt{-g}R =  - {1\over 2} \Big[ {1\over 2} \nabla_\alpha\htil_{\mu\nu} \nabla^\alpha \htil^{\mu\nu}
-{1\over 4} \nabla_\alpha \htil \nabla^\alpha\htil
-\nabla^\mu \htil_{\mu\nu} \nabla_\alpha \htil^{\nu\alpha}   \Big]+ \p^\mu\Big( \sqrt{-\eta}  \big[   \nabla^\nu h_{\mu\nu} - \nabla_\mu h  \big]\Big)  \cr
& + \p^\mu\Big( \sqrt{-\eta}  \big[ -{1\over 2} h \nabla_\mu h +  \nabla_\mu h_{\nu\alpha}h^{\nu\alpha} +  h_{\mu\nu} \nabla^\nu h + {1\over 2} \nabla^\nu h_{\mu\nu}h - {3\over 2}   h_{\mu\alpha}\nabla_\nu h^{\nu\alpha} -{1\over 2}   \nabla_\nu h_{\mu\alpha}h^{\nu\alpha}   \big] \Big) \cr
&  +O(h^3)  }
where we defined the trace reversed fluctuation 
\eq{hc17zz}{ \tilde{h}_{\mu\nu}  = h_{\mu\nu} -{1\over 2} h\eta_{\mu\nu}~, \quad  h= h^\lambda_\lambda}
and where indices are raised and lowered by $\eta_{\mu\nu}$ and its inverse. 

We note the presence of a linear term in $h_{\mu\nu}$. 
A linear term is potentially unwanted since it could contribute to a term in the partition function $Z[\overline{h}]$ at first order in $\overline{h}$, which should be absent, given that there is no ``1-point S-matrix".    Fortunately, if we evaluate this term on our  linearized solution \rf{aa}, as we do to extract the S-matrix, we see immediately that it vanishes since that solution obeys $\p^\mu \hb_{\mu\nu} = \hb{}^\mu_\mu=0$.  The partition function therefore has no term linear in $\overline{h}_{\mu\nu}$.

Now we turn to the quadratic terms.    We first of all notice that if we set $\xi=1$ in \rf{ACT2} then the gauge fixing term, expanded to quadratic order, will cancel the $\nabla^\mu \htil_{\mu\nu} \nabla_\alpha \htil^{\nu\alpha}  $ term in \rf{hc17}.  The graviton propagator will then take the  simple Feynman form.      Next we consider the total derivative terms at quadratic order, written on the second line of \rf{hc17}.  Converting these to boundary terms, it is straightforward to check all such terms either vanish under our assumed large $r$ falloffs or are total $u$  derivatives on $\Ic^+$  (or total $v$ derivatives on $\Ic^-$).   In line with our general philosophy, we disregard such total derivative terms on the boundary.  

The quadratic part of the Einstein--Hilbert action plus $\xi=1$  gauge fixing term is then 
\eq{xx99}{ \big[  I_{\rm EH} + I_{\rm gf} \big] \Big|_{h^2} = -   {1\over 32 \pi G}   \int\! d^4 x \Big[ {1\over 2} \nabla_\alpha\htil_{\mu\nu} \nabla^\alpha \htil^{\mu\nu}
-{1\over 4} \nabla_\alpha \htil \nabla^\alpha\htil \Big]~. }

We now consider an on-shell variation and demand the action be stationary under variations that respect our boundary conditions.   Computing the variation of \rf{xx99} we find an unwanted boundary term ${1\over 32\pi G } \int_{\Ic^+}\! \dr^3 x\sqrt{\gamma} \p_u \hb^{AB}_{1-} \delta h^{1+}_{AB}$, which we therefore cancel by taking\footnote{We define the measure on null infinity by $\sqrt{\gamma}\dr^3 x = \frac{2\dr u \dr z \dr\zb}{(1 + z\zb)^2}$.}
\eq{a34}{ I_{\rm bndy} & = -{1\over 32\pi G} \int_{\Ic^+} d^3x \sqrt{\gamma}  \p_u \hb_{1-}^{AB}  h^1_{AB}
~. }
Note that $h^1_{AB}$ contains both positive and negative frequency parts, and is thus local in the unfixed data   (unlike $h^{1+}_{AB}$), hence we are adding a local boundary term.

The resulting on-shell variation is 
\eq{c13}{ \delta_{\rm on-shell}  \big[  I_{\rm EH} + I_{\rm gf} + I_{\rm bndy} \big] = &{1\over 32\pi G}  \int_{\Ic^+} d^3x \sqrt{\gamma} \left( \p_u h_{1+}^{AB} \delta \hb^{1-}_{AB} - h_{1+}^{AB}\p_u\delta\overline h^{1-}_{AB}\right)\cr
- &{1\over 32\pi G} \int_{\Ic^-} d^3x \sqrt{\gamma} \left(\p_v h_{1-}^{AB} \delta \hb^{1+}_{AB} - h_{1-}^{AB}\p_v \delta \hb^{1+}_{AB}\right)~,}
where we added back in the contribution on $\Ic^-$.   Though derived from the quadratic part of the action, for reasons noted above  it holds to all orders under our assumed falloffs.   Similarly, the complete nonlinear  action is given by $I_{\rm EH}+ I_{\rm gf} + I_{\rm bndy}$ with $I_{\rm bndy}$ given by \rf{a34} (plus its counterpart on $\Ic^-$). 

So far we have not made explicit the role of Goldstone modes, which are activated by supertranslations and superrotations.  In particular, the fixed data contains both a hard part (the dynamical gravitons) and a soft part (the Goldstones),
\eq{c13a}{  \hb^{1-}_{AB} &=\hat{\hb}{}^{1-}_{AB} +C_{AB}^{\Ic^+} \cr
 \hb^{1+}_{AB} &=\hat{\hb}{}^{1+}_{AB} +C_{AB}^{\Ic^-}. }
The hatted data admits a Fourier transform to frequency space, while the Goldstone terms $C_{AB}$ are either constant (for supertranslations) or linear (for superrotations) in the null coordinates $u$ and $v$.   The supertranslation Goldstone can be treated just like the large gauge Goldstone for QED/YM \cite{Kraus:2025wgi}; in particular, as we discuss in section \ref{spatial}, we will demand that $C^{\Ic^+}_{AB}$ and   $C^{\Ic^-}_{AB}$ be antipodally related so as to maintain invariance of the action under supertranslations.

 One may note that the two terms of each line written in \eqref{c13} differ by a total $u$-derivative, so their coefficients amount to a choice of corner term. Using the $\ln u$ behavior at large $u$ of the unfixed data described in \cite{Kraus:2025wgi}, we see that this corner term only receives contributions from the Goldstones and not the hard data $\hat{\overline h}{}^{1-}_{AB}$. We discuss the particular choice of corners we will need later below \eqref{ACTGold}. A similar effect has been noted in gauge theory \cite{Kraus:2025wgi}. Though it may seem ad hoc to keep these particular corner terms while we have been so cavalier about dropping them, we shall see that these are required both to reproduce the correct soft theorems in section \ref{ward}, and then to show invariance of the partition function under supertranslations and superrotations using explicit Witten diagram computations in section \ref{diags}.

For the supertranslation Goldstone, the choice made here seems to be fixed by Lorentz invariance together with the numerical prefactor of the leading soft factor; see section \ref{Factors of 2}. The situation for superrotation Goldstones is not as well established; in particular their linear in $u$ (or $v$) behavior leads to divergent integrals which must be defined. While Poincar\'e invariance together with the leading soft theorem relates the ambiguity in these divergent integrals to the ambiguity in superrotation corner term above, an argument to fix them independently is currently lacking. Since this is a subtle point, we comment more explicitly on the interplay of these factors in section \ref{Factors of 2}.

\subsection{Boundary correlators and the S-matrix}

Now that we have arrived at an  acceptable action, the role of the boundary partition function as a generating functional for boundary correlators, and the latter's relation to the S-matrix,  follow from a  straightforward extension of the arguments previously employed  for scalar and gauge fields, so we can be brief.    To write these relations it is very convenient to set 
\eq{a20dd}{ \kappa^2  = 32\pi G =1~, }
since this renders the quadratic action canonically normalized.  We will at times restore $\kappa$. 
The expansion of  the partition function $Z[\hb]$ in terms of the fixed hard data reads,
\eq{apr56}{ Z[\hb]& =  \sum_{m,n} {1\over m! n!} \int_{\Ic^+} \int_{\Ic^-}  W^{A_1B_1 \cdots A_mB_m; C_1 D_1 \ldots C_n D_n  }(x_1,\ldots, x_m; y_1,\ldots y_n) \cr
& \quad\quad\quad\quad\quad\quad\quad \times \hat{\hb}{}^{1-}_{A_1 B_1} (x_1) \ldots  \hat{\hb}{}^{1-}_{A_m B_m} (x_m) \hat{\hb}{}^{1+}_{C_1 D_1}(y_1)  \ldots  \hat{\hb}{}^{1+}_{C_n D_n} (y_n) ~.  }
This relation defines the boundary correlators $W$.
Here $x_i$ and $y_i$ are coordinates on $\Ic^+$ and $\Ic^-$ respectively. The integration measure on $\Ic^+$ is   $\prod_{i=1}^m d^2 x_i^A du_i \sqrt{\gamma(x_i) } $ and likewise on $\Ic^-$.   Alternatively we can extract the boundary correlators  via functional derivatives, 
\eq{PART1}{
    W^{A_1B_1; \cdots; A_mB_m}(x_1,\ldots, x_m) &= \Bigg[ \frac{\delta}{\delta \hat{\overline h}{_{A_1 B_1}^{1-}}(x_1)} \cdots \frac{\delta}{\delta \hat{\overline h}{_{A_m B_m}^{1-}}(x_m)} Z \Bigg]\Bigg|_{\overline h = 0}
}
where for brevity we wrote only terms on $\Ic^+$. 

The relation of the boundary correlators $W$ to the S-matrix is simply stated in Fourier space.  
Our Fourier transform convention is $\tilde{f}(\omega) = \int_{-\infty}^\infty \! du f(u) e^{i\omega u}$.   For brevity we will only explicitly display dependence on outgoing particles.  We thus have the momentum space boundary correlators
\eq{apr91}{ &\tilde{W}^{A_1B_1 \cdots A_m B_m} (\omega_1 ,\xh_1; \ldots  ; \omega_m, \xh_m)\cr
&\qquad= \int\!\dr u_1\cdots \dr u_m W^{A_1B_1\cdots A_mB_m}(u_1,\hat x_1;\ldots;u_m,\hat x_m)e^{i\omega_1 u_1}\cdots e^{i\omega_m u_m}~.  }
Next, let 
\eq{apr88}{   \eps^{\alpha_1*}_{\mu_1\nu_1} (\omega_1 \xh_1) \ldots  \eps^{\alpha_m*}_{\mu_m \nu_m} (\omega_m \xh_m)\mathcal{A}^{\mu_1\nu_1 \ldots \mu_m \nu_m} \left(\omega_1 ,\hat{x}_1; \ldots; \omega_m ,\hat{x}_m  \right) }
be a standard momentum space amplitude, including the momentum conserving delta function, where the massless momenta have been labelled by their energy $\omega$ and direction $\xh$.  The $m$ polarization tensors can individually be positive or negative helicity. The relation between the two objects is
\eq{apr777}{ &  \vepsh^{\alpha_1}_{A_1 B_1} ( \xh_1) \cdots  \vepsh^{\alpha_m}_{A_m B_m} ( \xh_m) \tilde{W}^{A_1B_1 \ldots A_m B_m} (\omega_1 ,\xh_1; \ldots  ; \omega_m, \xh_m) \cr
& \quad  = \left[ \prod_{j=1}^m \left( {-i \omega_j \over 2\pi} \right)  \right]  \eps^{\alpha_1*}_{\mu_1\nu_1} (\xh_1) \cdots  \eps^{\alpha_m*}_{\mu_m \nu_m} (\xh_m)\mathcal{A}^{\mu_1\nu_1 \ldots \mu_m \nu_m} \left(\omega_1 ,\hat{x}_1; \ldots; \omega_m ,\hat{x}_m  \right) ~. \cr }
See \cite{Kraus:2025wgi} for a derivation of this result for Yang--Mills theory, which carries over essentially verbatim to the present context.    It is worth commenting on the origin of the prefactor $\prod_{j=1}^m \left( {-i \omega_j \over 2\pi} \right)   $.    This follows from \rf{PART1} applied to \rf{c13}. In particular, the relation
\eq{ACT5}{
    \frac{\delta}{\delta \hat{\overline h}{_{AB}^{1-}}(u, \hat x)} \ln Z  &= \frac{i}{32\pi G}\langle 2\p_u h_{1+}^{AB}(u, \hat x) \rangle_{\overline h}~,
}
shows that the operator conjugate to the ``source"    $\hat{\overline h}{}_{AB}^{1-}$   has a  $u$-derivative attached, which  turns into $-i\omega$ under Fourier transform.   The presence of the $\omega$ prefactor  has the convenient implication  that the correlators corresponding to soft emissions are finite as $\omega \rt 0$, as the soft poles in ${\cal A}$ are cancelled by the prefactor.\footnote{In the Carrollian language, one would say that \eqref{ACT5} is a matrix element of a first primary descendant operator. The improved IR properties of this object were previously noted in \cite{Mason:2023mti}.}      We also note that the  extension of \rf{apr777} to fields of other spin is the obvious one. 

We next consider variations with respect to the Goldstone modes, 
\eq{ACTGold}{
    \frac{\delta}{\delta C^0_{AB}(\hat x)}\ln Z &= \frac{i}{32\pi G}\int_{\scrI^+}\dr u \langle \p_u h^{AB}_{1+}(u, \hat x) \rangle_{\overline h}~,\cr
    \frac{\delta}{\delta C^1_{AB}(\hat x)}\ln Z &= \frac{i}{32\pi G}\int_{\Ic^+}\dr u \langle 2u\p_u h^{AB}_{1+}(u, \hat x) - \p_u (uh_{1+}^{AB}) \rangle_{\overline h}~,
}
where we have written $C_{AB} = C^0_{AB} + u C^1_{AB}$ and suppressed the antipodally matched contributions on $\scrI^-$.   The relative factor of $2$ between the first line and the $u$-integral of \rf{ACT5} will play a role in the supertranslation Ward identity.  Turning to the second line of \rf{ACTGold} we encounter a known subtlety.  The desired superrotation Ward identity is obtained by discarding the total derivative term, thereby reproducing \rf{ACT5} after multiplying the latter by $u$ and integrating.   However, as discussed in \cite{Compere:2018ylh} there is no obvious justification for dropping this term since its vanishing would require that $h_{1+}^{AB}$  fall off faster than ${1\over |u|}$, in contradiction with the existence of gravitational memory and our explicit examples in section \ref{diags}.   We will show explicitly that keeping only the first term leads to the subleading soft graviton theorem, and that this is essentially fixed by Poincar\'e invariance in section \ref{poincare}, but a better understanding of why this works is clearly desirable.

In the language of generating functions, symmetries manifest as invariances of the partition function. For example if
\eq{ACT6}{
    Z[\overline\phi_- + \delta\overline\phi_-, \overline\phi_+ + \delta\overline\phi_+] = Z[\overline\phi_-, \overline\phi_+]
}
for some infinitesimal transformations $\delta\overline\phi_\pm$ of the fixed data, then we may write
\eq{c28qr}{  \int   \delta \phib_-(y) {\delta \over \delta \phib_- (y)}  Z[\phib_-,\phib_+]=  -   \int   \delta \phib_+(y) {\delta \over \delta \phib_+  (y)}  Z[\phib_-,\phib_+]~.}
On the one hand, expanding $Z$ as in \eqref{apr56} yields a Ward identity relating the boundary correlators. On the other hand, the variation on each side of \eqref{c28qr} defines a charge conjugate to the source (i.e.~boundary condition), so that \rf{c28qr} reads as a statement of charge conservation, $Q_{\Ic^+} = Q_{\Ic^-} $. Finally, under the usual canonical formalism, the charges generate the symmetry transformation via the Poisson bracket.

\section{Asymptotic symmetries}
\label{asymp}

Asymptotic symmetries correspond in this work to coordinate transformations that preserve the asymptotic falloffs while acting nontrivially on the fixed boundary data.  We consider supertranslations and superrotations, which together form the extended BMS algebra \cite{Barnich:2009se}. The story for supertranslations is straightforward, while for the superrotations, described in terms of meromorphic vector fields on the celestial sphere, we encounter well-known subtleties associated to the singularities of these functions.  In particular, the asymptotic falloff conditions are not  preserved at the poles of the celestial sphere.   Nonetheless, for some purposes one can largely ignore this complication and still land on one's feet, as will be seen in what follows.

Throughout this section, we will primarily focus on the fields at $\scrI^+$, since the story at $\scrI^-$ is completely analogous. However, this statement in part assumes that the parameters of the supertranslations and superrotations are antipodally matched across spatial infinity. This assumption is well-known to be necessary to obtain the Ward identities which ultimately reproduce the leading and subleading soft theorems \cite{Strominger:2017zoo}, but from the path integral perspective at null infinity, the need for such a matching remains obscure, as discussed in \cite{Kraus:2025wgi}. Nevertheless, analysis of the variational principle at spatial infinity implies that, unless we antipodally match, operators supported at spatial infinity will be generated by our supertranslations and superrotations. To avoid these, we will always choose to antipodally match our large gauge parameters. See section \ref{spatial} for discussion on this point.

\subsection{Supertranslations}
\label{strans} 

Supertranslations are generated by vector fields whose Cartesian components go to a constant at large $r$ while preserving the fall-off conditions \rf{g65} near $\scrI^\pm$. The condition that the leading order metric \eqref{BC7} be unchanged implies that the vector field components are independent of $u$ at leading order. Further insisting that no $h_{r\mu}$ or $h_{z \zb}$ terms in the metric are generated at first subleading order leads to the following asymptotic form near $\Ic^+$, labeled by a function on the sphere,  $T(\hat x)$,
\eq{c24}{ \xi_T \approx T(\hat x)\p_u +{1\over 2 } D^A D_A T(\hat x) \p_r -{1\over r} D^A T(\hat x)  \p_A+ \cdots~, }
where $D_A$ is the covariant derivative with respect to $\gamma_{AB}$ on the sphere.
It is sometimes useful to write 
\eq{c25}{ T(z,\zb) = {p(z,\zb)\over 1+z\zb}~.}
In particular, an ordinary translation, acting on the Cartesian coordinates as $x^\mu \rt x^\mu  +b^\mu$, then corresponds to
\eq{c26}{  p(z,\zb) = b^0-b^3-(b^1-ib^2)z -(b^1+ib^2)\zb +(b^0+b^3)z\zb~. }
A supertranslation acts on the fixed asymptotic metric data as 
\eq{c27}{ \delta_T \hb^{1}_{zz} & = T(z,\zb) \p_u \hb^1_{zz} -2 D_z^2 T(z,\zb) \cr
   \delta_T \hb^{1}_{\zb\zb} & = T(z,\zb) \p_u \hb^1_{\zb\zb} -2 D_{\zb}^2 T(z,\zb)~.}
The action on asymptotic scalar field data \eqref{eq:scalarData} is
\eq{c28}{ \delta_T \phib_{-1} = T(z,\zb) \p_u \phib_{-1}~.}

\subsection{Superrotations}
\label{srots} 

The superrotation vector fields can be obtained in a similar way, extending the ansatz to allow for vector fields with Cartesian components that diverge linearly with $r$. Requiring the leading terms in the metric \eqref{BC7} to remain unchanged and the fall-off conditions \eqref{g65} to be respected leads to 
\eq{a49}{ 
\xi & = \frac{u}{2}\left(D_z Y^z+D_{\zb} Y^{\zb}\right)  \partial_u-\frac{r+u}{2}\left(D_z Y^z+D_{\zb} Y^{\zb}\right) \partial_r  \cr
&+\left[Y^z+\frac{u}{2 r}\left(Y^z-D^z D_{\zb} Y^{\zb}\right)\right] \partial_z+\left[Y^{\zb}+\frac{u}{2 r}\left(Y^{\zb}-D^{\zb} D_z Y^z\right)\right] \partial_{\zb} + \cdots
}
where $Y^z=Y^z(z)$ and $Y^{\zb}=Y^{\zb}(\zb)$. The six Lorentz Killing vectors of Minkowski space are obtained by restricting to $\p_z^3 Y^z = \p_{\zb}^3 Y^{\zb}=0$, corresponding to the six conformal Killing vectors on $S^2$.  Together with the four translation vector fields in \rf{c26} these generate the Poincar\'e algebra. 

Whereas these are the only smooth vector fields satisfying the condition $\partial_\zb Y^z = 0 = \partial_z Y^\zb$ globally on $S^2$, the extended BMS algebra is obtained by allowing any (anti)-meromorphic functions. Upon acting with such a superrotation, one finds that the asymptotic falloffs are preserved away from the poles of $Y^{z,\zb}$. However, the appearance of $\delta$-function localized terms involving $\p_{\zb}Y^z$ and $\p_z Y^{\zb}$ leads to variations that are incompatible with our assumed falloff conditions. We approach this by defining a modified superrotation transformation on the asymptotic data, defined by working out the transformation under the vector fields \rf{a49} and then simply dropping any contributions containing $\p_{\zb} Y^z $ or $\p_z Y^{\zb}$, or derivatives thereof. 
As written down momentarily, the asymptotic data will then transform under a representation of the extended BMS algebra.   The price to be paid for this maneuver is that since our modified transformation is not a true diffeomorphism, we have no reason to expect that it defines an invariance of the action, a point we will revisit later.

Under this rule we have
\eq{a50}{ 
    \delta \hb^{1}_{zz} & ={1\over 2}  D_AY^A  u\p_u    \hb^{1}_{zz} +  Y^A D_A \hb^{1}_{zz} + {1+s\over 2} D_z Y^z \hb^{1}_{zz} + {1-s\over 2} D_{\zb}Y^{\zb} \hb^{1}_{zz}    -u \p_z^3Y^z  \cr
    \delta \hb^{1}_{\zb\zb} & = {1\over 2}  D_AY^A  u\p_u    \hb^{1}_{\zb\zb} +  Y^A D_A \hb^{1}_{\zb\zb} + {1-s\over 2} D_z Y^z   \hb^{1}_{\zb\zb} + {1+s\over 2} D_{\zb}Y^{\zb}    \hb^{1}_{\zb\zb}   -u \p_{\zb}^3Y^{\zb}
}
with 
\eq{a51}{ s=2~.}
We write the transformation in this form since it generalizes naturally to the transformation of a massless spin-$s$ symmetric traceless tensor field, after dropping the inhomogeneous terms $  u \p_z^3Y^z $ and $u \p_{\zb}^3Y^{\zb}$.  In particular, for the scalar field $(s=0)$  we have
\eq{a52}{ \delta \overline{\phi}_{-1} = {1\over 2} D_A Y^A u\p_u  \phib_{-1} +Y^A D_A \phib_{-1}+ {1\over 2}D_A Y^A \overline{\phi}_{-1} ~. }

\subsection{Extended BMS transformation}

We  collect the transformations using the following notation.  Under a supertranslation corresponding to the function $T(z,\zb)$ we write the transformation as $\delta \Phi = V_{T}[T]\Phi$, where $\Phi$ denotes some asymptotic data $\overline h_{AB}$ and $\overline \phi$. (From here on, we will often suppress the label denoting the power of $r$ fall-off.)   Similarly for superrotations we write $\delta \Phi = V_{R}[Y^z]\Phi$ and $\delta \Phi = \Vb_{R}[Y^{\zb}] \Phi$. In this notation, we have for the scalar field
\eq{a53}{ V_T[T] \phib & = T \p_u \phib \cr
V_R[Y^z]\phib & = {1\over 2} D_z Y^z u \p_u \phib+ Y^z \p_z \phib +{1\over 2} D_z Y^z \phib \cr
\Vb_R[Y^{\zb}]\phib & = {1\over 2} D_{\zb} Y^{\zb} u \p_u \phib+ Y^{\zb} \p_{\zb} \phib +{1\over 2} D_{\zb} Y^{\zb} \phib~,}
and for the metric
\eq{a54}{ V_T[T] \hb_{zz} & = T \p_u \hb_{zz} -2 D_z^2 T \cr
 V_R[Y^z]\hb_{zz} &  = {1\over 2}  D_zY^z  u\p_u    \hb_{zz} +  Y^z D_z \hb_{zz} + {3\over 2} D_z Y^z \hb_{zz}    -u \p_z^3Y^z\cr
 \Vb_R[Y^{\zb}]\hb_{zz} &  = {1\over 2}  D_{\zb}Y^{\zb}  u\p_u    \hb_{zz} +  Y^{\zb} D_{\zb} \hb_{zz} - {1\over 2} D_{\zb} Y^{\zb} \hb_{zz}  }
along with
\eq{a54x}{ V_T[T] \hb_{\zb\zb} & = T \p_u \hb_{\zb\zb} -2 D_{\zb}^2 T\cr
 V_R[Y^{z}]\hb_{\zb\zb} &  = {1\over 2}  D_{z}Y^{z}  u\p_u    \hb_{\zb\zb} +  Y^{z} D_{z} \hb_{\zb\zb} - {1\over 2} D_{z} Y^{z} \hb_{\zb\zb} \cr
 \Vb_R[Y^{\zb}]\hb_{\zb\zb} &  = {1\over 2}  D_{\zb}Y^{\zb}  u\p_u    \hb_{\zb\zb} +  Y^{\zb} D_{\zb} \hb_{\zb\zb} + {3\over 2} D_{\zb} Y^{\zb} \hb_{\zb\zb}    -u \p_{\zb}^3Y^{\zb}~.  }
These transformations furnish representations of the extended BMS algebra,
\eq{a55}{ \big[V_T[T_1],V_T[T_2]\big] & =0  \cr
     \big[V_R[Y^z_1],V_R[Y^z_2]\big] &= V_R[ Y^z_1 \p_z Y^z_2- \p_z Y^z_1Y^z_2] \cr
     \big[\Vb_R[Y^{\zb}_1],\Vb_R[Y^{\zb}_2]\big] &= \Vb_R[ Y^{\zb}_1\p_{\zb} Y^{\zb}_2-\p_{\zb} Y^{\zb}_1 Y^{\zb}_2] \cr
      \big[V_R[Y^z_1],\Vb_R[Y^{\zb}_2]\big]& = 0 + {\rm pole~terms} \cr
    \big[ V_R[Y^z],V_T[T]\big] & = V_T[ Y^z \p_z p-{1\over 2} \p_z Y^z p]\cr
    \big[ \Vb_R[Y^{\zb}],V_T[T]\big] & = V_T[ Y^{\zb} \p_{\zb} p-{1\over 2} \p_{\zb} Y^{\zb} p]
}
where on the right-hand side of the last two lines we adopted the notation of \rf{c25}. 
As indicated, we in fact don't quite get the extended BMS algebra: the poles of $Y^z$ and $Y^{\zb}$ yield non-cancelling contributions in the $[V_R,\Vb_R]$ commutator.
This complication will not cause any difficulties since we will only be considering single soft particle emission and therefore we will not be using the BMS commutators in the rest of this work.  But it would be important when considering multiple, consecutive soft emissions.

\subsection{BMS invariance of the gravitational action}

To establish that the action is invariant under the BMS transformations, it is useful to consider the bulk part \eqref{ACT1} separately from the boundary terms.   Setting $32\pi G=1$, the bulk action \eqref{ACT1}
\eq{c32}{
     I_{\text{bulk}} = 2 \int\! d^4x \sqrt{-g} R  -  \int\! d^4x \sqrt{-\overline g} \overline{\nabla}^\mu \tilde{\mathsf{h}}_{\mu\nu} \overline{\nabla}_\alpha \tilde{\mathsf{h}}^{\nu\alpha} ~,
}
where $\tilde{\mathsf{h}}_{\mu\nu}$ is the trace-reversed fluctuating part of the metric, transforms under a general coordinate transformation \rf{hd4} as 
\eq{c33}{ 
    \delta_\xi  I_{\text{bulk}} & = 2\int\! d^4x \p_\mu\left(\xi^\mu  \sqrt{-g} R\right)  -  \int\! d^4x \p_\lambda \left(\xi^\lambda\sqrt{-\overline{g}} \overline{\nabla}^\mu \tilde{\mathsf{h}}_{\mu\nu} \overline{\nabla}_\alpha \tilde{\mathsf{h}}^{\nu\alpha} \right)   \cr
    &= 2 \int\! d^3x  \xi^r \sqrt{-g}  R  - \int\! d^3x \xi^r  \sqrt{-\overline{g}}  \overline{\nabla}^\mu \tilde{\mathsf{h}}_{\mu\nu} \overline{\nabla}_\alpha \tilde{\mathsf{h}}^{\nu\alpha}~,
}
with the integral taken over a surface of constant $r$, with $r$ subsequently taken to infinity.\footnote{The precise shape of the boundary surface depends on the specific way the spacetime manifold is IR regulated and the way in which this regulator is taken away. Depending on this choice, the dynamical data at null infinity can be captured on spacelike, timelike, or null surfaces. However, we expect physical statements, such as invariance of the action, to be independent of this procedure. In the present example, one can verify that different choices of boundary surfaces at $\Ic^\pm$ will lead to expressions that differ only by total $u$ (or $v$) derivatives.}
For a superrotation we have $\xi^r \sqrt{-g} \sim \xi^r \sqrt{-\overline{g}} \sim r^3$.   With our falloffs it's easy to check that $ \overline{\nabla}^\mu \tilde{\mathsf{h}}_{\mu\nu} \overline{\nabla}_\alpha \tilde{\mathsf{h}}^{\nu\alpha} \sim r^{-4}$ so the second  term  in \rf{c33} does not survive the large $r$ limit (even more strongly, the gauge fixing term vanishes on-shell). 

Next, under our falloffs one finds that $R \sim \frac1{2r^2} \p_u^2 h_{rr}^{(-2)} + O(r^{-3})$.
Discarding the total $u$ derivative, this automatically yields invariance under supertranslations, where $\xi^r \sim r^0$. For superrotations with $\xi^r \sim r$ we get a finite result in \rf{c33}.   To establish that the variation actually vanishes we need to use the field equations.  For gravity coupled to a scalar field we have $R \propto T^\mu_\mu$, where $T_{\mu\nu}$ is the stress tensor of the scalar field.   Taking the scalar field to have a potential $V(\phi) = g_3 \phi^3 + g_4 \phi^4 +\ldots$,  by using the field equations we can always bring the trace to the form $T^\mu_\mu = c_1 \phi \nabla^2 \phi + c_2 (\nabla \phi)^2 + O(\phi^4)$.   Since $\phi \sim r^{-1}$ the terms quartic or higher power in $\phi$ won't contribute.  The remaining terms   give total $u$ derivatives, since
\eq{g73}{  \phi \nabla^2 \phi \sim  (\nabla \phi)^2 \sim    r^{-3} \p_u \big( (  \phi_{-1})^2 \big) + O(r^{-4} ) ~.}
Following our general procedure, we drop these total derivative  terms on the boundary.     This same argument implies that the scalar action $I^{\text{bulk}}_{\text{scalar}} = \int\! d^4x \sqrt{-g} \big(- {1\over 2}(\nabla \phi)^2 -V(\phi)\big)$ is invariant under supertranslations and superrotations up to total $u$ derivatives.

This leaves the boundary terms.    These are readily seen to be invariant under supertranslations, up to total derivatives on the boundary. In particular, \eqref{a34} which encodes the boundary data, transforms under \eqref{c27} as
\begin{equation}
    \delta_{\xi_T} \int_{\Ic^+} d^3x \sqrt{\gamma} \, \p_u \hb_{1-}^{AB} h^1_{AB}
    = \int_{\Ic^+} d^3 x \sqrt{\gamma} \, \partial_u \left( T \p_u \hb_{1-}^{AB} h_{AB}^1 - 2 \hb_{1-}^{AB} D_A D_B T \right) \ .
\end{equation}
The story for superrotations is a bit more subtle.
We find that, up to total $u$ derivatives and total derivatives on the sphere, the AFS boundary terms in the action transform as
\begin{align}
\label{eq:deltaSR-SAFS}
    &
    \delta_{\xi_R} \int_{\Ic^+} d^3x \sqrt{\gamma} \, \p_u  \hb_{1-}^{AB} h^1_{AB}
    \\
    &= \frac14 \int_{\Ic^+} d^3 x \, (1 + z \zb)^2 \left( h_{\zb \zb}^{1+} \p_z^3 Y^z + h_{zz}^{1+} \p_\zb^3 Y^\zb \right) + \p_u(\ldots) + \p_A(\ldots)
    \ , \nonumber
\end{align}
seemingly nonzero due to the inhomogeneous part of the metric transformation under (non-Lorentz) superrotations. However, a careful analysis of the background gauge condition \eqref{ACT2}, $\overline{\nabla}^\mu \tilde{\mathsf{h}}_{\mu\nu} = 0$, and the leading order equations of motion in the $1/r$ expansion shows that these terms are total derivatives on-shell.
Indeed, we expand the $u$ component of the gauge condition to linear order in $h_{\mu\nu}$ and in $1/r$ to express $h_{uu}^1$ in terms of $h_{uA}^1$,
\begin{align}
    \nabla^\mu h_{\mu u}^1 - \frac12 \nabla_u h^1 &= 0
    & &\Rightarrow
    & h_{uu}^1 &= \frac12 (1 + z \zb)^2 (\p_\zb h_{uz}^1 + \p_z h_{u\zb}^1) + \p_u(\ldots)
    \ ,
\end{align}
and we solve the angular components to express $h_{uA}^1$ in terms of $h_{zz}^1$ and $h_{\zb \zb}^1$,
\begin{align}
    \nabla^\mu h_{\mu z}^1 - \frac12 \nabla_z h^1 &= 0
    & &\Rightarrow
    & h_{uz}^1 &= \frac14 (1 + z \zb)^2 \p_\zb h_{zz}^1 + \p_u(\ldots)
    \ .
\end{align}
Plugging this into the leading order equations of motion (which are not impacted by the matter contributions in \eqref{g73})
\begin{align}
    \nabla^2 h_{u\zb} &= 0
    & &\Rightarrow
    & \p_\zb h_{uu}^1 = \frac12 \p_\zb \left[ (1 + z \zb)^2 \p_z h_{u \zb}^1 \right] + \p_u(\ldots)
    \ ,
\end{align}
leads to the statement that 
\begin{align}
    \p_\zb^3 \left[ (1 + z \zb)^2 h_{zz}^1 \right] &= \p_u(\ldots)
    \ , &
    \p_z^3 \left[ (1 + z \zb)^2 h_{\zb \zb}^1 \right] &= \p_u(\ldots)
    \ .
\end{align}
Altogether this shows that \eqref{eq:deltaSR-SAFS} is a total derivative on-shell and concludes the argument that the action $I_{\text{bulk}}+ I_{\text{bndy}}$ is invariant under the BMS transformations, modulo potential contributions from spatial infinity, as discussed in section \rff{spatial}. 

\section{Spatial infinity and antipodal matching}
\label{spatial}

 To this point we have focused entirely on null infinity, since that is where the boundary data relevant to the Carrollian partition function lives.  However, our derivation of the Ward identities will  require invariance of the action under asymptotic symmetry transformations.  Under such transformations the action varies by boundary terms, including a boundary term at spatial infinity. This boundary term at spatial infinity must vanish for the desired Ward identities to hold. 
Another issue is that our asymptotic symmetry transformations are characterized by  their limiting form on the two components of null infinity, but we expect that the two limiting forms are correlated; were they independent we would, for example, obtain two copies of Poincar\'e symmetry.  
Instead, we expect there to exist a condition imposing the antipodal identification of the transformation parameters. In gauge theory,  a careful treatment of spatial infinity \cite{Campiglia:2017mua},  adapted to our general setup in \cite{Kraus:2025wgi},   shows that  demanding the absence of the unwanted boundary term  implies antipodal identification.      As we now discuss, there is an analogous story for gravity, at least for the case of supertranslations, though with some additional subtleties. 

To establish the needed result we can essentially follow the analysis in \cite{Capone:2022gme,Compere:2023qoa}, which in turn draws on results from  \cite{Mann:2005yr,Mann:2006bd, Mann:2008ay,Compere:2011ve}.  The outcome is that
the boundary terms at spatial infinity due to a supertranslation vanish provided that the supertranslation parameter at spatial infinity obeys a wave equation that in turn implies equality (up to antipodal reflection) of the  parameters on null infinity,
\eq{AM0}{
 T(\hat{x})\bigg|_{\mathcal{I^+}}=T(-\hat{x})\bigg|_{\mathcal{I^-}}~.
}
More precisely, this relies on some assumptions that we note below. 
A similar understanding for superrotations is more subtle and not considered herein.

We begin by writing the metric for four-dimensional asymptotically flat spacetimes in Beig-Schmidt coordinates $(\rho,\tau, {x}^A)$ \cite{Beig:1982ifu}. The metric in these coordinates takes the form

\eq{BSMet}{ds^2=\left(1+{2\sigma\over\rho}\right)d\rho^2+\left({N^a\over\rho}\right)dx^a d\rho + \rho^2 \left(h_{ab}^{(0)}+{1\over\rho}h_{ab}^{(1)}\right) dx^a dx^b +\cdots~.
}
Here, $h_{ab}^{(0)}$ is the metric for a unit hyperboloid
\eq{dsMet}{
ds^2_{\mathcal{H}}=-d\tau^2+\cosh^2{\tau}d\Omega_2^2~.
}
In Minkowski space, the metric can be written in this form by taking
\eq{AM2}{
    t=\rho \sinh{\tau},\quad r=\rho \cosh{\tau}~,
}
which covers the Rindler wedge $r^2 -t^2 \geq 0$.   Since the Beig-Schmidt patch extends to part of null infinity we can, under suitable regularity assumptions, put the asymptotically flat metric discussed in previous sections into Beig-Schmidt form at large $r$ via a coordinate change given by \rf{AM2} modified by subleading corrections; we can ignore the corrections  for the purposes of this section.  

An important role in what follows is played by a parity operation  that acts on the hyperboloid $\mathcal{H}$  as $\tau \rt -\tau$ combined with antipodal reflection on $S^2$.   

We now follow the discussion in  \cite{Compere:2011ve}. 
Assuming the falloffs in \rf{BSMet}, the Einstein-Hilbert action has a variation that logarithmically diverges  at the future and past boundaries of spatial infinity.
These boundary variations can be set to zero by imposing parity conditions on $\sigma$ and $k_{ab}$, where
\eq{kab}{k_{ab}=h_{ab}^{(1)}+2\sigma h_{ab}^{(0)}~.}
However, this choice of parity conditions results in a vanishing supertranslation charge. If instead we use the relaxed parity conditions of \cite{Compere:2011ve}, appropriate boundary terms can be added which remove the logarithmic divergence and result in an on-shell variation at the spatial hyperbolic boundary
\eq{svar}{\delta S\big|_{\rho=\Lambda}=\frac{1}{16\pi G}\int_{\mathcal{H}}d^3x\sqrt{-h^{(0)}}\sigma \delta k~.}
We define spatial supertranslations, parameterized by a scalar function on the unit hyperboloid $\omega(x^a),$ to send
\eq{spistr}{
\rho\to\rho+\omega(x^{a})+\mathcal{O}(\rho^{-1})~,\quad\quad x^a\to x^a+{1\over\rho}\mathcal{D}^{a}\omega(x^b)+\mathcal{O}(\rho^{-2})~,
}
where $\mathcal{D}^a$ is the covariant derivative defined with respect to the unit hyperboloid metric  \rf{dsMet}.
This transformation acts as  $\delta k= 2\left(\mathcal{D}^2+3\right)\omega~.$ We only consider supertranslations that are symmetries of the action and thus leave $k$ invariant. Admissible supertranslations then obey
\eq{AM1}{
    \left(\mathcal{D}^2+3\right)\omega = 0~.
}
Furthermore, following \cite{Compere:2023qoa} we will demand that  $\omega$ have odd parity.  In \cite{Compere:2023qoa} the stated motivation for this parity choice is to obtain BMS as the asymptotic symmetry algebra and nothing more.  Whether a larger symmetry algebra could be realized physically remains an interesting open question.  

We now show that the wave equation \rf{AM1} supplemented by the odd parity condition implies the antipodal identification \rf{AM0} of the supertranslation parameters on null infinity, following the logic in \cite{Campiglia:2017mua}.
In $(\rho,\tau,\theta,\phi)$ coordinates, \rf{AM1} becomes
\eq{AM4}{
    \p_{\tau}^2\omega+2\tanh{\tau}\p_\tau\omega-\frac{1}{\cosh^2{\tau}}D^2\omega-3\omega=0~,
}
where $D$ is the covariant derivative on the sphere. It is useful to now decompose $\omega$ into spherical harmonics so that
\eq{AM5}{
    \omega=\sum_{l,m}B_{l,m}(\tau)Y_{l}^{m}(\hat x)~.
}
Substituting this into \rf{AM4} yields
\eq{AM6}{
    B''_{l,m}(\tau)+2\tanh{\tau}B'_{l,m}(\tau)+\left(\frac{l(l+1)}{\cosh^2{\tau}}-3\right)B_{l,m}(\tau)=0~.
}
This can be put into a more familiar form by making the substitutions
\eq{AM7}{
    B_{l,m}(\tau)=\frac{A_{l,m}(\tau)}{\cosh{\tau}}~,\quad\quad x=\tanh{\tau}~.
}
The result is the general Legendre equation with order $\mu=2$, 
\eq{AM8}{
    (1-x^2)A''_{lm}(x)-2x A'_{l,m}(x)+\left[l(l+1)-\frac{4}{1-x^2}\right]A_{l,m}(x)=0~.
}

In the dS slicing, $\mathcal{I}^{+}_{-}\ (\mathcal{I}^{-}_{+})$ can be reached by taking $\tau\to+\infty\ (\tau\to-\infty)$. In order to identify the spatial supertranslations with the supertranslations at null infinity, we only need to consider solutions to \rf{AM8} which are nonzero in the $\tau\to \pm \infty$ limit. We further need to respect the odd parity condition on $\omega$.  The relevant solutions are
\eq{AM9}{
    B_{l,m}(\tau)=\frac{C_{l,m}}{\cosh{\tau}}Q_{l}^{2}(\tanh{\tau})~.
}
The associated Legendre functions $Q^2_l(\tanh \tau)$  have parity $(-1)^{l+1}$, while
the spherical harmonics have parity $(-1)^l$. Thus, the spatial supertranslation function has the desired odd parity,
\eq{AM10}{
 \omega(\tau,\theta,\phi) = -\omega(-\tau,-\hat x)~.
}

Writing the BMS supertranslation \rf{c24} on $\mathcal{I}^{+}$ in the coordinates \eqref{AM2} gives
\eq{AM11}{
\xi_T &\approx \left(-\sinh{\tau}\ T(\hat x)+\frac{1}{2}e^{-\tau}D^{A}D_{A}T(\hat x)\right)\p_{\rho}\cr
&\quad +\frac{1}{\rho}\left(\cosh{\tau}\ T(\hat x)+\frac{1}{2}e^{-\tau}D^{A}D_{A}T(\hat x)\right) \p_{\tau} \cr
&\quad -\frac{1}{\rho} \text{sech }\tau D^A T(\hat x)  \p_A + \cdots~.
}
Evaluating this as $\tau \rt \infty$ and comparing with \rf{spistr} we obtain the dictionary relating the supertranslation functions on $\Ic^+$ and $i^{0}$: $ T(\hat x)= - \lim_{\tau\to\infty}{\omega(\tau,\hat x)\over \sinh \tau}$.    Imposing this matching condition produces  a smooth supertranslation vector field upon passing between $\Ic^+$ and $i^{0}$.
A similar calculation  identifies the supertranslation function on
  $\mathcal{I}^-$ as $ T(\hat x)= - \lim_{\tau\to-\infty}{\omega(\tau,\hat x)\over \sinh \tau}$.  
  As $\sinh(\tau)$ has odd parity, these smoothness requirements and \rf{AM10} combine to give the antipodal matching condition  \rf{AM0}  on $T(\hat{x})$.

\section{Ward identities}
\label{ward}

Suppose the action is invariant under some asymptotic symmetry transformation.  This implies an invariance of the partition function, $\delta Z[\hb,\phib]=0$.  Expanding  $\delta Z$ in terms of boundary correlators $W$  as in \rf{apr56},  applying integration by parts to bring all derivatives onto the $W$'s,   it follows that invariance of $Z$ implies equations obeyed by the correlators.  We refer to these relations as Ward identities.  

We will be interested in two types of Ward identities.   First we have the unbroken symmetries  that act homogeneously on the sources and so imply relations for correlators with a fixed number of legs.  Here, these will be the Poincar\'e symmetries and the output is the Poincar\'e invariance of the correlators and hence of the S-matrix.   On the other hand, spontaneously broken symmetries include a piece that acts by changing the graviton source, and thus imply relations between processes with $k$ and $k+1$ external gravitons.   These correspond to soft graviton theorems. 

In this section we first discuss the Poincaré symmetries and the associated conservation laws for the S-matrix. 
We then turn to the Ward identities for supertranslations and superrotations, whose solutions will imply the leading and subleading soft graviton theorems.

\subsection{Unbroken asymptotic symmetries}

\subsubsection{Translation Ward identities}

As discussed in section \rff{strans} ordinary Poincar\'e translations act as 
\eq{c39}{ \delta_T \overline{\Phi} = T(z,\zb) \p_u \overline{\Phi} }
where $T(z,\zb)$ takes the form \rf{c25}-\rf{c26}.    The transformation \rf{c39} holds for  sources $\overline{\Phi}$ of arbitrary spin, including the graviton.    Invariance of $Z$ \eqref{apr56} under this transformation implies, upon integration by parts in $u$, 
\eq{c40}{   \sum_{i=1}^n T(z_i,\zb_i) \p_{u_i}  W^{AB\ldots}[x_1, \ldots x_n] = 0~.}
Here  $W$ can carry any collection of indices,  indicated generically by $AB\ldots$. 
We write this as
\eq{c42s}{  \sum_{i=1}^n  L_T(x_i)  W^{AB\ldots}[x_1, \ldots x_n] = 0~.}
 This implies momentum conservation of the S-matrix.
 
\subsubsection{Lorentz Ward identities}
 
Under a Lorentz transformation a spin-$s$ source transforms as described in section \rff{srots}.  We write this as 
\eq{c41}{ \delta_Y \phib^{(s)}_{z\ldots z } & =  {1\over 2}  D_AY^A  u\p_u  \phib^{(s)}_{z\ldots z }+  Y^A D_A   \phib^{(s)}_{z\ldots z } + {1+s\over 2} D_z Y^z \phib^{(s)}_{z\ldots z }+ {1-s\over 2} D_{\zb}Y^{\zb}   \phib^{(s)}_{z\ldots z }~.}
Here $\p_z^3 Y^z = \p_{\zb}^3 Y^{\zb}  =0$ so as to give a Lorentz transformation, as opposed to a more general superrotation. 
Under  the  source-operator pairing\footnote{Introduction of the operators ${\cal O}$ here is really just a notational device for  dealing with each argument of $W(x_1,\ldots , x_n)$ one at a time.} $\int\! d^3y \sqrt{\gamma} \Oc_{(s)}^{z\ldots z} \phi^{(s)}_{z\ldots z} $ this induces the Lorentz transformation of the operator 
 \eq{c43}{ \delta \Oc_{(s)}^{z\ldots z}& = -\Big[ {1\over 2} D_A Y^A u\p_u  \Oc_{(s)}^{z\ldots z}+ \left( 1 - {s\over 2}\right) D_z Y^z \Oc_{(s)}^{z\ldots z} + \left( 1 + {s\over 2}\right) D_{\zb} Y^{\zb}   \Oc_{(s)}^{z\ldots z}+ Y^A D_A \Oc_{(s)}^{z\ldots z} \Big]  }
We  write
\eq{a88}{  \delta \Oc_{(s)}^{z\ldots z} = L_Y^{(s)}  \Oc_{(s)}^{z\ldots z} ~,\quad  \delta \Oc_{(-s)}^{\zb\ldots \zb} = L_Y^{(-s)}  \Oc_{(-s)}^{\zb\ldots \zb} }
which defines the $L^{(s)}_Y$ operators via \rf{c43}.    Correlation functions thus obey
\eq{c42r}{  \sum_{i=1}^n  L^{(s_i)}_Y(y_i)  W^{z\ldots z }[y_1, \ldots y_n] = 0~,}
and similarly for the $\zb$ components.    This implies Lorentz invariance of the S-matrix.

\subsection{Spontaneously broken symmetries}

\subsubsection{Supertranslation Ward identity}

We now allow for the supertranslation function $T(\hat x)$ to be a general smooth function on the sphere.    Non-gravitational sources transform the same as for ordinary translations, 
\eq{c44hh}{  
   \delta_T \phib &= T \p_u \phib~,}
but the new feature is that the graviton sources transform inhomogeneously \eqref{c27}, 
\eq{c44ii}{  \delta_T \hb_{AB} & = T \p_u \hb_{AB} -2 \left(D_A D_B - {1\over 2} \gamma_{AB} D^C D_C\right)   T~.}
We consider terms in the partition function with and without an external graviton.  In a schematic notation the corresponding contributions are written as,
\eq{c43v}{ \ln Z & = {1\over n!} \int d^3y \sqrt{\gamma(y)} \left[ \prod_{i=1}^n d^3y_i \sqrt{\gamma(y_i)}\right]  W^{AB}[y,y_1, \ldots y_n] \hat{\hb}_{AB}(y) \phib(y_1) \ldots \phib(y_n) \cr 
&\quad  + {1\over n!} \int \left[ \prod_{i=1}^n d^3y_i \sqrt{\gamma(y_i)}\right]  W[y_1, \ldots y_n] \phib(y_1) \ldots \phib(y_n) ~. }
Invariance of $\ln Z$ then implies  
\eq{c45}{  \int\ d^3y \sqrt{\gamma(y) }  T(y^A)  D_A D_B W^{AB}[y,y_1, \ldots y_n]    = -{1\over 2} \sum_i L_T(y_i^A) W[y_1, \ldots y_n] ~,}
where we used that $W^{AB}$ is traceless and
\eq{c45ttt}{ L_T(y_i^A) = T(y_i^A) \p_{u_i}~. }
We should comment on two potentially confusing factor of two issues that are obscured by our schematic notation.  First, the top line of \rf{c43v} includes terms for graviton sources on both $\Ic^-$ and $\Ic^+$, each of which shifts according to \rf{c44ii}.   However in \rf{c45} the left-hand side is the correlator for the graviton on $\Ic^+$ only, where we have used the fact that the incoming and outgoing gravitons contribute equally to the Ward identity. Second, we know that the inhomogeneous part of \eqref{c44ii} really acts on the Goldstone, not the hard data appearing in \eqref{c43v}. From \eqref{ACT5}, the contribution from the Goldstone is simply half what we would have naively obtained by acting \eqref{c44ii} on the hard data. Together, these factors combine to form the coefficient in \eqref{c45}. See \cite{Kraus:2025wgi} for more detail.

\subsubsection{Superrotation Ward identity}

Under  superrotation a  matter source transforms as in \rf{c41}, but now allowing $\big(Y^z(z),Y^{\zb}(\zb)\big)$  to be arbitrary (anti)meromorphic functions on the sphere.   The new feature is the inhomogeneous term in the transformation of the graviton source, 
\eq{c45yyya}{  \delta \hb_{zz} & = {\rm homogeneous} -u\p_z^3 Y^z \cr
 \delta \hb_{\zb\zb} & ={\rm homogeneous} -u\p_{\zb}^3 Y^{\zb}~.   }
Imposing invariance of $\ln Z$ under the superrotation, for the term in the second line of \rf{c43v} we apply the homogeneous transformation, which after integration by parts generates the $L_Y^{(s)} $ operators defined in  \rf{c43}-\rf{a88},
 \eq{c43ppp}{ L_Y^{(s)} & = -\Big[ {1\over 2} D_A Y^A u\p_u  + \left( 1 - {s\over 2}\right) D_z Y^z  + \left( 1 + {s\over 2}\right) D_{\zb} Y^{\zb}  + Y^A D_A \Big]~.
}

Invariance of $\ln Z$ under a superrotation generated by $Y^z(z)$   (setting $Y^{\zb}=0$) implies
\eq{c50gg}{   \int\ d^3y \sqrt{\gamma(y) }  \p_z^3 Y^z(z)  u W^{zz} [y,y_1, \ldots y_n]    ={1\over 2}  \sum_i L_Y^{(s)}(y_i)  W[y_1, \ldots y_n] ~,}
where the factor of $1/2$ comes from taking into account the (equal) variations of the graviton piece on $\Ic^+$ and $\Ic^-$.
The notation here is schematic.  Each argument of $W[y_1,\ldots, y_n]$ is associated to a spin$-s_i$ particle whose indices have been suppressed, and the covariant derivative $D_A$ appearing in \rf{c43ppp} should be understood as being defined in the spin$-s_i$ representation when acting on $y_i$.

\section{Soft theorems from Ward identities}
\label{Soft-thm}

We now turn to the implications of the Ward identities for the S-matrix. The case of ordinary translations and Lorentz transformations have the obvious implications of momentum conservation and Lorentz invariance of the S-matrix, so we do not discuss them further.   The spontaneously broken supertranslations and superrotations are more interesting, as they imply the leading and subleading soft graviton theorems.  
Before discussing them in general, we first write down the simplest examples of momentum space amplitudes exhibiting leading and subleading soft behavior and extract the corresponding soft factors, as these will be useful for comparison. 
We then identify certain Green's functions on the sphere which are related to the supertranslation and superrotation Ward identities derived in subsequent sections.
Finally, we show that the Ward identities are satisfied if the appropriate S-matrix elements are related by soft theorems.

\subsection{Simple momentum space examples}
\label{mom}

To set the stage for what follows, here we write down some simple momentum space amplitudes and extract their soft graviton behavior.   Subsequent sections consider these same amplitudes in terms of the Carrollian partition function.   We use these results in the next section to verify the symmetry claims made in our abstract discussion, along with their precise links to the soft theorems.    We use the standard notation for the reduced amplitude ${\cal M}$,
\eq{ssxx}{  {\cal A}(p_i) = -i {\cal M}(p_i) (2\pi)^4 \delta^{(4)}\left(\sum p_i\right)~.}
Here our examples will focus on computing $\mathcal{M}$. As pointed out in \cite{Broedel:2014fsa}, both $\mathcal{A}$ and $\mathcal{M}$ obey the same leading and subleading soft theorems, so the delta may be ignored without loss for our purposes.  We come back to this point in section \rff{reduced}.

\subsubsection{Leading soft example}

The scalar action is
\eq{gq4}{ I_\phi = \int\! d^4x \sqrt{g} \Big( -{1\over 2}  (\nabla \phi)^2 -{\lambda\over 4!} \phi^4 \Big)~. }
Writing $g_{\mu\nu} = \eta_{\mu\nu}+ h_{\mu\nu}$ the term linear in $h_{\mu\nu}$ is
\eq{gq5}{ I_h & = \int\! d^4x \sqrt{g} \Big( {1\over 2} h^{\mu\nu} \p_\mu \phi \p_\nu \phi + {1\over 2} {\cal L}  \eta^{\mu \nu} h_{\mu\nu}   \Big) }
where the second term, proportional to the scalar Lagrangian density ${\cal L}$, can be dropped  due to tracelessness of the graviton polarization as far as the amplitude below is concerned.   The scalar $2\rt 2$ amplitude is ${\cal M}_{\phi^2 \rt \phi^2}(p_i)  =\lambda$.  Adding a graviton to the final state gives
\eq{c59}{  {\cal M}_{\phi^2 \rt \phi^2h}(q,p_i)  = {\kappa \over 2}  \eps^*_{\mu\nu}(q) \sum_i \eta_i {p_i^\mu p_i^\nu \over p_i \cdot q} {\cal M}_{\phi^2 \rt \phi^2}(p_i)  }
where $\eta_i = +1(-1)$ for incoming(outgoing) scalars.  The prefactor on the right-hand side defines the leading soft factor, which is universal.

\subsubsection{Subleading soft example}

 We take the matter action
\eq{gp1}{ I  = \int\! d^4x \sqrt{g} \Big(  -{1\over 2} (\nabla  \phi)^2 -{1\over 2}(\nabla \psi)^2 - {\lambda \over 4} \psi^2 (\nabla \phi)^2 \Big)~. }
The derivative interaction  gives subleading soft behavior coming from emission from the $\phi$ lines.     The scalar  $2\rt 2$  amplitude is
\eq{gp2}{ {\cal M}_{\psi^2 \rt \phi^2} = -  \lambda p_1 \cdot p_2 }
where $p_{1,2}$ denote the $\phi$ momenta.

Now consider the amplitude where we add  an outgoing graviton. The graviton can be emitted from the $\phi$ line, the $\psi$ line, or the 5pt vertex, the latter coming from the term in the action  $  {\lambda \over 4}\int\!d^4x \sqrt{g} \psi^2 h^{\mu\nu} \nabla_\mu \phi \nabla_\nu \phi $.   Emission from the $\psi$ line will not give any subleading behavior so we  will omit it.   We then get 
\eq{gp3c}{   {\cal M}_{\psi^2 \rt \phi^2 h} &=-\lambda \kappa \eps^*_{\mu\nu}(q) \Big[  {  (p_1+q) \cdot p_2 p_1^\mu p_1^\nu  \over 2p_1 \cdot q}  +   {p_1 \cdot (p_2+q) p_2^\mu p_2^\nu \over 2 p_2 \cdot q}  - p_1^\mu p_2^\nu \Big]~+~\psi~ {\rm emission}  }
Expanding as  $q^\mu \rt 0$ and displaying  just the $O(q^0)$ term, 
\eq{gp3}{&  {\cal M}_{\psi^2 \rt \phi^2 h}   \approx  O(q^{-1}) - \lambda \kappa \eps^*_{\mu\nu} \Bigg[  {p_2 \cdot q p_1^\mu p_1^\nu \over 2 p_1 \cdot q} +   {p_1 \cdot q p_2^\mu p_2^\nu\over 2 p_2 \cdot q}- p_1^\mu p_2^\nu  \Bigg]  + O(q)~.     }
The subleading soft operator is 
\eq{gp4}{ S_{(1)}= -{i\over 2} \kappa  \sum_{i=1}^n \frac{\epsilon^*_{\mu \nu}(q) p_i^\mu q_\lambda J_i^{\nu \lambda}}{q \cdot p_i}, \quad J_i^{\mu\nu}  =i\left(p_i^\mu \frac{\partial}{\partial p_{i \nu}}-p_i^\nu \frac{\partial}{\partial p_{i \mu}}\right)}
We verify this by applying it   to  ${\cal M}_{\psi^2 \rt \phi^2} $,
\eq{gp5}{ S_{(1)}{\cal M}_{\psi^2 \rt \phi^2}  & =- {\kappa \over 2}{1\over p_1 \cdot q} \eps^*_{\mu\nu} p_1^\mu q_\lambda  \left( p_1^\nu {\p \over \p p^1_\lambda} -  p_1^\lambda {\p \over \p p^1_\nu} \right)  \lambda p_1 \cdot p_2    + (p_1 \leftrightarrow p_2) \cr
&  =- \lambda \kappa \eps^*_{\mu\nu} \Bigg[  {p_2 \cdot q p_1^\mu p_1^\nu \over 2 p_1 \cdot q} +   {p_1 \cdot q p_2^\mu p_2^\nu\over 2 p_2 \cdot q}- p_1^\mu p_2^\nu  \Bigg]      }
which  agrees with \rf{gp3} thus verifying the subleading soft theorem.

Note that for the amplitude \rf{gp3c} the full $q$ expansion only has terms at order $q^{-1}$ and $q^0$, so the full  graviton emissions amplitude is obtained by acting with the leading and subleading soft operators on the amplitude with no graviton emission.

\subsection{Green's functions on sphere}

\subsubsection{Symmetric traceless Green's function}

Let $G_{AB}(y,y_i)$ be  a symmetric traceless tensor obeying
\eq{a72}{  D_A D_B   G^{AB} (\yh,\yh_i) =  \delta^2(\yh,\yh_i)- {\rm background~charge}~,
 }
where the delta function obeys $\int\! d^2\yh \sqrt{\gamma(\yh)} \delta^2(\yh,\yh_i) = 1$, and the background charge is added so that both sides vanish when integrated over the sphere. We take the background charge to be a delta function at infinity (i.e.~north pole of the sphere.)   This equation may be solved by choosing the ``purely electric" ansatz,
\eq{c35}{ G_{AB} = (D_A D_B - {1\over 2}\gamma_{AB} D^C D_C) \Phi~.}
and we find
\eq{a74}{ G_{zz}(\yh,\yh_i) & = {1\over 2 \pi} {1\over (1+z\zb)(1+z_i\zb_i) }{ \zb-\zb_i \over z-z_i} \cr
 G_{\zb\zb}(\yh,\yh_i) & = {1\over 2 \pi} {1\over (1+z\zb)(1+z_i\zb_i) }{ z-z_i \over \zb-\zb_i} }
The relation to the leading soft graviton factor is provided by the relation 
\eq{a77}{ G^{AB}(\hat{y} ; \hat{y}_i) & = - \frac{1}{4 \pi} \sum_{\alpha = \pm} \frac{\hat{\veps}_\alpha^{* AB}(\hat{y}) \epsilon_{\mu\nu}^\alpha(\hat{y}) n^\mu(\hat{y}_i) n^\nu(\hat{y}_i)   }{n(\hat{y}) \cdot n(\hat{y}_i)}     }   
where $n(\hat y)$ is defined as in \eqref{jj1}.

\subsubsection{Subleading soft Green's functions}

To solve the superrotation Ward identity we will need solutions of the following equations
\eq{c35r}{ \int\! d^2z \p_z^3 Y^z(z) P(z,\zb) & = 2\pi Y^z(z_i) \cr
\int\! d^2z \p_z^3 Y^z(z) Q(z,\zb) & = 2\pi  D_z Y^z(z_i)~.}
The solutions are 
\eq{c36}{ P(z,\zb) & = -{ (z-z_i)^2 \over 2 (\zb-\zb_i)} \cr
Q(z,\zb) &=     {(z-z_i)(1+z\zb_i) \over (1+z_i \zb_i) (\zb-\zb_i)}~.     }
More precisely these hold if one integrates by parts in \rf{c35r}, ignoring boundary terms, and using the standard formula $\int\! d^2z  \p_z {1\over \zb-\zb_i} f(z,\zb)= 2\pi f(z_i,\zb_i)$.
 
\subsection{Leading soft graviton theorem from the supertranslation Ward identity}

In \rf{c45} we obtained the Ward identity  
\eq{c45yyy}{  \int\ d^3y \sqrt{\gamma(y) }  T(y^A)  D_A D_B W^{AB}[y,y_1, \ldots y_n]    = -{1\over 2} \sum_i L_T(y_i^A) W[y_1, \ldots y_n] ~,}
with $L_T(y_i^A) = T(y_i^A) \p_{u_i}$.
We now show that this is satisfied if amplitudes with a soft graviton are given by the leading soft graviton theorem.
In terms of the Green's function \rf{a72} we can write \rf{c45yyy} as,
\eq{c46}{   \int\! du W^{AB}[u,\yh ;y_1, \ldots y_n]  = -{1\over 2} \sum_i G^{AB}(\yh,\yh_i)\p_{u_i} W[y_1, \ldots y_n]~.}
Note that the background charge gives no contribution by virtue of the translation Ward identity.
\rf{c46} gives  an expression for the leading soft factor via the definition
\eq{c46pp}{  \int\! du W^{AB}[u,\yh ;y_1, \ldots y_n]   = {1\over 4\pi} S_{(0)}^{AB}(\yh;u_i,\yh_i) W[y_1, \ldots y_n]~.  }

Fourier transforming from $u$ to $\omega$, the left-hand side of \rf{c46} reads 
\eq{c47q}{  \int\! du W^{AB}[u,\yh ;y_1, \ldots y_n]   = {1\over 2} \lim_{\omega \rt 0^+} W^{AB}[\omega,\yh ;y_1, \ldots y_n] ~,}
where the factor of $1/2$ comes from a careful treatment of the conditionally convergent $u$-integral, which we define using the prescription \eqref{eq:cosReg}.
Indeed, we can argue that the function $W^{AB}[u, \ldots]$ is analytic in the lower-half $u$ plane, since the conjugate operators on $\Ic^+$ are purely positive frequency. (See also the explicit examples below, such as \eqref{gq11z}.) Assuming it also falls of fast enough for large $|u|$, we can write $\int du \, W^{AB}[u, \dots] \equiv \lim_{\omega \to 0^+} \int du \, \frac12 (e^{i \omega u} + e^{-i \omega u}) W^{AB}[u, \dots]$ as the sum of two contour integrals, one for each exponential. The second vanishes due to analyticity in the lower half plane, whereas the first one gives the right-hand side of \eqref{c47q} with the correct factor of $1/2$. Similarly, for soft gravitons coming in from $\Ic^-$, only the second term contributes.
We therefore have 
\eq{c48pp}{ \lim_{\omega \rt 0^+} W^{AB}[\omega,\yh ;y_1, \ldots y_n]  = - \sum_i G^{AB}(\yh,\yh_i)\p_{u_i} W[y_1, \ldots y_n]~.}
Using \rf{a77} this translates into the standard leading soft graviton theorem 
\eq{c49}{    \eps_\alpha^{*\mu\nu} (q) {\cal A}_{\mu\nu} (q,p_1,\ldots ,p_n) \approx   {\kappa \over 2} \sum_{i=1}^n {\eps_\alpha^{*\mu\nu }(q) p^i_\mu p^i_\nu \over p_i\cdot q}  {\cal A} (p_1,\ldots ,p_n) ~,\quad q^0 \rt 0}
where the factor of $2\pi$ in $G^{AB}$ is absorbed when passing from $W^{AB}$ to ${\cal A}$ according to \rf{apr777}. 

\subsection{Subleading soft graviton theorem from the superrotation Ward identity}
\label{sub-soft-thm}

For simplicity we first consider the case of scalar amplitudes with a single external graviton, leaving the essentially trivial generalization to spinning amplitudes to section \rff{subspin}.    For this case, the superrotation Ward identity \rf{c50gg} reads (setting $s=0$ and $Y^{\zb}=0$)
\eq{c50ggz}{   \int\ d^3y \sqrt{\gamma(y) }  \p_z^3 Y^z(z)  u W^{zz} [y,y_1, \ldots y_n]    = {1\over 2} \sum_i L_Y^{(0)}(y_i)  W[y_1, \ldots y_n] ~,}
with
 \eq{c43pppz}{ L_Y^{(0)} & = -\Big[ {1\over 2} D_A Y^A u\p_u   +D_z Y^z    + Y^z D_z \Big]  }

Using \rf{c35}-\rf{c36} we rewrite \rf{c50ggz} as 
\eq{c52}{ \int\! du u   W^{zz} [y,y_1, \ldots y_n] &= {1\over 4\pi} S_{(1)}^{zz}[\yh,y_1,\ldots, y_n] W[y_1,\ldots, y_n]  }
with 
\eq{c52pp}{ 
S_{(1)}^{zz}[\yh,y_1,\ldots, y_n]  &= {1\over 2\sqrt{\gamma(\yh)} } \sum_i    { 1 \over \zb-\zb_i} \Big[ (z-z_i)^2  \p_{z_i} -{(z-z_i) (1+\zb_i z) \over 1+z_i \zb_i}  (2+ u_i\p_{u_i} ) \Big]~.   }
We now  convert  to Fourier space using 
\eq{c53}{ &    -(2+u\p_u) ~\leftrightarrow ~ \omega \p_{\omega}-1 \cr
 & \int\! du u W^{zz} [u,y^A;y_1, \ldots y_n]  = -{i\over 2} \lim_{\omega \rt 0^+} \p_\omega W^{zz} [\omega,y^A;y_1, \ldots y_n]~.}
The second relation is formal in the sense that the integral on the left-hand side is divergent. The factor of $1/2$ follows from our prescription \eqref{eq:cosReg} as follows:
\eq{c53xx}{ \int\! du u W^{zz} [u,y^A;y_1, \ldots y_n] & \equiv  \lim_{\omega \rt 0^+} \int\! du u \cos(\omega u) W^{zz} [u,y^A;y_1, \ldots y_n]  \cr
& = {1\over 2}  \lim_{\omega \rt 0^+} \int\! du u e^{i\omega u} W^{zz} [u,y^A;y_1, \ldots y_n]  \cr 
& = -{i\over 2} \lim_{\omega \rt 0^+} \p_\omega   \int\! du  e^{i\omega u} W^{zz} [u,y^A;y_1, \ldots y_n]   \cr 
&  = -{i\over 2} \lim_{\omega \rt 0^+} \p_\omega W^{zz} [\omega,y^A;y_1, \ldots y_n]    }
%
As we'll see in section \rff{poincare}  the coefficient in the subleading soft theorem is best thought of as being fixed by Poincar\'e  invariance. This arises again in our explicit computation in section \ref{diags}; see the discussion following \eqref{g112}. We then have 
\eq{c58vv}{
 &  \lim_{\omega \rt 0^+} \p_\omega W^{zz} [\omega,y^A;y_1, \ldots y_n]  =- {i\over 2\pi}   S^{zz}_{(1)}[\yh;\omega_1, y^A_1, \ldots  \omega_n ,y^A_n]  W[\omega_i, y^A_i, \ldots  \omega_n ,y^A_n] ~.   }
with 
\eq{c53kk}{  S^{zz}_{(1)}[\yh;\omega_1, y^A_1, \ldots  \omega_n ,y^A_n]  = -{1\over 2\sqrt{\gamma(\yh)} }  \sum_i  { z-z_i \over \zb-\zb_i} \Big[ (z-z_i)  \p_{z_i} +{1+\zb_i z \over 1+z_i \zb_i}  (\omega_i \p_{\omega_i} -1)\Big] ~. }
We then convert the above into a relation obeyed by the conventional amplitudes, related to the Carrollian correlators via the dictionary \rf{apr777}. 
Noting that $(\omega_i \p_{\omega_i}-1)(\omega_i {\cal A} )= \omega^2_i \p_{\omega_i} {\cal A} $  and that $\epsh_{zz} = \sqrt{\gamma}$, applying the dictionary gives the soft relation
\eq{c53ks}{  \lim_{\omega \rt 0^+} \p_\omega\Big( \omega \eps^{+*}_{\mu\nu} {\cal A}^{\mu\nu}[\yh;\omega_1, y^A_1, \ldots  \omega_n ,y^A_n] \Big) & = \sum_i S_{(1)}^+[\yh;\omega_1, y^A_1, \ldots  \omega_n ,y^A_n]  {\cal A}[\omega_1, y^A_1, \ldots  \omega_n ,y^A_n]  }
with
\eq{c54ks}{S_{(1)}^+[\yh;\omega_1, y^A_1, \ldots  \omega_n ,y^A_n] =  -{1\over 2 } \sum_i   { 1 \over \zb-\zb_i} \Big[ (z-z_i)^2  \p_{z_i} +{(z-z_i) (1+\zb_i z) \over 1+z_i \zb_i}  \omega_i \p_{\omega_i}\Big] ~.   }

\subsection{Rewriting subleading soft factor in terms of Lorentz generators}

Our next task is to rewrite \rf{c54ks} in terms of the standard subleading soft factor. 
 Writing a massless particle momentum as $p^\mu = \omega n^\mu$, with $n^\mu$ given in \rf{jj1},  Lorentz transformations act as 
\eq{c55}{ z \rt {az+b\over cz+d}~,\quad \omega  \rt { |az+b|^2+|cz+d|^2 \over 1+z\zb} ~   \omega~,\quad ad-bc=1~. }
The corresponding generators are 
\eq{c56}{
L_{-1} & =-\partial_z-\frac{\zb}{1+z \zb} \omega \partial_\omega~, & \Lb_{-1} & =-\partial_{\zb}-\frac{z}{1+z \zb} \omega \partial_\omega~, \\
L_0 & =-z \partial_z+\frac{1-z \zb}{2(1+z \zb)} \omega \partial_\omega~, & \Lb_0 & =-\zb \partial_{\zb}+\frac{1-z \zb}{2(1+z \zb)} \omega \partial_\omega~, \\
L_1 & =-z^2 \partial_z+\frac{z}{1+z \zb} \omega \partial_\omega~, & \Lb_1 & =-\zb^2 \partial_{\zb}+\frac{\zb}{1+z \zb} \omega \partial_\omega~,
 }
which obey $[L_m,L_n]=(m-n)L_{m+n}$ along with the barred version. The relation to the usual $J_{\mu\nu}$ labeling is 
\eq{c57}{ J_{12}& = -i(L_0-\Lb_0) \cr
J_{23} &= -{i\over 2} (L_1-\Lb_1-L_{-1}+\Lb_{-1}) \cr
J_{31}& =- {1\over 2} (L_1+\Lb_1+L_{-1}+\Lb_{-1}) \cr
J_{01}& =- {1\over 2} (L_1+\Lb_1- L_{-1}-\Lb_{-1}) \cr
J_{02}& = {i\over 2}( L_1-\Lb_1+L_{-1}-\Lb_{-1}) \cr
J_{03}& = -(L_0 + \Lb_0) }
The soft factor in \rf{c54ks} then takes the form 
\eq{c57pp}{  S^+_{(1)}[\yh;\omega_1, y^A_1, \ldots  \omega_n ,y^A_n]  =  {1\over 2 }\sum_{i=1}^n {1 \over \zb-\zb_i}\Big(  L^{(i)}_1 -2 zL_0^{(i)} +  z^2 L_{-1}^{(i)}\Big)~.}
The combination of generators  corresponds to $L_1^{(i)}$ transported from the origin to the point $z$.   
Finally we  can  reexpress the soft factor  in terms of $J_{\mu\nu}$ and also convert from $(\omega,z,\zb)$ space to $p^\mu$ space. This gives 
\eq{a86}{  S^+_{(1)}[\yh;\omega_1, y^A_1, \ldots  \omega_n ,y^A_n]  & =-i  {\kappa \over 2} \sum_i{ [ \epsilon^{+}_{\mu \nu}(q)]^* p_i^\mu q_\lambda J_i ^{\nu \lambda} \over q\cdot p_i}  }
which  is the standard momentum space subleading soft factor; see \rf{gp4}.  This establishes that the subleading soft factor satisfies the superrotation Ward identity.   However, we have not shown that this is the unique solution.   In the   next section we show Lorentz invariance picks this out as the unique solution given some mild assumptions. 

\subsection{Subleading soft theorem with spinning hard particles}
\label{subspin}

It is simple to extend the analysis of the subleading soft theorem to the case of hard particles with arbitrary spin, again with a single soft graviton.    We just need to replace the $L_Y^{(0)}$ generators  by the $L_Y^{(s)}$ generators in \rf{c43ppp}, which simply changes the coefficient of $D_z Y^z $  (as usual we set $Y^{\zb}=0$.)     
The solution of the Ward identity then proceeds exactly as before with the result 
\eq{a86x3}{& S_{(1)}^{zz}[\yh, y_1, \ldots , y_n]  =  {1\over 2\sqrt{\gamma(\yh)} } \sum_i    { 1 \over \zb-\zb_i} \Big[ (z-z_i)^2  D_{z_i} -{(z-z_i) (1+\zb_i z) \over 1+z_i \zb_i}  (2-s_i+ u_i\p_{u_i} ) \Big]~.}

\subsection{Comment on numerical factors}
\label{Factors of 2}

It is worth emphasizing  that we require three factors to come together correctly in order to obtain the correct coefficient in front of the soft factors \eqref{c49} and \eqref{a86}. This point is delicate, and seems to have implications for the problem of corner terms on null infinity, which we have not fixed in this work.

 The first is the factor of $\frac{1}{2}$ appearing in the conjugate operator to the supertranslation Goldstone; see the comment below \eqref{ACTGold}. This numerical factor was the result of keeping the particular corner terms in \eqref{c13} and discussed below \eqref{c13a}. On the other hand, no such factor arises for the superrotation Goldstone if we follow the rule proposed below \rf{ACTGold}.

The second is a factor of $2$ obtained by equating the contributions to the Ward identity from soft gravitons on future and past null infinity. This was used to obtain \eqref{c45} and \eqref{c50gg}, and discussed below \eqref{c45}. This is essentially equivalent to assuming that  the future/past contributions to the Goldstone conjugates in \eqref{ACTGold} are equal, and writing the conjugate only in terms of the future operator insertion.  This assumption may be verified in explicit examples, as in section \rff{diags}.

Third, we have the factor that comes from needing to define integrals which are not properly convergent, i.e. \eqref{c48pp} and \eqref{c53}. As these are not convergent integrals, they can be made to take any value. At least in the case of \eqref{c48pp}, it was argued in \cite{Kraus:2025wgi} that selecting a Lorentz-invariant regularization of the integral fixes this value. We hope that a similar argument might be able to be made for the relation \eqref{c53}.

While the second of these is non-negotiable, the first and third are where subtleties may lie. For the leading soft theorem, the factor of $\frac{1}{2}$ in the conditionally convergent integral seems to be fixed by Lorentz invariance, so reproducing the correct coefficient of the leading soft theorem fixes the prefactor in the conjugate to the supertranslation Goldstone \eqref{ACTGold}. In turn, this would seem to have immediate implications for what corner terms must ultimately appear in a more detailed analysis.

For the subleading soft theorem, since we lack a similar independent argument that \eqref{c53} is fixed by Lorentz invariance, only the product of the first and third factors is fixed by demanding we reproduce the subleading soft factor. One would require a justification for the rule \eqref{c53} independent of the soft theorems in order to deduce implications for the set of corner terms which ultimately must appear.

\section{Subleading soft factor from Poincar\'e invariance}
\label{poincare}

To recap the outcome of the previous section,
we  defined the subleading soft factor as
 \eq{c42}{  \int\! du u W^{AB}(u,\yh;u_i,\yh_i)   & =  {1\over 4\pi} S_{(1)}^{AB}(\yh;u_i,\yh_i)  W(y_i) }
where 
\eq{c42yy}{  S_{(1)}^{zz}(\yh;u_i,\yh_i)   = \epsh_+^{*zz}(\yh) S^{(1)}(y,y_i) }
and showed that the superrotation Ward identity was satisfied by taking
\eq{c42z}{ S_{(1)}^{zz}(\yh;u_i,\yh_i)  & =  {1\over 2\sqrt{\gamma(\yh)} } \sum_i  { 1 \over \zb-\zb_i} \Big[ (z-z_i)^2  \p_{z_i} -{(z-z_i) (1+\zb_i z) \over 1+z_i \zb_i}  (2+ u_i\p_{u_i} ) \Big] ~.    }
In this section we ask if we can derive \rf{c42z}   from Poincar\'e invariance and knowledge of the leading soft factor.    We expect this to be the case based on  amplitudes arguments showing that the subleading factor follows from gauge invariance and other mild assumptions, along with the tight link between gauge invariance and Poincar\'e invariance, e.g. \cite{Bern:2014vva, Broedel:2014fsa}.  For simplicity we restrict to the case of scalar matter particles.

\subsection{Lorentz invariance}
 
Lorentz invariance of the amplitudes implies (see \rf{a88})
\eq{c44g}{  \sum_i L_Y^{(0)}(y_i)  W(y_i) & =0 \cr
\Big[  L^{(\pm 2)} _Y(y) +  \sum_i L_Y^{(0)}(y_i) \Big]  W^{AB} (y,y_i) & =0    }
where we take $s=+2$ for $AB=zz$ and $s=-2$ for $AB=\zb\zb$. 

The idea is to apply the sum of scalar generators to both sides of \rf{c42}. Using \rf{c44g}  the right-hand side becomes
\eq{c45p}{  {1\over 4\pi}    \Big[  \sum_i L_Y^{(0)}(y_i) ,S^{AB}_{(1)}(\yh;u_i,\yh_i) \Big] W(y_i)    }
where we note that the soft operator is a sum of terms, each term only acting on a single $y_i$.  Now consider the left-hand side. We have
\eq{c46x}{ \sum_i L_Y^{(0)}(y_i)    \int\! du u W^{AB}(u,\yh;u_i,\yh_i)  & =    \int\! du u \sum_i L_Y^{(0)}(y_i)   W^{AB}(u,\yh;u_i,\yh_i)  \cr
&  = -      \int\! du u  L_Y^{(\pm 2)} (y)   W^{AB}(u,\yh;u_i,\yh_i)~.  }
Now, when we use the explicit form of $ L_Y^{(\pm 2)}(y) $ in \rf{c41} we can pull the terms without $u$'s outside the integral.  For the term with $u$ we integrate by parts so that $u^2 \p_u $  turns into $- 2u$.     Focusing on  $AB=zz$,  \rf{c46x}  becomes
\eq{c47}{  & \Big[ -D_A Y^A(\yh)   + \left( 1 - {2\over 2}\right) D_z Y^z(\yh)  + \left( 1 + {2\over 2}\right) D_{\zb} Y^{\zb} (\yh)  + Y^A D_A  \Big] \int\! du u W^{zz}(u,\yh;u_i,\yh_i) \cr
& ={1\over 4\pi}   \Big[ -D_A Y^A(\yh)  + 2 D_{\zb} Y^{\zb} (\yh)  +Y^A(\yh)  D^{(\yh)} _A  \Big]    S^{zz}_{(1)}(\yh;u_i,\yh_i)  W(y_i) }
Equating this to \rf{c45p} and temporarily suppressing $W$,   we have 
\eq{c48}{ &     \Big[  \sum_i L_Y^{(0)}(y_i) ,S^{zz}_{(1)}(\yh;u_i,\yh_i) \Big]  =    \Big[ -D_A Y^A(\yh)    +2 D_{\zb} Y^{\zb} (\yh)  +Y^A(\yh) D_A   \Big]   S^{zz}_{(1)}(\yh;u_i,\yh_i)~.  \cr }
We  rewrite this as
\eq{c48p}{ Y^A(\yh) D_A      S^{zz}_{(1)}(\yh;u_i,\yh_i)-
\Big[  \sum_i L_Y^{(0)}(y_i) ,S^{zz}_{(1)}(\yh;u_i,\yh_i) \Big]  =    \Big[ D_A Y^A(\yh)    -2 D_{\zb} Y^{\zb} (\yh) \Big] S^{zz}_{(1)}(\yh;u_i,\yh_i) }
Note that, in $ Y^A(\yh) D_A      S^{zz}_{(1)}(\yh;u_i,\yh_i)$,  $D_A$ is the covariant derivative acting on a  rank-$2$ tensor.

Rewriting and reinstating  $W$, 
\eq{c48q}{& Y^A(\yh) \p_A  S^{zz}_{(1)}(\yh;u_i,\yh_i)W(y_i)-
    \Big[  \sum_i L_Y^{(0)}(y_i) ,S^{zz}_{(1)}(\yh;u_i,\yh_i) \Big] W(y_i) \cr
    & \quad\quad =    \Big[ D_A Y^A(\yh)    -2 \p_{\zb} Y^{\zb} (\yh) +{4\over 1+z\zb} \big( \zb Y^z(z)+z Y^{\zb}(\zb) \big)  \Big] S^{zz}_{(1)}(\yh;u_i,\yh_i) W(y_i)~.
}
The condition \rf{c48q} is that the subleading soft factor respects Lorentz invariance.  Note that $W(y_i)$ is not an arbitrary function since it is annihilated by the sum of scalar generators.  Hence, if we strip off the $W$'s from \rf{c48q} we should only demand equality up to a sum of scalar generators.  Indeed the known solution has this property.

\subsection{Translation invariance}

We now show that  imposing translation invariance yields a relation between the leading and subleading soft factors.    Translation invariance implies that  correlators obey
\eq{c42g6}{  \sum_i L_T^0(y_i)  W(y_i) & =0 \cr
    \Big[  L^2_T(y) +  \sum_i L_T^0(y_i) \Big]  W^{zz} (y,y_i) & =0
}
where 
\eq{a89}{ L_T^{(s)} = T(y)\p_{u} ~.}
For a translation we have 
\eq{c42g4}{ T(z,\zb) = { c_{0,0}+ c_{1,0} z  + c_{0,1} \zb+ c_{1,1} z\zb \over 1+z\zb} ~. }
We start from the definitions of the leading and subleading soft factors,  
\eq{c42g5}{ \int\! du  W^{zz}(u,\yh;u_i,\yh_i)   & = {1\over 4\pi}   S_{(0)}^{zz}(\yh;u_i,\yh_i)  W(y_i)   \cr  \int\! du u W^{zz}(u,\yh;u_i,\yh_i)   & = {1\over 4\pi}   S_{(1)}^{zz}(\yh;u_i,\yh_i)  W(y_i)~. }
Hitting the left-hand side of the second line in \rf{c42g5} with $ \sum_i L_T^{(0)}(y_i)$, using \rf{c42g6}, and integrating by parts  gives
\eq{c42g7}{{\rm LHS} & =  - \int \! du  u L^{(2)}_T(y) W^{zz}(u,\yh,u_i,\yh_i)\cr
& = {1\over 4\pi}  T(y) S_{(0)}^{zz}(\yh;u_i,\yh_i)  W(y_i) ~. }
On the other hand, applying $ \sum_i L_T^0(y_i)  W(y_i)$  to the right-hand side of \rf{c42g5} gives
\eq{c42g8}{ {\rm RHS} = {1\over 4\pi}  \Big[  \sum_i L_T^0(y_i),S_{(1)}^{zz}(\yh;u_i,\yh_i)\Big] W(y_i) ~.}
Equating the two sides gives
\eq{c42g9}{   T(y) S_{(0)}^{zz}(\yh;u_i,\yh_i)  W(y_i)  =  \Big[  \sum_i L_T^{(0)} (y_i),S_{(1)}^{zz}(\yh;u_i,\yh_i)\Big] W(y_i) }
yielding the desired relation between the two soft factors.   

More explicitly,
\eq{jj1x}{   T(y) S_{(0)}^{zz}(\yh;u_i,\yh_i)  W(y_i)  =  \Big[ \sum_i T(y_i) \p_{u_i} ,S_{(1)}^{zz}(\yh;u_i,\yh_i)\Big] W(y_i)~. }
Now, from the relation between \rf{c46} and \rf{c46pp} $S_{(0)}^{zz}(\yh;u_i,\yh_i)$ is given by
\eq{a90}{  S_{(0)}^{zz}(\yh;u_i,\yh_i)  &=-2\pi \sum_iG^{zz}(\hat{y},\hat{y}_i) \p_{u_i}\cr 
&=-{1\over 4} \sum_i  { (1+y\yb)^3\over 1+y_i\yb_i } {y-y_i\over \yb-\yb_i }  \p_{u_i} ~. }
Using this \rf{jj1x} becomes 
\eq{jj2}{  -{1\over 4} T(z,\zb)   { (1+z\zb)^3 (z-z_i) \over (1+z_i\zb_i)(\zb-\zb_i)} \p_{u_i}W(y_i)   =  \Big[ T(y_i) \p_{u_i} ,S_{(1)}^{zz}(\yh;u_i,\yh_i)\Big] W(y_i) }

\subsection{Solving for the subleading soft factor}

We now solve \rf{jj2} under a suitable ansatz.  First, dimensional analysis implies that $u_i$ and $\p_{u_i}$ must appear in the combination $u_i \p_{u_i}$.    We expect $S_{(1)}^{zz}$ to involve at most first derivatives given that only these appear on the left-hand side of \rf{jj2}.    We thus have an initial ansatz 
\eq{jj3x}{ S^{zz}_{(1)} = f_1(y,y_i)\p_{z_i}  +f_2(y,y_i)u_i \p_{u_i} +f_3(y,y_i) \p_{\zb_i} + f_4(y,y_i) ~. }
This ansatz is unwieldy to work with, so we impose an additional condition.  In momentum space
the soft factor is by design a first order differential operator.
Recall from \rf{c53}  that $\omega \p_\omega \leftrightarrow -(2+u\p_u)$, so we assume that $f_4 =2 f_2$.

In this case we get
\eq{jj4x}{
&-\frac14 T(z,\zb) \frac{(1+z\zb)^3 (z-z_i)}{(1+z_i\zb_i)(\zb-\zb_i)} \p_{u_i}W(y_i) \\
&= \left[ f_2(y,y_i) T(y_i) - f_1(y,y_i) \p_{z_i} T(y_i) - f_3(y, y_i) \p_{\zb_i} T(y_i) \right] \p_{u_i} W(y_i)
}
and we recall
\eq{jj5}{ T(z,\zb) = { c_1+ c_2 z  + c_3 \zb+ c_4 z\zb \over 1+z\zb}~.  }
Recall that we only need to satisfy this up to a term proportional to the translation Ward identity $\tilde{T}(y_i) \p_{u_i} W(y_i)=0$, where $\tilde{T}$ is a function of the form \rf{jj5}. In other words we need to solve
\eq{jj6}{  T(y_i) f_2(y,y_i) - f_1(y,y_i) \p_{z_i} T(y_i) - f_3(y, y_i) \p_{\zb_i} T(y_i) =  -{1\over 4} T(z,\zb)   { (1+z\zb)^3 (z-z_i) \over (1+z_i\zb_i)(\zb-\zb_i)} + \tilde{T}(y_i)~. }
Note that $\tilde{T}(y_i)$ is allowed to depend on $(z, \zb)$ and $T(y_i)$, but $f_1$ and $f_2$ are fixed functions not depending on the parameters in $T(y_i)$.

We now sketch the solution of this equation. Given the form \eqref{jj5} of $T(z, \zb)$, we see that $\tilde T$ in \eqref{jj6} has to take the form $(d_1 + d_2 z_i + d_3 \zb_i + d_4 z_i \zb_i) / (1 + z_i \zb_i)$ where the coefficients $d_i$ are allowed to depend on $(z, \zb)$ and can at most be linear in one of the $c_i$. 
Equation \eqref{jj6} then leads to a system of algebraic equations, as the coefficient of each of the $c_i$ and each power of $z_i$ and $\zb_i$ has to vanish separately.
The result is the following expression,
\begin{align}
\label{jj7}
    f_1 &= \frac14 (1 + z \zb)^2 \frac{(z - z_i)^2}{\zb - \zb_i} - p_3 + (p_1 - p_4) z_i + p_2 z_i^2
    \ , \nonumber \\
    f_2 &= -\frac14 (1 + z \zb)^2 \frac{1 + z \zb_i}{1 + z_i \zb_i} \frac{z - z_i}{\zb - \zb_i} + \frac{p_1 + p_5 + (p_2 + p_6) z_i + (p_3 + p_7) \zb_i + p_4 z_i \zb_i}{1 + z_i \zb_i}
    \ , \nonumber \\
    f_3 &= -p_6 + p_5 \zb_i + p_7 \zb_i^2
    \ ,
\end{align}
where $p_n=p_n(z,\zb)$ are free functions.  We observe  that the   first term in each line reproduces the known soft factor in \rf{c42z}. 
The corresponding $\tilde{T}$ function is
\begin{align}
\label{j8}
    (1 + z_i \zb_i) \tilde T 
    &= \frac14 (1 + z \zb)^2 (z - z_i) (c_3 + c_4 z) + (p_1 + p_5) c_1 + p_3 c_2 + p_6 c_3
    \nonumber  \\
    &\quad + \left[ p_2 c_1 + (p_4 + p_5) c_2 + p_6 c_4 \right] z_i
    \nonumber \\
    &\quad  + \left[ p_7 c_1 + p_1 c_3 + p_3 c_4 \right] \zb_i + \left[ p_7 c_2 + p_2 c_3 + p_4 c_4 \right] z_i \zb_i
    \ .
\end{align}
Now consider the  free parameters $p_n$.    It's easy to check that the parts proportional to $p_2, p_3, p_5, p_6, p_7$ and $p_1=-p_4$ give Lorentz generators.  Hence these contributions to the soft factor annihilate the scalar amplitude so can be dropped.  The only thing to consider is then the $p_1=p_4$ combination which has $f_1=0$ and $f_2=1$, corresponding to   $\Delta S_1^{zz} = a(z,\zb) (2+u_i\p_{u_i})$. Indeed it's clear that if we have a solution of \rf{jj6} then we can shift $f_2$ by a $y_i$ independent value and still satisfy the equation after shifting $\tilde{T}(y_i)$.
Since $f_2$ has spin $2$ the general ambiguity  is really 
\eq{a92}{ \Delta S_1^{zz} = a(z\zb)z^2 (2+u_i \p_{u_i})}
for some unfixed $a(z\zb)$.   However, by substituting this into  the Lorentz Ward identity \rf{c48q} it's straightforward to deduce that $a(z\zb)=0$. 

This completes the solution. 
Given our ansatz, we have thus succeeded in deriving the subleading soft factor from Poincar\'e invariance and the leading soft theorem.

\section{Verification of Ward identities via Witten diagram contributions to the Carrollian partition function} 
\label{diags}

In this section we compute contributions to the partition function that capture the leading and subleading soft graviton effects, and we use this to perform an explicit check of our Ward identities.  This proceeds in a straightforward fashion for supertranslations, while for superrotations we encounter some divergent integrals that we define using the rule \eqref{eq:cosReg}.

We calculate contributions to the partition function using flat space Witten diagrams, analogous to how one computes boundary correlators in AdS.  Namely we evaluate Witten diagrams using the appropriate bulk-boundary and bulk-bulk propagators; we start by reviewing the form of these propagators.

\subsection{Propagators}

The scalar bulk-boundary propagator is defined by writing a free field solution in the form  
\eq{tta1}{ \phib(x) = \int_{\scrI^+}\dr^3 y \sqrt{\gamma} K_+(x; y) \phib_{-1}^-(y) + \int_{\scrI^-}\dr^3 y \sqrt{\gamma} K_-(x; y) \phib_{-1}^+(y)
}
and demanding that it obey the asymptotic boundary conditions 
\eq{tta2}{  \phib(x) = \left\{ \begin{array}{cc}  {1\over r} \phib_{-1}^-(u,\hat x) + {\rm O}(r^{-2}) &  {\rm as}~~  x^\mu \rt \Ic^+ \cr  {1\over r} \phib_{-1}^+(v,\hat x) + {\rm O}(r^{-2}) & {\rm as}~~  x^\mu \rt \Ic^- \end{array} 
\right.~.    }
These conditions are satisfied by
\eq{gq1a}{
\begin{tikzpicture}[baseline=-3pt]
	\draw (0,1) -- (1,0) -- (0,-1) -- (-1,0) -- cycle;
	\coordinate (v) at (0, 0);
	\filldraw (v) circle (0.02);
    \draw[dashed] (v) node[left] {$x$} -- (.5, .5) circle (0.02) node[above right] {$y$};
\end{tikzpicture}
&= K_{+}\left(x ; y\right)=\frac{i}{(2 \pi)^2} \frac{1}{\left(u_y + n\left(\hat{y}\right) \cdot x-i \epsilon\right)^2} \\
\begin{tikzpicture}[baseline=-3pt]
	\draw (0,1) -- (1,0) -- (0,-1) -- (-1,0) -- cycle;
	\coordinate (v) at (0, 0);
	\filldraw (v) circle (0.02);
    \draw[dashed] (v) node[left] {$x$} -- (.5, -.5) circle (0.02) node[below right] {$y$};
\end{tikzpicture}
&= K_{-}\left(x ; y\right)=\frac{i}{(2 \pi)^2} \frac{1}{\left(v_y + n\left(-\hat{y}\right) \cdot x+i \epsilon\right)^2}~,  }
as shown in \cite{Kraus:2024gso}. Similarly,  a solution of the linearized Einstein equations obeying  
\eq{tta3}{  \hb_{AB}(x) = \left\{ \begin{array}{cc}  r\hb_{AB,1}^-(u,\hat x) + {\rm O}(r^{0}) &  {\rm as}~~  x^\mu \rt \Ic^+ \cr  r\hb_{AB,1}^+(v,\hat x) + {\rm O}(r^{0}) & {\rm as}~~  x^\mu \rt \Ic^- \end{array} 
\right.    }
 may be written 
\eq{PART6}{
    \overline h_{\mu\nu}(x) = \int_{\scrI^+}\dr^3 y \sqrt{\gamma} K_{\scrI^+\mu\nu}^{AB}(x; y) \overline h_{AB}(y) + \int_{\scrI^-}\dr^3 y \sqrt{\gamma} K_{\scrI^-\mu\nu}^{AB}(x; y) \overline h_{AB}(y)
}
with
\eq{tta4}{
\begin{tikzpicture}[baseline=-3pt]
	\draw (0,1) -- (1,0) -- (0,-1) -- (-1,0) -- cycle;
    \coordinate (o1) at (.5, .5);
	\coordinate (v) at (0, 0);
    \filldraw (o1) circle (0.02) node[above right] {$y$};
	\filldraw (v) circle (0.02) node[left] {$x$};
    \draw[graviton] (v) -- (o1);
\end{tikzpicture}
= K_{+\mu\nu}^{AB}(x; y) &= \epsilon^\alpha_{\mu\nu}(\hat y)\hat\varepsilon^{*AB}_\alpha(\hat y) K_{+}\left(x ; y\right)  ~,\cr 
\begin{tikzpicture}[baseline=-3pt]
	\draw (0,1) -- (1,0) -- (0,-1) -- (-1,0) -- cycle;
	\coordinate (v) at (0, 0);
    \coordinate (i1) at (.5, -.5);
	\filldraw (v) circle (0.02) node[left] {$x$};
    \filldraw (i1) circle (0.02) node[below right] {$y$};
    \draw[graviton] (v) -- (i1);
\end{tikzpicture}
= K_{-\mu\nu}^{AB}(x; y) &= \epsilon^\alpha_{\mu\nu}(-\hat y)\hat\varepsilon^{*AB}_\alpha(-\hat y)K_{-}\left(x ; y\right)~.
}
We note that $\eta^{\mu\nu} K_{+\mu\nu}^{AB}=0$ and $\nabla^\mu_x  K_{+\mu\nu}^{AB}=0$ by virtue of $\eta^{\mu\nu}\epsilon_{\mu\nu} = \eps(\hat{x})\cdot n(\hat{x})=0$ and the form of the scalar propagator.

The scalar bulk-bulk propagator is the usual Feynman propagator.  Its momentum space form, along with that of the scalar bulk-boundary propagator is
\eq{gq7}{
G_F\left(x ; x_0\right) & =-i \int \frac{d^4 p}{(2 \pi)^4} \frac{e^{i p \cdot\left(x-x_0\right)}}{p^2-i \epsilon} \\
K_{ \pm}\left(x ; y\right) & =-\frac{i}{(2 \pi)^2} \int_0^{\infty} d \omega \omega e^{\mp i \omega\left(u_y+n\left( \pm \hat{y} \right) \cdot x \mp i \epsilon\right)}~.
}
The momentum space versions will be used in the computations that follow. 

\subsection{Verification of supertranslation Ward identity}

We take the bulk action to be \rf{gq4} and compute the $\phi \phi \rt \phi \phi $ and $\phi \phi \rt \phi \phi  h$ contributions to the partition function at tree level 

The pure scalar diagram gives
\eq{gp8}{
W_{\phi \phi \rt \phi \phi }(y_1,y_2,y_3,y_4) 
&= 
\begin{tikzpicture}[baseline=-3pt]
	\draw (0,1) -- (1,0) -- (0,-1) -- (-1,0) -- cycle;
	\coordinate (v) at (0, 0);
	\filldraw (v) circle (0.02);
	\draw (v) -- (-.5, .5) circle (0.02) node[above left] {$y_1$};
	\draw (v) -- (.5, .5) circle (0.02) node[above right] {$y_2$};
	\draw (v) node[left] {$x_0$} -- (-.5, -.5) circle (0.02) node[below left] {$y_3$};
	\draw (v) -- (.5, -.5) circle (0.02) node[below right] {$y_4$};
\end{tikzpicture}
\nonumber \\
& = -i\lambda \int\! d^4x_0 \sqrt{g_0} K_-^\phi (x_0;y_3) K_-^\phi (x_0;y_4) K^\phi_+(x_0,y_1)  K^\phi_+(x_0,y_2)~, }
and contributes to the partition function as 
\eq{gp8z}{  \ln Z_{\phi \phi \rt \phi \phi } & = {1\over 4}  \int\! \left( \prod_{i=1}^4d^3 y_i  \sqrt{\gamma(y_i)} \right)  W_{\phi \phi \rt \phi \phi }(y_1,y_2,y_3,y_4)  \phib_-(y_1)  \overline{\phi}_-(y_2)  \phib_+(y_3)  \overline{\phi}_+(y_4) ~.}

Next we have the diagrams for graviton emission.   There are four diagrams corresponding to emission from any of the four scalar legs, though we only write out a single diagram for emission from an outgoing $\phi$ line, 
\eq{gq6}{
W^{AB}_{\phi \phi \rt \phi \phi  h}
&=
\begin{tikzpicture}[baseline=-3pt]
	\draw (0,1) -- (1,0) -- (0,-1) -- (-1,0) -- cycle;
	\coordinate (v1) at (0, 0);
	\coordinate (v2) at (.25, .25);
	\filldraw (v1) circle (0.02);
	\filldraw (v2) circle (0.02);
	\draw[graviton] (v2) -- (-.25, .75) node[above] {$y$};
	\draw (v1) -- (-.5, .5) circle (0.02) node[above left] {$y_1$};
    \draw (v2) -- (.5, .5) circle (0.02) node[above right] {$y_2$};
	\draw (v1) -- (v2) node[right] {$x$};
	\draw (v1) node[left] {$x_0$} -- (-.5, -.5) circle (0.02) node[below left] {$y_3$};
	\draw (v1) -- (.5, -.5) circle (0.02) node[below right] {$y_4$};
\end{tikzpicture}
+ 3~{\rm more} \nonumber \\
& = \int\! d^4x_0\sqrt{g_0} \, d^4x\sqrt{g} K_+(x_0;y_2)K_-(x_0;y_3)K_-(x_0;y_4) \cr
&\qquad \times \nabla_x^\mu G_F(x,x_0) \nabla_x^\nu K_+(x,y_1)  K_{+\mu\nu}^{AB}\left(x ; y\right) \cr
& \quad + 3~{\rm more} }
This is evaluated to give  
\eq{gq11z}{ W^{AB}_{\phi \phi \rt \phi \phi  h} &=   {\lambda\over 2 (2\pi)^2}\epsh^{*AB}(\yh){[\eps(\yh)\cdot n(\yh_1)]^2\over n(\yh_1)\cdot n(\yh)}\cr
& \quad\quad \times \int\! d^4x_0 \sqrt{g_0}K_+(x_0;y_2)K_-(x_0;y_3)K_-(x_0;y_4)   {\p_{u_1} K_+(x_0,y_1)\over u_y+n(\yh)\cdot x_0-i\eps} \cr
& \quad + 3~{\rm more}  }
One of the ``three more" corresponds to simply interchanging the outgoing scalar legs via $y_1 \leftrightarrow y_2$.   The other two involve incoming scalar legs, and involve some sign flips and such.

Using \rf{a77},
we can write  
\eq{a96}{ W^{AB}_{\phi \phi \rt \phi \phi  h}  &=   -{\lambda \over 2\pi}G^{AB}(\yh,\yh_1)\int\! d^4x_0  \sqrt{g_0}K_+(x_0;y_2)K_-(x_0;y_3)K_-(x_0;y_4)   {\p_{u_1} K_+(x_0,y_1)\over u_y+n(\yh)\cdot x_0-i\eps} \cr
& \quad + 3~{\rm more}~.  }
The contribution to the partition function is 
\eq{a97}{ Z_{\phi \phi \rt  \phi \phi h} = {1\over 4} \int d^3y \sqrt{\gamma(y)}\left( \prod_{i=1}^4d^3 y_i  \sqrt{\gamma(y_i)} \right) W^{AB}_{\phi \phi \rt \phi \phi  h}(y,y_i) \hat{{\hb}}^-_{AB}(y)\phib_-(y_1)  \overline{\phi}_-(y_2)  \phib_+(y_3)  \overline{\phi}_+(y_4)~.  }
To this we must add the contribution from graviton absorption, $Z_{\phi \phi h \to \phi \phi}$,  which takes an analogous form and will   contribute equally to the Ward identity.    Note that we have included a hat on the graviton source, defined via the decomposition  of $\hb_{AB}$ in \rf{c13a}  into  the Fourier transformable (hatted) part and the Goldstone part $C_{AB}$.  

Now we apply a supertranslation.  We focus on the inhomogeneous part, which will cancel the pure scalar variation.    A supertranslation shifts the Goldstone as  $\delta C^{\Ic^+}_{AB}(y) = -2D_A D_B T(y) - {\rm trace}$.  As follows from  \rf{ACTGold}   the resulting variation of $Z$ is equal to half that of the variation  under  $\delta \hat{\hb}{}^{1-}_{AB}(y) = -2D_A D_B T(y) - {\rm trace}$; see the discussion following \rf{c45ttt}. 

Substituting in the source  variation to get an expression for $\delta Z_{\phi \phi \rt  \phi \phi h }$, we can first of all perform the $du_y$ integral using 
\eq{a98}{ \int^\infty_{-\infty}\! {du_y \over u_y +n(\hat{y})\cdot x_0 -i\eps} = i\pi~,}
which follows from the rule \eqref{eq:cosReg}.
Using this, and integrating by parts on the sphere, we have
\eq{a99}{ \delta Z_{\phi \phi \rt  \phi \phi h} & =- {1\over 2}\cdot {i \lambda  \over  4}  \int\! d^4x_0  \sqrt{g_0} \, d^2y \sqrt{\gamma(y)} \prod_{i=1}^4d^3 y_i \sqrt{\gamma(y_i)}  T(y) D_A D_B G^{AB} (\yh,\yh_i) \cr
& \quad \times K_+(x_0;y_2)K_-(x_0;y_3)K_-(x_0;y_4)  \p_{u_1} K_+(x_0,y_1)\cr
&  + 3  {\rm ~more}~,  }
where the ${1\over 2}$ came from the argument in the previous paragraph. 
Using \rf{a72} we can perform the $d^3y$ integral to get, after integrating by parts in $u_i$,  
\eq{b2}{\delta Z_{\phi \phi \rt  \phi \phi h}  &= {1\over 2} \cdot {i\lambda \over 4} \int\! d^4x_0  \sqrt{g_0} \, \prod_{i=1}^4d^3 y_i \sqrt{\gamma(\yh_i)} 
  K_+(x_0;y_2)K_-(x_0;y_3)K_-(x_0;y_4)  K_+(x_0,y_1) \cr
& \quad \times \left[  T(y_1) \p_{u_1} + \ldots +  T(y_4) \p_{u_4} \right]  \phib_-(y_1) \ldots \overline{\phi}_-(y_4)~.  }
The total variation is  a sum of contributions from ingoing and outgoing gravitons,
\eq{b2qqq}{\delta Z_{\phi \phi \rt  \phi \phi h}  +\delta Z_{\phi \phi h \rt  \phi \phi }  = 2  \delta Z_{\phi \phi \rt  \phi \phi h} ~. }
The result then  precisely cancels the supertranslation variation of $Z_{\phi \phi \to \phi \phi}$, thus verifying supertranslation invariance of the partition function. 

\subsection{Verification of superrotation Ward identity}
\label{Verify SRot}

We now start from the bulk action \rf{gp1}.   The pure scalar diagram is 
\begin{align}
\label{gp8xf}
& W_{\psi \psi  \rt \phi \phi}(y_1,y_2,y_3,y_4) 
= 
\begin{tikzpicture}[baseline=-3pt]
	\draw (0,1) -- (1,0) -- (0,-1) -- (-1,0) -- cycle;
	\coordinate (v) at (0, 0);
	\filldraw (v) circle (0.02);
	\draw[dashed] (v) -- (-.5, .5) circle (0.02) node[above left] {$y_1$};
	\draw[dashed] (v) -- (.5, .5) circle (0.02) node[above right] {$y_2$};
	\draw (v) node[left] {$x_0$} -- (-.5, -.5) circle (0.02) node[below left] {$y_3$};
	\draw (v) -- (.5, -.5) circle (0.02) node[below right] {$y_4$};
\end{tikzpicture}
\\
& = -i \lambda \int\! d^4x_0 \sqrt{g_0} K_-^\psi (x_0;y_3) K_-^\psi (x_0;y_4)g^{\mu\nu}(x_0) \nabla^{x_0}_\mu K^\phi_+(x_0,y_1)  \nabla^{x_0}_\nu K^\phi_+(x_0,y_2) 
\nonumber
\end{align}

Next we have the diagram describing graviton emission from the $\lambda $  vertex, corresponding to the term in the action 
\eq{gp9}{ I_v = {\lambda \over 4}\int\!d^4x \sqrt{g} \psi^2 h^{\mu\nu} \nabla_\mu \phi \nabla_\nu \phi~. }
The diagram gives 
\begin{align}
\label{gp10x}
& W_{\psi \psi  \rt \phi \phi h}^{\text{ver};AB}(y;y_i)
=
\begin{tikzpicture}[baseline=-3pt]
	\draw (0,1) -- (1,0) -- (0,-1) -- (-1,0) -- cycle;
	\coordinate (v) at (0, 0);
	\filldraw (v) circle (0.02);
	\draw[graviton] (v) -- (-.25, .75) node[above] {$y$};
	\draw[dashed] (v) -- (-.5, .5) circle (0.02) node[above left] {$y_1$};
    \draw[dashed] (v) -- (.5, .5) circle (0.02) node[above right] {$y_2$};
	\draw (v) node[left] {$x_0$} -- (-.5, -.5) circle (0.02) node[below left] {$y_3$};
	\draw (v) -- (.5, -.5) circle (0.02) node[below right] {$y_4$};
\end{tikzpicture}
\\
& =i \lambda \int\! d^4x_0 \sqrt{g_0} K_-^\psi (x_0;y_3) K_-^\psi (x_0;y_4) K^{AB}_{+\mu\nu}(x_0;y) \nabla^\mu_{x_0} K^\phi_+(x_0,y_1)   \nabla^\nu_{x_0} K^\phi_+(x_0,y_2)
\end{align}
We will not need to perform  the  integration over the vertex location $x_0$  so we  leave the result in this form.

Finally we have emission from the $\phi$ legs (the contribution from the $\psi$ legs satisfies the Ward identity independently and so will be omitted from the discussion).   The diagram is 
\begin{align}
\label{gp11}
&W_{\psi \psi  \rt \phi \phi h}^{\phi;AB} (y;y_i)
=
\begin{tikzpicture}[baseline=-3pt]
	\draw (0,1) -- (1,0) -- (0,-1) -- (-1,0) -- cycle;
	\coordinate (v1) at (0, 0);
	\coordinate (v2) at (.25, .25);
	\filldraw (v1) circle (0.02);
	\filldraw (v2) circle (0.02);
	\draw[graviton] (v2) -- (-.25, .75) node[above] {$y$};
	\draw[dashed] (v1) -- (-.5, .5) circle (0.02) node[above left] {$y_1$};
    \draw[dashed] (v2) -- (.5, .5) circle (0.02) node[above right] {$y_2$};
	\draw[dashed] (v1) -- (v2) node[right] {$x$};
	\draw (v1) node[left] {$x_0$} -- (-.5, -.5) circle (0.02) node[below left] {$y_3$};
	\draw (v1) -- (.5, -.5) circle (0.02) node[below right] {$y_4$};
\end{tikzpicture}
\cr
& = \lambda   \int\! d^4x_0\sqrt{g_0} \, d^4x \sqrt{g} K_-^\psi (x_0;y_3) K_-^\psi (x_0;y_4) \nabla^{x_0}_\sigma K^\phi_+(x_0,y_2) \nabla_{x_0}^\sigma \nabla^\mu_x G_F(x,x_0) 
\nonumber \\
&\qquad \times \nabla^\nu_x K^\phi_+(x,y_1)K^{AB}_{\mu\nu}(x,y) \cr
& \quad +  (y_1 \leftrightarrow y_2)~.
\end{align}
Writing the propagators in Fourier space and performing the integration over the emission point $x$ gives 
\eq{gp18}{& W_{\psi^2 \rt \phi^2 h}^{\phi;AB} (y;y_i) =\cr
&  -\lambda  {\epsh^{*AB}(\yh)\over 2 (2\pi)^2} {[\eps(\yh)\cdot n(\yh_1)]^2\over n(\yh_1)\cdot n(\yh)}  \int\! d^4x_0\sqrt{g_0}  
\begin{aligned}[t]
    &K^\psi_-(x_0;y_3)K^\psi_-(x_0;y_4)\nabla_\sigma^{x_0}K^\phi_+(x_0;y_2) 
    \\
    & \times \nabla^\sigma_{x_0} \left( { \p_{u_1} K^\phi_+(x_0,y_1)\over u_y+n(\yh)\cdot x_0-i\eps} \right) \quad + (y_1 \leftrightarrow y_2) ~.
\end{aligned}
}

Next we need to work out the variation of the graviton emission contributions to $Z$ under the superrotation shift $\delta \hb^-_{zz} (y)= -u_y\p_z^3 Y^z$, along with the shift of the graviton source on $\Ic^-$.   

We first establish that the vertex emission diagram has no variation under the superrotation.   We first perform the integration over $u_y$ (the $u$-coordinate of the vertex).  The only $u_y$ dependence in \rf{gp10x}   comes from the graviton bulk-boundary propagator, and the corresponding integral is, again following \eqref{eq:cosReg}
\eq{gp19}{ \int_{-\infty}^\infty \! du_y  {u_y\over (u_y + n(\hat{y})\cdot x_0 -i\eps)^2}=i\pi~. }
Therefore
\eq{gp20}{ & \int\!du_y  u_y W_{\psi \psi  \rt \phi \phi h}^{\text{ver};AB}(y;y_i) \cr
& = {\lambda\over 4\pi}    \epsilon_{\mu\nu}^\alpha\left(\hat{y}\right) \hat{\veps}_\alpha^{* AB}\left(\hat{y}\right) \int\! d^4x_0 \sqrt{g_0} K_-^\psi (x_0;y_3) K_-^\psi (x_0;y_4) \nabla^\mu_{x_0} K^\phi_+(x_0,y_1)   \nabla^\nu_{x_0} K^\phi_+(x_0,y_2)  \cr }
Next, we want to integrate by parts the $\p_z$ derivatives in $\p_z^3 Y^z(z)$.    Using \rf{gp20} and taking into account the measure factor $\sqrt{\gamma(y) } $, the $y^A=(z,\zb)$ dependence in $Z_{\text{ver}}$ is given by 
\eq{g46}{ \sqrt{\gamma(\yh)}  \epsilon^\alpha(\yh) \cdot n(\yh_1)\epsilon^\alpha(\yh) \cdot n(\yh_2)  \hat{\veps}_\alpha^{* zz}(\yh) = 2 {(z-z_1)(z-z_2) \over (1+z_1\zb_1) (1+z_2\zb_2)} ~. }
Since this is annihilated by $\p_z^3$ we find vanishing variation under the superrotation as claimed.  The fact that the entire superrotation variation comes from emission from external lines is related to the universal properties of the subleading soft graviton theorem.

We next turn to the superrotation variation coming from the diagram \rf{gp18}.  Here we encounter the $u_y$ integral 
\eq{g46kk}{  \int_{-\infty}^\infty \! du_y  {u_y \over u_y+n(\yh)\cdot x_0-i\eps} }
which is divergent and defined using \eqref{eq:cosReg} to be
\eq{g112}{   \int_{-\infty}^\infty \! du_y  {u_y \over u_y+n(\yh)\cdot x_0-i\eps}  & \equiv {1\over 2} 
 \lim_{\omega \rt 0^+} \int_{-\infty}^\infty \! du_y  {u_y ( e^{i\omega u_y}+e^{-i\omega u_y} )  \over u_y+n(\yh)\cdot x_0-i\eps}  \cr 
& = -{i\over 2}  \lim_{\omega \rt 0^+}  \p_\omega \int_{-\infty}^\infty \! du_y  { e^{i\omega u_y}  \over u_y+n(\yh)\cdot x_0-i\eps} \cr
& =  \pi  \lim_{\omega \rt 0^+}  \p_\omega  e^{-i\omega n(\yh)\cdot x_0} \cr
&=  - \pi i n(\yh)\cdot x_0 ~.}
It should be clear that from the reasoning given the overall numerical coefficient in this result is on shaky ground;\footnote{For example, we  equally well could have  written the first line as $ \int_{-\infty}^\infty \! du_y  {u_y \over u_y+n(\yh)\cdot x_0-i\eps}   \equiv \lim_{\omega \rt 0^+} \int_{-\infty}^\infty \! du_y  {u_y \big(\beta  e^{i\omega u_y}  +(1-\beta) e^{-i\omega u_y} \big) \over u_y+n(\yh)\cdot x_0-i\eps}   $  for any coefficient $\beta$, since the term proportional to $1-\beta$ vanishes.}  however, recall that we previously   showed  how the normalization is fixed by Poincar\'e invariance.  Compare the second line of \eqref{g112} with \eqref{c53}.  Finally note that by differentiating \rf{g112} we correctly  reproduce \rf{gp19}.

Using this rule we have 
\eq{g113}{   &   \int du_y   u_y  W_{\psi \psi  \rt \phi \phi h}^{\phi;AB} (y_1,\ldots, y_4,y)  
=   {i \lambda \over 8\pi } \frac{\hat{\veps}_\alpha^{* A}(\hat{y})\hat{\veps}_\alpha^{* B}(\hat{y}) (\epsilon^\alpha(\hat{y}) \cdot n(\hat{y}_1) )^2 }{n(\hat{y}) \cdot n(\hat{y}_1)} \cr
& \quad \times  \int\! d^4x_0 \sqrt{g_0} K^\psi_-(x_0;y_3)K^\psi_-(x_0;y_4)\nabla_\sigma^{x_0}K^\phi_+(x_0;y_2)  \nabla^\sigma_{x_0} \Big[ n(\yh)\cdot x_0 \p_{u_1} K^\phi_+(x_0,y_1) \Big] \cr
&\quad + (y_1 \leftrightarrow y_2)~. }

The next step is to perform the $\int d^2 y  $ integral.  Taking $AB=zz$, the $y$ dependence of the integrand is  
\eq{g117}{ \sqrt{\gamma(\yh)}  \epsh^{*zz}_- {[\eps^-(\yh)\cdot n(\yh_1)]^2\over n(\yh_1)\cdot n(\yh)} n(\yh)\cdot x_0  = -{(1+z\zb)(z-z_1) \over  (1+z_1\zb_1)(\zb-\zb_1)}  n(\yh)\cdot x_0~.  }
We have the integral 
\eq{ga9}{ & \int\! d^2z  \p_z^3 Y^z(z)  {-(z-z_1)(1+z\zb)n(\yh)\cdot x_0  \over \zb-\zb_1}\cr
& \quad  = - 2\pi  \p_z Y^z(z_1)(1+z_1\zb_1)n(\yh_1)\cdot x_0 + 4\pi Y^z(z_1) \p_{z_1} \big[ (1+z_1\zb_1) n(\yh_1)\cdot x_0  \big]
}
obtained by applying integration by parts and ignoring boundary terms at infinity. 
Next, using the explicit form of the scalar propagator we have the identity
\eq{ga12}{    & Y^z(z_1)   {\p_{z_1} \big[ (1+z_1\zb_1) n(\yh_1)\cdot x_0  \big] \over 1+z_1\zb_1} \p_{u_1} K(x_0,y_1) \cr  
& \quad  =Y^z(z_1) \p_{z_1} K(x_0,y_1) +{\zb_1 \over 1+z_1\zb_1} Y^z(z_1) n(\yh_1)\cdot x_0  \p_{u_1} K(x_0,y_1)~. }
Using this along with 
\eq{ga14}{ n(\yh_1) \cdot x_0 \p_{u_1}  K(x_0,y_1) = -u_1 \p_{u_1}  K(x_0,y_1) -2  K(x_0,y_1) }
\rf{ga9} gives 
\eq{ga15}{  &{1\over 4\pi} {1\over 1+  z_1\zb_1} \int\! d^2z  \p_z^3 Y^z(z)  {-(z-z_1)(1+z\zb)  \over \zb-\zb_1} \nabla_{x_0}^\sigma \Big[ n(\yh)\cdot x_0\p_{u_1} K(x_0,y_1)\Big] \cr
& = \Big[  Y^z(z_1)  \p_{z_1}  +{1\over 2} D_z Y^z(z_1)u_1 \p_{u_1}  + D_z Y^z(z_1) \Big] \nabla_{x_0}^\sigma K(x_0,y_1)~.  }
The superrotation variation of the partition function then takes the form 
\eq{ga16}{\delta Z^{\phi~ext}_{\psi \psi  \rt \phi \phi h} &= - {1\over 2}\cdot {\lambda \over 4} \int \left(  \prod_{i=1}^4 d^2y_i du_i \sqrt{\gamma(\yh_i)}\right)\phib_1(y_1) \phib_1(y_2) \psib_1(y_3)\psib_1(y_4) \cr& \quad\quad \times \sum_{i=1}^n \left(D_A Y^A(z_i) +Y^A(z_i) D_A+{1\over 2} D_AY^A(z_i) u_i \p_{u_i}   \right) W_{\psi \psi \to \phi \phi}\left(y_1, \ldots, y_4\right)\cr }
where $ W_{\psi \psi \rt  \phi \phi}\left(y_1, \ldots, y_4\right)$ is the pure scalar diagram \rf{gp8xf}.
So after integrating by parts to bring the variation to act on the sources, and including the contribution from the incoming graviton,  we find 
\eq{b7}{ \delta Z_{\psi \psi  \rt \phi \phi h }+\delta Z_{\psi \psi  h \rt \phi \phi  } = - \delta Z_{ \psi \psi  \rt \phi \phi  }  }
thus verifying invariance under superrotations. 

\subsection{Soft theorems for reduced correlators}
\label{reduced}

In momentum space the full amplitude ${\cal A}$ may be decomposed  in terms of the reduced amplitude ${\cal M}$ and a momentum conserving delta function, as in \rf{ssxx}.   The full and reduced amplitudes clearly obey the same leading order soft theorems, but at subleading order one has to take into account the soft expansion of the delta function.  Nonetheless, it can be shown that the subleading soft theorem again takes the same form in terms of the full and reduced amplitudes \cite{Broedel:2014fsa}.   In this section we show how this works in terms of the boundary correlators, using our explicit examples.  

Given a correlator $W(y_1, \ldots, y_n)$  a reduced correlator $\Wc(y_1, \ldots, y_n)$ may be defined via the relation 
\eq{hhh1}{W(y_1, \ldots, y_n) = \int \![dg] \Wc(y^{(g)}_1, \ldots, y^{(g)}_n)~,   }
where $g \in  \mathbb{R}^{1,3}$ is an element of the group of translations in Minkowski space, and $y^{(g)}$ denotes its action on the boundary coordinates.  Up to  customary factors of $\omega/2\pi$, the reduced correlator $\Wc$ is then the Fourier transform of the reduced momentum space amplitude ${\cal M}$.    In the case where the correlators are obtained from  bulk Witten diagrams there is a simple way of implementing this:  to obtain $\Wc$ rather than $W$ simply omit the integration over the location of one of the vertices.  For our examples we implement this by setting $x_0 =0$.  We now show how the resulting correlators obey the leading and subleading soft theorems.   

\subsubsection{No-derivative contact interaction}

Here we consider the scalar action \rf{gq4} and its associated correlators \rf{gp8} and \rf{gq6}. 
Using the expressions \rf{gq1a} for the  bulk-boundary propagators and setting $x_0=0$,  \rf{gp8} becomes 
\eq{hhh2}{ \Wc_{\phi\phi \rt \phi\phi} =- {i\lambda \over (2\pi)^8}  {1\over (u_1 -i\eps)^2  (u_2 -i\eps)^2  (v_3 +i\eps)^2  (v_4 +i\eps)^2  }~. }
Similarly \rf{a96} becomes
\eq{hhh3}{\Wc^{AB}_{\phi\phi \rt \phi\phi h} & =  {\lambda \over 2\pi} G^{AB}(\yh,\yh_1) {2 \over (2\pi)^8} { 1\over (u_1 -i\eps)^3  (u_2 -i\eps)^2  (v_3 +i\eps)^2  (v_4 +i\eps)^2 (u_y-i\eps)   }\cr
& \quad + 3 {\rm ~more}~. }
We now use the integrals \rf{a98} and   \rf{g112} to  compute 
\eq{hhh4}{ \int\! du_y   \Wc^{zz}_{\phi\phi \rt \phi\phi h}& =  {\lambda \over 2\pi} G^{zz}(\yh,\yh_1) {2\pi i \over (2\pi)^8} { 1\over (u_1 -i\eps)^3  (u_2 -i\eps)^2  (v_3 +i\eps)^2  (v_4 +i\eps)^2   }   + 3 {\rm ~more}
\cr 
 \int\! du_y u_y    \Wc^{zz}_{\phi\phi \rt \phi\phi h}& = 0~.   }
Using  the expressions for the soft factors in \rf{c46}-\rf{c46pp} and \rf{c52pp}   we readily verify the soft relations
\eq{hh5}{ \int\! du_y   \Wc^{zz}_{\phi\phi \rt \phi\phi h}& = {1\over 4\pi} S_{(0)}^{zz}  \Wc_{\phi\phi \rt \phi\phi} \cr
\int\! du_y  u_y  \Wc^{zz}_{\phi\phi \rt \phi\phi h}& ={1\over 4\pi} S_{(1)}^{zz}  \Wc_{\phi\phi \rt \phi\phi}~.  }
Here the second line reads simply $0=0$, but had we used the full amplitude  $W$  the equality would involves nonzero quantities on the two sides, originating from the momentum conserving delta functions.  

\subsubsection{Derivative contact interaction}

Starting from the action   \rf{gp1} and the corresponding correlators \rf{gp8xf},  \rf{gp10x} and \rf{gp11} we can again verify the relations \rf{hh5}, though the computation is now more involved.  
The reduced amplitudes work out to be
\eq{hh6}{ \Wc_{\phi\phi \rt \psi \psi} & =  {8i\lambda \over (2\pi)^8}  { (z_1-z_2)(\zb_1-\zb_2) \over (1+z_1\zb_1)(1+z_2\zb_2)  } {1\over (u_1-i\eps)^3  (u_2-i\eps)^3  (v_3+i\eps)^2 (v_4+i\eps)^2 }   \cr
\Wc^{ver;zz}_{\psi \psi \rt \phi\phi h} & =  -{4 \lambda \over (2\pi)^{10}} { (1+z\zb)^2(z-z_1)(z-z_2) \over (1+z_1\zb_1)(1+z_2\zb_2) } {1\over  (u_y-i\eps)^2 (u_1-i\eps)^3  (u_2-i\eps)^3  (v_3+i\eps)^2 (v_4+i\eps)^2 }  \cr
 \Wc_{\psi \psi \rt \phi\phi h}^{\phi;zz} & =  {16\pi \lambda \over (2\pi)^{10}} G^{zz}(\yh,\yh_1)  {1\over  (u_1-i\eps)^3  (u_2-i\eps)^3  (v_3+i\eps)^2 (v_4+i\eps)^2 } \cr
 & \quad \times \left[ {3(z_1-z_2)(\zb_1-\zb_2) \over (1+z_1\zb_1)(1+z_2\zb_2)}  {1\over (u_1-i\eps)(u_y-i\eps) }  +  {(z-z_2)(\zb-\zb_2) \over (1+z\zb)(1+z_2\zb_2)}  {1\over (u_y-i\eps)^2 }   \right]\cr
 & \quad + (y_1\leftrightarrow y_2) ~.  }
To verify the first relation in \rf{hh5},  we note that neither $\Wc^{ver;zz}_{\psi \psi \rt \phi\phi h}  $   nor the second term in the  square brackets of $\Wc_{\psi \psi \rt \phi\phi h}^{\phi;zz} $ contribute since they give zero upon integration over $u_y$.  Keeping just the first term in the square brackets, it's easy to see that the soft relation is satisfied.  For the second, subleading soft, relation in \rf{hh5} we have the reverse situation: the first term in square brackets of $\Wc_{\psi \psi \rt \phi\phi h}^{\phi;zz} $   does not contribute while  the second term in square brackets does, as does  $\Wc^{ver;zz}_{\psi \psi \rt \phi\phi h}  $.   Some straightforward  algebra then leads to verification of the subleading soft relation.

\section{Conserved charges}
\label{charges}

The on-shell variation of the action gives boundary terms on $\Ic^+$ and $\Ic^-$.   Taking these variations to be asymptotic symmetries, and assuming invariance of the action (keeping in mind the issues discussed in the previous section), cancellation between the two boundary terms becomes the statement of charge conservation.   In more physical terms, these quantities represent the flux of charge through null infinity.   Using the field equations and assuming vacuum boundary conditions in the far future, the flux through $\Ic^+$ can be reexpressed in terms of a charge at $\Ic^+_-$, the latter being the canonical charge.     Here we write out the corresponding charges expressed as flux integrals.    

The  variation of the action takes the form  
\eq{mm3}{ \delta I_{\Ic^+} &= {1\over 32 \pi G}    \int_{\Ic^+} d^3x \sqrt{\gamma}  \Big[ \p_u h_{1+}^{AB} \delta \hb^{1-}_{AB} - h_{1+}^{AB}   \p_u \delta \hb^{1-}_{AB}  \Big] +2   \int_{\Ic^+} d^3x \sqrt{\gamma}   \p_u   \phi^+_{-1} \delta  \phib^{~\!-}_{-1} ~, }
where we reinstated the gravitational coupling.
For the gravitational contribution we have  written out two separate terms which are naively related by integration by parts. See the discussion below \eqref{ACTGold} for how we treat integration by parts in $u$.  Doing so would give the wrong numerical factor for the  soft part of the supertranslation charge, as will be seen below;  see \cite{Kraus:2025wgi} for a more systematic treatment in the Yang--Mills context.   For superrotations, the rules for including or discarding boundary terms are not clear a priori.

\subsection{Supertranslation charge}

Inserting a supertranslation variation, \rf{c44hh}-\rf{c44ii}, we find  the corresponding charge
\eq{mm4}{  Q^+_T & = {1\over 16\pi G}  \int_{\Ic^+} d^3x \sqrt{\gamma}  T  \Big[ -  D_A D_B \p_u h_{1+}^{AB}+\p_u h_{1+}^{AB} \p_u  \hb^{1-}_{AB}  \Big] + 2  \int_{\Ic^+} d^3x \sqrt{\gamma}  T   \p_u   \phi_{-1,+} \p_u \phib_{-1, -} ~. }
At least formally this reproduces the standard expression, e.g. in  \cite{Strominger:2017zoo}.  We say formally because agreement involves using the relation $ \p_u h_{1+}^{AB} \p_u  \hb^{1-}_{AB}  = {1\over 2} \p_u h_{1}^{AB} \p_u  h^{1}_{AB} $ corresponding to the vanishing of integrals involving two  same-sign-frequency factors.   With the proper in-out asymptotics appropriate to the S-matrix this is not automatically guaranteed.  The discrepancy appears to be analogous to a normal ordering constant, but this deserves to be better understood. The same issue was previously noted in the asymptotic charge appearing in gauge theories \cite{Kraus:2025wgi}.

\subsection{Superrotation charge}

The linear in $u$ behavior of superrotation charges makes the rules for integration by parts on the boundary even more  unclear than for the supertranslation case discussed above.   To match expressions in the literature we will allow ourselves to freely integrate by parts, though we again emphasize that since these existing results in the literature are not careful about using proper in-out  asymptotics the fully correct form of the charges relevant for the S-matrix  remains to be established.   As in the supertranslation case,  we also assume the vanishing of same sign frequency integrals.

In this section, in order to avoid inessential complications involving sphere derivatives we will adopt the flat boundary slicing, which in the formulas below will amount  to using $\p_z \sqrt{\gamma}=0$, allowing us to easily integrate by parts on the sphere.  

We start with the scalar contribution, recalling 
\eq{mn1}{  \delta_Y \phib_- &=  {1\over 2} D_z Y^z u \p_u \phib_-+ Y^z \p_z \phib_- +{1\over 2} D_z Y^z \phib_- ~,}
where here and below we drop the falloff subscripts on the fields for brevity.  Also, we are restricting attention to the $Y^z$ contribution; that of $Y^{\zb}$ may be obtained by adding the complex conjugate part at the end.   We have 
\eq{mn2}{ 
 & 2\int\! d^3x \sqrt{\gamma} \p_u \phi_+ \delta \phib_-  =2 \int\! d^3x \sqrt{\gamma} \p_u \phi_+ \left(     {1\over 2} D_z Y^z u \p_u \phib_- + Y^z \p_z \phib_- +{1\over 2} D_z Y^z \phib_- \right) \cr
&=  \int\! d^3x \sqrt{\gamma} Y^z \Big(  - u \p_z\p_u \phi_+  \p_u \phib_-  - u\p_u \phi_+ \p_u \p_z \phib_- +  \p_u \phi_+ \p_z \phib_- - \p_z  \p_u \phi_+ \phib_-   \Big)~.  }
In terms of the stress tensor components 
\eq{mn3}{  T_{uu} = (\p_u \phi)^2~,\quad T_{uz} = \p_u \phi  \p_z\phi }
we have the scalar contribution to the charge 
\eq{mn4}{ Q_{Y,\phi}^{+}=  2\int\! d^3x \sqrt{\gamma} \p_u \phi_+ \delta \phib_-  &= \int\! d^3x \sqrt{\gamma} Y^z \Big(  -{u\over 2} \p_z T_{uu}  +T_{uz}  \Big)~. }
This agrees (up to a convention dependent sign) with the result in \cite{Donnay:2022wvx}.
 
Next we have the gravitational contribution, where we recall
\eq{mn5}{ \delta_Y \hb^-_{zz}& =  {1\over 2}  D_zY^z  u\p_u    \hb_{zz} +  Y^z D_z \hb^-_{zz} + {3\over 2} D_z Y^z \hb^-_{zz}    -u \p_z^3Y^z~, \cr
\delta_Y \hb^-_{\zb\zb} &  = {1\over 2}  D_{z}Y^{z}  u\p_u    \hb^-_{\zb\zb} +  Y^{z} D_{z} \hb^-_{\zb\zb} - {1\over 2} D_{z} Y^{z} \hb^-_{\zb\zb}~.   }
For the terms in the charge quadratic in fields we will assume that we can freely integrate by parts in $u$.  Under this assumption a straightforward computation yields 
\eq{mn6}{ Q_{Y,h}^{+} \big|_{h^2} & =  {1\over 16\pi G} \int_{\Ic^+} \! d^3x \sqrt{\gamma}  Y^z \Big(   -{u\over 2}  \p_z\p_u h_+^{AB}  \p_u  \hb^-_{AB}  -{u\over 2} \p_uh_+^{AB} \p_u \p_z  \hb^-_{AB} \cr
&\quad\quad    -{1\over 2} \p_u h_+^{zz} \p_z \hb^-_{zz} -{1\over 2} \p_z h_+^{\zb\zb} \p_u \hb^-_{\zb\zb}   +{3\over 2} \p_z h_+^{zz} \p_u \hb^-_{zz}  +{3\over 2} \p_u h_+^{\zb\zb} \p_z \hb^-_{\zb\zb} \Big)~.  }

Finally, we have the contribution from  the variation $ \delta_Y \hb^-_{zz} =  -u \p_z^3Y^z$.  In order to match  results in the literature we should first  integrate by parts  in \rf{mm3} to combine the two metric terms, dropping the surface term as discussed below \eqref{ACTGold}. Doing so gives 
\eq{mn7}{  Q^+_{Y,h}\big|_{h^1} & =   {1\over 16\pi G} \int_{\Ic^+} \! d^3x \sqrt{\gamma}  Y^z u \p_z^3 \p_u h_+^{zz}~.}
Both \rf{mn6} and \rf{mn7}    agree with \cite{Donnay:2022wvx} after some integration by parts.

\section*{Acknowledgments}

We thank Temple He,  Prahar Mitra, Romain Ruzziconi, and Sasha Zhiboedov for discussions.  P.K. is supported in part by  National Science Foundation grant PHY-2209700. R.M.M is supported by the Heising-Simons Foundation “Observational Signatures of Quantum Gravity” collaboration.

\bibliographystyle{bibstyle2017}
\bibliography{collection}

@article{Kraus:2025wgi,
    author = "Kraus, Per and Myers, Richard M.",
    title = "{Carrollian partition function for bulk Yang-Mills theory}",
    eprint = "2503.00916",
    archivePrefix = "arXiv",
    primaryClass = "hep-th",
    doi = "10.1007/JHEP08(2025)180",
    journal = "JHEP",
    volume = "08",
    pages = "180",
    year = "2025"
}

@article{Kapec:2014opa,
    author = "Kapec, Daniel and Lysov, Vyacheslav and Pasterski, Sabrina and Strominger, Andrew",
    title = "{Semiclassical Virasoro symmetry of the quantum gravity $ \mathcal{S}$-matrix}",
    eprint = "1406.3312",
    archivePrefix = "arXiv",
    primaryClass = "hep-th",
    doi = "10.1007/JHEP08(2014)058",
    journal = "JHEP",
    volume = "08",
    pages = "058",
    year = "2014"
}

@article{Campiglia:2015yka,
    author = "Campiglia, Miguel and Laddha, Alok",
    title = "{New symmetries for the Gravitational S-matrix}",
    eprint = "1502.02318",
    archivePrefix = "arXiv",
    primaryClass = "hep-th",
    doi = "10.1007/JHEP04(2015)076",
    journal = "JHEP",
    volume = "04",
    pages = "076",
    year = "2015"
}

@article{Compere:2018ylh,
    author = "Comp{\`e}re, Geoffrey and Fiorucci, Adrien and Ruzziconi, Romain",
    title = "{Superboost transitions, refraction memory and super-Lorentz charge algebra}",
    eprint = "1810.00377",
    archivePrefix = "arXiv",
    primaryClass = "hep-th",
    doi = "10.1007/JHEP11(2018)200",
    journal = "JHEP",
    volume = "11",
    pages = "200",
    year = "2018",
    note = "[Erratum: JHEP 04, 172 (2020)]"
}

@article{Campiglia:2021bap,
    author = "Campiglia, Miguel and Laddha, Alok",
    title = "{BMS Algebra, Double Soft Theorems, and All That}",
    eprint = "2106.14717",
    archivePrefix = "arXiv",
    primaryClass = "hep-th",
    month = "6",
    year = "2021"
}

@article{Ruzziconi:2024kzo,
    author = "Ruzziconi, Romain and Saha, Amartya",
    title = "{Holographic Carrollian currents for massless scattering}",
    eprint = "2411.04902",
    archivePrefix = "arXiv",
    primaryClass = "hep-th",
    doi = "10.1007/JHEP01(2025)169",
    journal = "JHEP",
    volume = "01",
    pages = "169",
    year = "2025"
}

@article{Agrawal:2023zea,
    author = "Agrawal, Shreyansh and Donnay, Laura and Nguyen, Kevin and Ruzziconi, Romain",
    title = "{Logarithmic soft graviton theorems from superrotation Ward identities}",
    eprint = "2309.11220",
    archivePrefix = "arXiv",
    primaryClass = "hep-th",
    doi = "10.1007/JHEP02(2024)120",
    journal = "JHEP",
    volume = "02",
    pages = "120",
    year = "2024"
}

@article{Upadhyay:2025ged,
    author = "Upadhyay, Shivam",
    title = "{On symmetries of gravitational on-shell boundary action at null infinity}",
    eprint = "2501.07136",
    archivePrefix = "arXiv",
    primaryClass = "hep-th",
    month = "1",
    year = "2025"
}

@article{Broedel:2014fsa,
    author = "Broedel, Johannes and de Leeuw, Marius and Plefka, Jan and Rosso, Matteo",
    title = "{Constraining subleading soft gluon and graviton theorems}",
    eprint = "1406.6574",
    archivePrefix = "arXiv",
    primaryClass = "hep-th",
    reportNumber = "HU-EP-14-20",
    doi = "10.1103/PhysRevD.90.065024",
    journal = "Phys. Rev. D",
    volume = "90",
    number = "6",
    pages = "065024",
    year = "2014"
}

@article{Kim:2023qbl,
    author = "Kim, Seolhwa and Kraus, Per and Monten, Ruben and Myers, Richard M.",
    title = "{S-matrix path integral approach to symmetries and soft theorems}",
    eprint = "2307.12368",
    archivePrefix = "arXiv",
    primaryClass = "hep-th",
    doi = "10.1007/JHEP10(2023)036",
    journal = "JHEP",
    volume = "10",
    pages = "036",
    year = "2023"
}

@article{Gubser:1998bc,
    author = "Gubser, S. S. and Klebanov, Igor R. and Polyakov, Alexander M.",
    title = "{Gauge theory correlators from noncritical string theory}",
    eprint = "hep-th/9802109",
    archivePrefix = "arXiv",
    reportNumber = "PUPT-1767",
    doi = "10.1016/S0370-2693(98)00377-3",
    journal = "Phys. Lett. B",
    volume = "428",
    pages = "105--114",
    year = "1998"
}

@article{Witten:1998qj,
    author = "Witten, Edward",
    title = "{Anti-de Sitter space and holography}",
    eprint = "hep-th/9802150",
    archivePrefix = "arXiv",
    reportNumber = "IASSNS-HEP-98-15",
    doi = "10.4310/ATMP.1998.v2.n2.a2",
    journal = "Adv. Theor. Math. Phys.",
    volume = "2",
    pages = "253--291",
    year = "1998"
}

@article{Strominger:2017zoo,
    author = "Strominger, Andrew",
    title = "{Lectures on the Infrared Structure of Gravity and Gauge Theory}",
    eprint = "1703.05448",
    archivePrefix = "arXiv",
    primaryClass = "hep-th",
    month = "3",
    year = "2017"
}

@article{Papadimitriou:2007sj,
    author = "Papadimitriou, Ioannis",
    title = "{Multi-Trace Deformations in AdS/CFT: Exploring the Vacuum Structure of the Deformed CFT}",
    eprint = "hep-th/0703152",
    archivePrefix = "arXiv",
    reportNumber = "DESY-06-218, ZMP-HH-06-17",
    doi = "10.1088/1126-6708/2007/05/075",
    journal = "JHEP",
    volume = "05",
    pages = "075",
    year = "2007"
}

@article{Jevicki:1987ax,
    author = "Jevicki, A. and Lee, Choon-kyu",
    title = "{The S Matrix Generating Functional and Effective Action}",
    reportNumber = "BROWN-HET-634",
    doi = "10.1103/PhysRevD.37.1485",
    journal = "Phys. Rev. D",
    volume = "37",
    pages = "1485",
    year = "1988"
}

@article{Arefeva:1974jv,
    author = "Arefeva, I. Ya. and Faddeev, L. D. and Slavnov, A. A.",
    title = "{Generating Functional for the s Matrix in Gauge Theories}",
    reportNumber = "SACLAY-DPH-T-74-44",
    doi = "10.1007/BF01038094",
    journal = "Teor. Mat. Fiz.",
    volume = "21",
    pages = "311--321",
    year = "1974"
}

@article{Campiglia:2017mua,
    author = "Campiglia, Miguel and Eyheralde, Rodrigo",
    title = "{Asymptotic $U(1)$ charges at spatial infinity}",
    eprint = "1703.07884",
    archivePrefix = "arXiv",
    primaryClass = "hep-th",
    doi = "10.1007/JHEP11(2017)168",
    journal = "JHEP",
    volume = "11",
    pages = "168",
    year = "2017"
}

@article{Compere:2011ve,
    author = "Compere, Geoffrey and Dehouck, Fran",
    title = "{Relaxing the Parity Conditions of Asymptotically Flat Gravity}",
    eprint = "1106.4045",
    archivePrefix = "arXiv",
    primaryClass = "hep-th",
    doi = "10.1088/0264-9381/28/24/245016",
    journal = "Class. Quant. Grav.",
    volume = "28",
    pages = "245016",
    year = "2011",
    note = "[Erratum: Class.Quant.Grav. 30, 039501 (2013)]"
}

@proceedings{Balian:1976vq,
    author = "Balian, R. and Zinn-Justin, Jean",
    title = "{Methods in Field Theory. Les Houches Summer School in Theoretical Physics. Session 28, July 28-September 6, 1975}",
    year = "1976"
}

@article{Raclariu:2021zjz,
    author = "Raclariu, Ana-Maria",
    title = "{Lectures on Celestial Holography}",
    eprint = "2107.02075",
    archivePrefix = "arXiv",
    primaryClass = "hep-th",
    month = "7",
    year = "2021"
}

@inproceedings{Pasterski:2021raf,
    author = "Pasterski, Sabrina and Pate, Monica and Raclariu, Ana-Maria",
    title = "{Celestial Holography}",
    booktitle = "{Snowmass 2021}",
    eprint = "2111.11392",
    archivePrefix = "arXiv",
    primaryClass = "hep-th",
    month = "11",
    year = "2021"
}

@article{McLoughlin:2022ljp,
    author = "McLoughlin, Tristan and Puhm, Andrea and Raclariu, Ana-Maria",
    title = "{The SAGEX review on scattering amplitudes chapter 11: soft theorems and celestial amplitudes}",
    eprint = "2203.13022",
    archivePrefix = "arXiv",
    primaryClass = "hep-th",
    reportNumber = "SAGEX-22-12, CPHT-RR016.032022, HU-EP-22/13, TCDMATH 22-02",
    doi = "10.1088/1751-8121/ac9a40",
    journal = "J. Phys. A",
    volume = "55",
    number = "44",
    pages = "443012",
    year = "2022"
}

@article{Susskind:1998vk,
    author = "Susskind, Leonard",
    editor = "Burgess, C. P. and Myers, Robert C.",
    title = "{Holography in the flat space limit}",
    eprint = "hep-th/9901079",
    archivePrefix = "arXiv",
    doi = "10.1063/1.1301570",
    journal = "AIP Conf. Proc.",
    volume = "493",
    number = "1",
    pages = "98--112",
    year = "1999"
}

@article{Polchinski:1999ry,
    author = "Polchinski, Joseph",
    title = "{S matrices from AdS space-time}",
    eprint = "hep-th/9901076",
    archivePrefix = "arXiv",
    reportNumber = "NST-ITP-99-02",
    month = "1",
    year = "1999"
}

@article{Giddings:1999jq,
    author = "Giddings, Steven B.",
    title = "{Flat space scattering and bulk locality in the AdS / CFT correspondence}",
    eprint = "hep-th/9907129",
    archivePrefix = "arXiv",
    doi = "10.1103/PhysRevD.61.106008",
    journal = "Phys. Rev. D",
    volume = "61",
    pages = "106008",
    year = "2000"
}

@article{Gary:2009mi,
    author = "Gary, Mirah and Giddings, Steven B.",
    title = "{The Flat space S-matrix from the AdS/CFT correspondence?}",
    eprint = "0904.3544",
    archivePrefix = "arXiv",
    primaryClass = "hep-th",
    reportNumber = "CERN-PH-TH-2009-051",
    doi = "10.1103/PhysRevD.80.046008",
    journal = "Phys. Rev. D",
    volume = "80",
    pages = "046008",
    year = "2009"
}

@article{Gary:2009ae,
    author = "Gary, Mirah and Giddings, Steven B. and Penedones, Joao",
    title = "{Local bulk S-matrix elements and CFT singularities}",
    eprint = "0903.4437",
    archivePrefix = "arXiv",
    primaryClass = "hep-th",
    reportNumber = "CERN-PH-TH-2009-035, NSF-KITP-09-35",
    doi = "10.1103/PhysRevD.80.085005",
    journal = "Phys. Rev. D",
    volume = "80",
    pages = "085005",
    year = "2009"
}

@article{Fitzpatrick:2011jn,
    author = "Fitzpatrick, A. Liam and Kaplan, Jared",
    title = "{Scattering States in AdS/CFT}",
    eprint = "1104.2597",
    archivePrefix = "arXiv",
    primaryClass = "hep-th",
    reportNumber = "SLAC-PUB-14507",
    month = "4",
    year = "2011"
}

@article{Fitzpatrick:2011ia,
    author = "Fitzpatrick, A. Liam and Kaplan, Jared and Penedones, Joao and Raju, Suvrat and van Rees, Balt C.",
    title = "{A Natural Language for AdS/CFT Correlators}",
    eprint = "1107.1499",
    archivePrefix = "arXiv",
    primaryClass = "hep-th",
    reportNumber = "SLAC-PUB-14506, HRI-ST-1107",
    doi = "10.1007/JHEP11(2011)095",
    journal = "JHEP",
    volume = "11",
    pages = "095",
    year = "2011"
}

@article{Duval:2014uva,
    author = "Duval, C. and Gibbons, G. W. and Horvathy, P. A.",
    title = "{Conformal Carroll groups and BMS symmetry}",
    eprint = "1402.5894",
    archivePrefix = "arXiv",
    primaryClass = "gr-qc",
    doi = "10.1088/0264-9381/31/9/092001",
    journal = "Class. Quant. Grav.",
    volume = "31",
    pages = "092001",
    year = "2014"
}

@article{Duval:2014uoa,
    author = "Duval, C. and Gibbons, G. W. and Horvathy, P. A. and Zhang, P. M.",
    title = "{Carroll versus Newton and Galilei: two dual non-Einsteinian concepts of time}",
    eprint = "1402.0657",
    archivePrefix = "arXiv",
    primaryClass = "gr-qc",
    doi = "10.1088/0264-9381/31/8/085016",
    journal = "Class. Quant. Grav.",
    volume = "31",
    pages = "085016",
    year = "2014"
}

@article{Bagchi:2019clu,
    author = "Bagchi, Arjun and Basu, Rudranil and Mehra, Aditya and Nandi, Poulami",
    title = "{Field Theories on Null Manifolds}",
    eprint = "1912.09388",
    archivePrefix = "arXiv",
    primaryClass = "hep-th",
    doi = "10.1007/JHEP02(2020)141",
    journal = "JHEP",
    volume = "02",
    pages = "141",
    year = "2020"
}

@article{Ciambelli:2019lap,
    author = "Ciambelli, Luca and Leigh, Robert G. and Marteau, Charles and Petropoulos, P. Marios",
    title = "{Carroll Structures, Null Geometry and Conformal Isometries}",
    eprint = "1905.02221",
    archivePrefix = "arXiv",
    primaryClass = "hep-th",
    reportNumber = "CPHT-RR025.052019, CPHT-RR010.022019",
    doi = "10.1103/PhysRevD.100.046010",
    journal = "Phys. Rev. D",
    volume = "100",
    number = "4",
    pages = "046010",
    year = "2019"
}

@article{Ciambelli:2018wre,
    author = "Ciambelli, Luca and Marteau, Charles and Petkou, Anastasios C. and Petropoulos, P. Marios and Siampos, Konstantinos",
    title = "{Flat holography and Carrollian fluids}",
    eprint = "1802.06809",
    archivePrefix = "arXiv",
    primaryClass = "hep-th",
    reportNumber = "CPHT-RR049.082017, CERN-TH-2017-229",
    doi = "10.1007/JHEP07(2018)165",
    journal = "JHEP",
    volume = "07",
    pages = "165",
    year = "2018"
}

@article{Hartong:2015xda,
    author = "Hartong, Jelle",
    title = "{Gauging the Carroll Algebra and Ultra-Relativistic Gravity}",
    eprint = "1505.05011",
    archivePrefix = "arXiv",
    primaryClass = "hep-th",
    doi = "10.1007/JHEP08(2015)069",
    journal = "JHEP",
    volume = "08",
    pages = "069",
    year = "2015"
}

@article{Mason:2023mti,
    author = "Mason, Lionel and Ruzziconi, Romain and Yelleshpur Srikant, Akshay",
    title = "{Carrollian amplitudes and celestial symmetries}",
    eprint = "2312.10138",
    archivePrefix = "arXiv",
    primaryClass = "hep-th",
    doi = "10.1007/JHEP05(2024)012",
    journal = "JHEP",
    volume = "05",
    pages = "012",
    year = "2024"
}

@article{Hartong:2015usd,
    author = "Hartong, Jelle",
    title = "{Holographic Reconstruction of 3D Flat Space-Time}",
    eprint = "1511.01387",
    archivePrefix = "arXiv",
    primaryClass = "hep-th",
    doi = "10.1007/JHEP10(2016)104",
    journal = "JHEP",
    volume = "10",
    pages = "104",
    year = "2016"
}

@article{Saha:2023hsl,
    author = "Saha, Amartya",
    title = "{Carrollian approach to 1 + 3D flat holography}",
    eprint = "2304.02696",
    archivePrefix = "arXiv",
    primaryClass = "hep-th",
    doi = "10.1007/JHEP06(2023)051",
    journal = "JHEP",
    volume = "06",
    pages = "051",
    year = "2023"
}

@article{Bagchi:2016bcd,
    author = "Bagchi, Arjun and Basu, Rudranil and Kakkar, Ashish and Mehra, Aditya",
    title = "{Flat Holography: Aspects of the dual field theory}",
    eprint = "1609.06203",
    archivePrefix = "arXiv",
    primaryClass = "hep-th",
    doi = "10.1007/JHEP12(2016)147",
    journal = "JHEP",
    volume = "12",
    pages = "147",
    year = "2016"
}

@article{Donnay:2022wvx,
    author = "Donnay, Laura and Fiorucci, Adrien and Herfray, Yannick and Ruzziconi, Romain",
    title = "{Bridging Carrollian and celestial holography}",
    eprint = "2212.12553",
    archivePrefix = "arXiv",
    primaryClass = "hep-th",
    doi = "10.1103/PhysRevD.107.126027",
    journal = "Phys. Rev. D",
    volume = "107",
    number = "12",
    pages = "126027",
    year = "2023"
}

@article{deGioia:2024yne,
    author = "de Gioia, Leonardo Pipolo and Raclariu, Ana-Maria",
    title = "{Celestial amplitudes from conformal correlators with bulk-point kinematics}",
    eprint = "2405.07972",
    archivePrefix = "arXiv",
    primaryClass = "hep-th",
    month = "5",
    year = "2024"
}

@article{Alday:2024yyj,
    author = "Alday, Luis F. and Nocchi, Maria and Ruzziconi, Romain and Yelleshpur Srikant, Akshay",
    title = "{Carrollian Amplitudes from Holographic Correlators}",
    eprint = "2406.19343",
    archivePrefix = "arXiv",
    primaryClass = "hep-th",
    month = "6",
    year = "2024"
}

@article{Bagchi:2023cen,
    author = "Bagchi, Arjun and Dhivakar, Prateksh and Dutta, Sudipta",
    title = "{Holography in Flat Spacetimes: the case for Carroll}",
    eprint = "2311.11246",
    archivePrefix = "arXiv",
    primaryClass = "hep-th",
    month = "11",
    year = "2023"
}

@article{Bagchi:2023fbj,
    author = "Bagchi, Arjun and Dhivakar, Prateksh and Dutta, Sudipta",
    title = "{AdS Witten diagrams to Carrollian correlators}",
    eprint = "2303.07388",
    archivePrefix = "arXiv",
    primaryClass = "hep-th",
    doi = "10.1007/JHEP04(2023)135",
    journal = "JHEP",
    volume = "04",
    pages = "135",
    year = "2023"
}

@article{Bagchi:2022emh,
    author = "Bagchi, Arjun and Banerjee, Shamik and Basu, Rudranil and Dutta, Sudipta",
    title = "{Scattering Amplitudes: Celestial and Carrollian}",
    eprint = "2202.08438",
    archivePrefix = "arXiv",
    primaryClass = "hep-th",
    doi = "10.1103/PhysRevLett.128.241601",
    journal = "Phys. Rev. Lett.",
    volume = "128",
    number = "24",
    pages = "241601",
    year = "2022"
}

@article{Hijano:2019qmi,
    author = "Hijano, Eliot",
    title = "{Flat space physics from AdS/CFT}",
    eprint = "1905.02729",
    archivePrefix = "arXiv",
    primaryClass = "hep-th",
    doi = "10.1007/JHEP07(2019)132",
    journal = "JHEP",
    volume = "07",
    pages = "132",
    year = "2019"
}

@article{Hijano:2020szl,
    author = "Hijano, Eliot and Neuenfeld, Dominik",
    title = "{Soft photon theorems from CFT Ward identites in the flat limit of AdS/CFT}",
    eprint = "2005.03667",
    archivePrefix = "arXiv",
    primaryClass = "hep-th",
    doi = "10.1007/JHEP11(2020)009",
    journal = "JHEP",
    volume = "11",
    pages = "009",
    year = "2020"
}

@article{Duary:2022pyv,
    author = "Duary, Sarthak and Hijano, Eliot and Patra, Milan",
    title = "{Towards an IR finite S-matrix in the flat limit of AdS/CFT}",
    eprint = "2211.13711",
    archivePrefix = "arXiv",
    primaryClass = "hep-th",
    month = "11",
    year = "2022"
}

@article{Komatsu:2020sag,
    author = "Komatsu, Shota and Paulos, Miguel F. and Van Rees, Balt C. and Zhao, Xiang",
    title = "{Landau diagrams in AdS and S-matrices from conformal correlators}",
    eprint = "2007.13745",
    archivePrefix = "arXiv",
    primaryClass = "hep-th",
    reportNumber = "CPHT-RR119.122020",
    doi = "10.1007/JHEP11(2020)046",
    journal = "JHEP",
    volume = "11",
    pages = "046",
    year = "2020"
}

@article{Li:2021snj,
    author = "Li, Yue-Zhou",
    title = "{Notes on flat-space limit of AdS/CFT}",
    eprint = "2106.04606",
    archivePrefix = "arXiv",
    primaryClass = "hep-th",
    doi = "10.1007/JHEP09(2021)027",
    journal = "JHEP",
    volume = "09",
    pages = "027",
    year = "2021"
}

@article{Marotta:2024sce,
    author = "Marotta, Raffaele and Skenderis, Kostas and Verma, Mritunjay",
    title = "{Flat space spinning massive amplitudes from momentum space CFT}",
    eprint = "2406.06447",
    archivePrefix = "arXiv",
    primaryClass = "hep-th",
    month = "6",
    year = "2024"
}

@article{Weinberg:1965nx,
    author = "Weinberg, Steven",
    title = "{Infrared photons and gravitons}",
    doi = "10.1103/PhysRev.140.B516",
    journal = "Phys. Rev.",
    volume = "140",
    pages = "B516--B524",
    year = "1965"
}

@article{Bern:2014vva,
    author = "Bern, Zvi and Davies, Scott and Di Vecchia, Paolo and Nohle, Josh",
    title = "{Low-Energy Behavior of Gluons and Gravitons from Gauge Invariance}",
    eprint = "1406.6987",
    archivePrefix = "arXiv",
    primaryClass = "hep-th",
    reportNumber = "UCLA-14-TEP-104, NORDITA-2014-78",
    doi = "10.1103/PhysRevD.90.084035",
    journal = "Phys. Rev. D",
    volume = "90",
    number = "8",
    pages = "084035",
    year = "2014"
}

@article{Pasterski:2021rjz,
    author = "Pasterski, Sabrina",
    title = "{Lectures on celestial amplitudes}",
    eprint = "2108.04801",
    archivePrefix = "arXiv",
    primaryClass = "hep-th",
    doi = "10.1140/epjc/s10052-021-09846-7",
    journal = "Eur. Phys. J. C",
    volume = "81",
    number = "12",
    pages = "1062",
    year = "2021"
}

@article{Donnay:2023mrd,
    author = "Donnay, Laura",
    title = "{Celestial holography: An asymptotic symmetry perspective}",
    eprint = "2310.12922",
    archivePrefix = "arXiv",
    primaryClass = "hep-th",
    doi = "10.1016/j.physrep.2024.04.003",
    journal = "Phys. Rept.",
    volume = "1073",
    pages = "1--41",
    year = "2024"
}

@article{Bondi:1962px,
    author = "Bondi, H. and van der Burg, M. G. J. and Metzner, A. W. K.",
    title = "{Gravitational waves in general relativity. 7. Waves from axisymmetric isolated systems}",
    doi = "10.1098/rspa.1962.0161",
    journal = "Proc. Roy. Soc. Lond. A",
    volume = "269",
    pages = "21--52",
    year = "1962"
}

@article{Sachs:1962zza,
    author = "Sachs, R.",
    title = "{Asymptotic symmetries in gravitational theory}",
    doi = "10.1103/PhysRev.128.2851",
    journal = "Phys. Rev.",
    volume = "128",
    pages = "2851--2864",
    year = "1962"
}

@article{Barnich:2010eb,
    author = "Barnich, Glenn and Troessaert, Cedric",
    title = "{Aspects of the BMS/CFT correspondence}",
    eprint = "1001.1541",
    archivePrefix = "arXiv",
    primaryClass = "hep-th",
    reportNumber = "ULB-TH-09-28",
    doi = "10.1007/JHEP05(2010)062",
    journal = "JHEP",
    volume = "05",
    pages = "062",
    year = "2010"
}

@article{Sahoo:2018lxl,
    author = "Sahoo, Biswajit and Sen, Ashoke",
    title = "{Classical and Quantum Results on Logarithmic Terms in the Soft Theorem in Four Dimensions}",
    eprint = "1808.03288",
    archivePrefix = "arXiv",
    primaryClass = "hep-th",
    doi = "10.1007/JHEP02(2019)086",
    journal = "JHEP",
    volume = "02",
    pages = "086",
    year = "2019"
}

@article{Nguyen:2023vfz,
    author = "Nguyen, Kevin and West, Peter",
    title = "{Carrollian Conformal Fields and Flat Holography}",
    eprint = "2305.02884",
    archivePrefix = "arXiv",
    primaryClass = "hep-th",
    doi = "10.3390/universe9090385",
    journal = "Universe",
    volume = "9",
    number = "9",
    pages = "385",
    year = "2023"
}

@article{deBoer:2023fnj,
    author = "de Boer, Jan and Hartong, Jelle and Obers, Niels A. and Sybesma, Watse and Vandoren, Stefan",
    title = "{Carroll stories}",
    eprint = "2307.06827",
    archivePrefix = "arXiv",
    primaryClass = "hep-th",
    reportNumber = "NORDITA-2023-036",
    doi = "10.1007/JHEP09(2023)148",
    journal = "JHEP",
    volume = "09",
    pages = "148",
    year = "2023"
}

@article{Barnich:2009se,
    author = "Barnich, Glenn and Troessaert, Cedric",
    title = "{Symmetries of asymptotically flat 4 dimensional spacetimes at null infinity revisited}",
    eprint = "0909.2617",
    archivePrefix = "arXiv",
    primaryClass = "gr-qc",
    reportNumber = "ULB-TH-09-24",
    doi = "10.1103/PhysRevLett.105.111103",
    journal = "Phys. Rev. Lett.",
    volume = "105",
    pages = "111103",
    year = "2010"
}

@article{Barnich:2010ojg,
    author = "Barnich, Glenn and Troessaert, Cedric",
    editor = "Anagnostopoulos, Konstantinos N. and Bahns, Dorothea and Grosse, Harald and Irges, Nikos and Zoupanos, George",
    title = "{Supertranslations call for superrotations}",
    eprint = "1102.4632",
    archivePrefix = "arXiv",
    primaryClass = "gr-qc",
    reportNumber = "ULB-TH-11-02",
    doi = "10.22323/1.127.0010",
    journal = "PoS",
    volume = "CNCFG2010",
    pages = "010",
    year = "2010"
}

@article{Campiglia:2014yka,
    author = "Campiglia, Miguel and Laddha, Alok",
    title = "{Asymptotic symmetries and subleading soft graviton theorem}",
    eprint = "1408.2228",
    archivePrefix = "arXiv",
    primaryClass = "hep-th",
    doi = "10.1103/PhysRevD.90.124028",
    journal = "Phys. Rev. D",
    volume = "90",
    number = "12",
    pages = "124028",
    year = "2014"
}

@article{Campiglia:2015kxa,
    author = "Campiglia, Miguel and Laddha, Alok",
    title = "{Asymptotic symmetries of gravity and soft theorems for massive particles}",
    eprint = "1509.01406",
    archivePrefix = "arXiv",
    primaryClass = "hep-th",
    doi = "10.1007/JHEP12(2015)094",
    journal = "JHEP",
    volume = "12",
    pages = "094",
    year = "2015"
}

@article{Campiglia:2020qvc,
    author = "Campiglia, Miguel and Peraza, Javier",
    title = "{Generalized BMS charge algebra}",
    eprint = "2002.06691",
    archivePrefix = "arXiv",
    primaryClass = "gr-qc",
    doi = "10.1103/PhysRevD.101.104039",
    journal = "Phys. Rev. D",
    volume = "101",
    number = "10",
    pages = "104039",
    year = "2020"
}

@article{Campiglia:2024uqq,
    author = "Campiglia, Miguel and Sudhakar, Adarsh",
    title = "{Gravitational Poisson brackets at null infinity compatible with smooth superrotations}",
    eprint = "2408.13067",
    archivePrefix = "arXiv",
    primaryClass = "gr-qc",
    doi = "10.1007/JHEP12(2024)170",
    journal = "JHEP",
    volume = "12",
    pages = "170",
    year = "2024"
}

@article{Cotler:2024xhb,
    author = "Cotler, Jordan and Jensen, Kristan and Prohazka, Stefan and Raz, Amir and Riegler, Max and Salzer, Jakob",
    title = "{Quantizing Carrollian field theories}",
    eprint = "2407.11971",
    archivePrefix = "arXiv",
    primaryClass = "hep-th",
    doi = "10.1007/JHEP10(2024)049",
    journal = "JHEP",
    volume = "10",
    pages = "049",
    year = "2024"
}

@article{Cotler:2025npu,
    author = "Cotler, Jordan and Dhivakar, Prateksh and Jensen, Kristan",
    title = "{Carrollian holographic duals are non-local}",
    eprint = "2512.05072",
    archivePrefix = "arXiv",
    primaryClass = "hep-th",
    month = "12",
    year = "2025"
}

@article{Ciambelli:2025unn,
    author = "Ciambelli, Luca and Jai-akson, Puttarak",
    title = "{Foundations of Carrollian Geometry}",
    eprint = "2510.21651",
    archivePrefix = "arXiv",
    primaryClass = "hep-th",
    reportNumber = "RIKEN-iTHEMS-Report-25",
    month = "10",
    year = "2025"
}

@article{Baulieu:2025itt,
    author = "Baulieu, Laurent and Ciambelli, Luca and Wetzstein, Tom",
    title = "{Extended-BMS Anomalies and Flat Space Holography}",
    eprint = "2504.10304",
    archivePrefix = "arXiv",
    primaryClass = "hep-th",
    month = "4",
    year = "2025"
}

@article{Donnay:2022aba,
    author = "Donnay, Laura and Fiorucci, Adrien and Herfray, Yannick and Ruzziconi, Romain",
    title = "{Carrollian Perspective on Celestial Holography}",
    eprint = "2202.04702",
    archivePrefix = "arXiv",
    primaryClass = "hep-th",
    doi = "10.1103/PhysRevLett.129.071602",
    journal = "Phys. Rev. Lett.",
    volume = "129",
    number = "7",
    pages = "071602",
    year = "2022"
}

@article{Campoleoni:2023fug,
    author = "Campoleoni, Andrea and Delfante, Arnaud and Pekar, Simon and Petropoulos, P. Marios and Rivera-Betancour, David and Vilatte, Matthieu",
    title = "{Flat from anti de Sitter}",
    eprint = "2309.15182",
    archivePrefix = "arXiv",
    primaryClass = "hep-th",
    reportNumber = "CPHT-RR054.082023",
    doi = "10.1007/JHEP12(2023)078",
    journal = "JHEP",
    volume = "12",
    pages = "078",
    year = "2023"
}

@article{Kraus:2024gso,
    author = "Kraus, Per and Myers, Richard M.",
    title = "{Carrollian Partition Functions and the Flat Limit of AdS}",
    eprint = "2407.13668",
    archivePrefix = "arXiv",
    primaryClass = "hep-th",
    month = "7",
    year = "2024"
}

@article{Ma:2023gir,
    author = "Ma, Yao and Sterman, George and Venkata, Aniruddha",
    title = "{Soft Photon Theorem in QCD with Massless Quarks}",
    eprint = "2311.06912",
    archivePrefix = "arXiv",
    primaryClass = "hep-ph",
    reportNumber = "YITP-SB-2023-35",
    doi = "10.1103/PhysRevLett.132.091902",
    journal = "Phys. Rev. Lett.",
    volume = "132",
    number = "9",
    pages = "091902",
    year = "2024"
}

@article{Distler:2018rwu,
    author = "Distler, Jacques and Flauger, Raphael and Horn, Bart",
    title = "{Double-soft graviton amplitudes and the extended BMS charge algebra}",
    eprint = "1808.09965",
    archivePrefix = "arXiv",
    primaryClass = "hep-th",
    reportNumber = "UTTG-16-18",
    doi = "10.1007/JHEP08(2019)021",
    journal = "JHEP",
    volume = "08",
    pages = "021",
    year = "2019"
}

@article{Capone:2022gme,
    author = "Capone, Federico and Nguyen, Kevin and Parisini, Enrico",
    title = "{Charge and antipodal matching across spatial infinity}",
    eprint = "2204.06571",
    archivePrefix = "arXiv",
    primaryClass = "hep-th",
    doi = "10.21468/SciPostPhys.14.2.014",
    journal = "SciPost Phys.",
    volume = "14",
    number = "2",
    pages = "014",
    year = "2023"
}

@article{Costello:2022wso,
    author = "Costello, Kevin and Paquette, Natalie M.",
    title = "{Celestial holography meets twisted holography: 4d amplitudes from chiral correlators}",
    eprint = "2201.02595",
    archivePrefix = "arXiv",
    primaryClass = "hep-th",
    doi = "10.1007/JHEP10(2022)193",
    journal = "JHEP",
    volume = "10",
    pages = "193",
    year = "2022"
}

@article{Costello:2022jpg,
    author = "Costello, Kevin and Paquette, Natalie M. and Sharma, Atul",
    title = "{Top-Down Holography in an Asymptotically Flat Spacetime}",
    eprint = "2208.14233",
    archivePrefix = "arXiv",
    primaryClass = "hep-th",
    doi = "10.1103/PhysRevLett.130.061602",
    journal = "Phys. Rev. Lett.",
    volume = "130",
    number = "6",
    pages = "061602",
    year = "2023"
}

@article{Costello:2023hmi,
    author = "Costello, Kevin and Paquette, Natalie M. and Sharma, Atul",
    title = "{Burns space and holography}",
    eprint = "2306.00940",
    archivePrefix = "arXiv",
    primaryClass = "hep-th",
    doi = "10.1007/JHEP10(2023)174",
    journal = "JHEP",
    volume = "10",
    pages = "174",
    year = "2023"
}

@article{Bern:2014oka,
    author = "Bern, Zvi and Davies, Scott and Nohle, Josh",
    title = "{On Loop Corrections to Subleading Soft Behavior of Gluons and Gravitons}",
    eprint = "1405.1015",
    archivePrefix = "arXiv",
    primaryClass = "hep-th",
    reportNumber = "UCLA-14-TEP-102",
    doi = "10.1103/PhysRevD.90.085015",
    journal = "Phys. Rev. D",
    volume = "90",
    number = "8",
    pages = "085015",
    year = "2014"
}

@article{Strominger:2013jfa,
    author = "Strominger, Andrew",
    title = "{On BMS Invariance of Gravitational Scattering}",
    eprint = "1312.2229",
    archivePrefix = "arXiv",
    primaryClass = "hep-th",
    doi = "10.1007/JHEP07(2014)152",
    journal = "JHEP",
    volume = "07",
    pages = "152",
    year = "2014"
}

@article{Beig:1982ifu,
    author = "Beig, R. and Schmidt, B. G.",
    title = "{Einstein's equations near spatial infinity}",
    doi = "10.1007/BF01211056",
    journal = "Commun. Math. Phys.",
    volume = "87",
    number = "1",
    pages = "65--80",
    year = "1982"
}

@article{Ruzziconi:2026bix,
    author = "Ruzziconi, Romain",
    title = "{Carrollian Physics and Holography}",
    eprint = "2602.02644",
    archivePrefix = "arXiv",
    primaryClass = "hep-th",
    month = "2",
    year = "2026"
}

@article{Ammon:2025avo,
    author = "Ammon, Martin and Capone, Federico and Sieling, Christoph",
    title = "{Flat Holography {\&} Holographic Renormalization: Scalar Field}",
    eprint = "2512.14818",
    archivePrefix = "arXiv",
    primaryClass = "hep-th",
    month = "12",
    year = "2025"
}

@article{Mann:2008ay,
    author = "Mann, Robert B. and Marolf, Donald and McNees, Robert and Virmani, Amitabh",
    title = "{On the Stress Tensor for Asymptotically Flat Gravity}",
    eprint = "0804.2079",
    archivePrefix = "arXiv",
    primaryClass = "hep-th",
    doi = "10.1088/0264-9381/25/22/225019",
    journal = "Class. Quant. Grav.",
    volume = "25",
    pages = "225019",
    year = "2008"
}

@article{Mann:2005yr,
    author = "Mann, Robert B. and Marolf, Donald",
    title = "{Holographic renormalization of asymptotically flat spacetimes}",
    eprint = "hep-th/0511096",
    archivePrefix = "arXiv",
    doi = "10.1088/0264-9381/23/9/010",
    journal = "Class. Quant. Grav.",
    volume = "23",
    pages = "2927--2950",
    year = "2006"
}

@article{Mann:2006bd,
    author = "Mann, Robert B. and Marolf, Donald and Virmani, Amitabh",
    title = "{Covariant Counterterms and Conserved Charges in Asymptotically Flat Spacetimes}",
    eprint = "gr-qc/0607041",
    archivePrefix = "arXiv",
    doi = "10.1088/0264-9381/23/22/017",
    journal = "Class. Quant. Grav.",
    volume = "23",
    pages = "6357--6378",
    year = "2006"
}

@article{AtulBhatkar:2021txo,
    author = "Atul Bhatkar, Sayali",
    title = "{Asymptotic conservation law with Feynman boundary condition}",
    eprint = "2101.09734",
    archivePrefix = "arXiv",
    primaryClass = "hep-th",
    doi = "10.1103/PhysRevD.103.125026",
    journal = "Phys. Rev. D",
    volume = "103",
    number = "12",
    pages = "125026",
    year = "2021"
}

@article{Bagchi:2025vri,
    author = "Bagchi, Arjun and Banerjee, Aritra and Dhivakar, Prateksh and Mondal, Saikat and Shukla, Ashish",
    title = "{The Carrollian Kaleidoscope}",
    eprint = "2506.16164",
    archivePrefix = "arXiv",
    primaryClass = "hep-th",
    month = "6",
    year = "2025"
}

@article{He:2014laa,
    author = "He, Temple and Lysov, Vyacheslav and Mitra, Prahar and Strominger, Andrew",
    title = "{BMS supertranslations and Weinberg{\textquoteright}s soft graviton theorem}",
    eprint = "1401.7026",
    archivePrefix = "arXiv",
    primaryClass = "hep-th",
    doi = "10.1007/JHEP05(2015)151",
    journal = "JHEP",
    volume = "05",
    pages = "151",
    year = "2015"
}

@article{Cachazo:2014fwa,
    author = "Cachazo, Freddy and Strominger, Andrew",
    title = "{Evidence for a New Soft Graviton Theorem}",
    eprint = "1404.4091",
    archivePrefix = "arXiv",
    primaryClass = "hep-th",
    month = "4",
    year = "2014"
}

@article{Capone:2023roc,
    author = "Capone, Federico and Mitra, Prahar and Poole, Aaron and Tomova, Bilyana",
    title = "{Phase space renormalization and finite BMS charges in six dimensions}",
    eprint = "2304.09330",
    archivePrefix = "arXiv",
    primaryClass = "hep-th",
    doi = "10.1007/JHEP11(2023)034",
    journal = "JHEP",
    volume = "11",
    pages = "034",
    year = "2023"
}

@article{Ciambelli:2025mex,
    author = "Ciambelli, Luca",
    title = "{Asymptotic limit of null hypersurfaces}",
    eprint = "2501.17357",
    archivePrefix = "arXiv",
    primaryClass = "hep-th",
    doi = "10.1088/1361-6382/ae22b5",
    journal = "Class. Quant. Grav.",
    volume = "42",
    number = "23",
    pages = "235020",
    year = "2025"
}

@article{Kapec:2016jld,
    author = "Kapec, Daniel and Mitra, Prahar and Raclariu, Ana-Maria and Strominger, Andrew",
    title = "{2D Stress Tensor for 4D Gravity}",
    eprint = "1609.00282",
    archivePrefix = "arXiv",
    primaryClass = "hep-th",
    doi = "10.1103/PhysRevLett.119.121601",
    journal = "Phys. Rev. Lett.",
    volume = "119",
    number = "12",
    pages = "121601",
    year = "2017"
}

@article{Kapec:2017gsg,
    author = "Kapec, Daniel and Mitra, Prahar",
    title = "{A $d$-Dimensional Stress Tensor for Mink$_{d+2}$ Gravity}",
    eprint = "1711.04371",
    archivePrefix = "arXiv",
    primaryClass = "hep-th",
    doi = "10.1007/JHEP05(2018)186",
    journal = "JHEP",
    volume = "05",
    pages = "186",
    year = "2018"
}

@article{Bagchi:2024gnn,
    author = "Bagchi, Arjun and Dhivakar, Prateksh and Dutta, Sudipta",
    title = "{3D Stress Tensor for Gravity in 4D Flat Spacetime}",
    eprint = "2408.05494",
    archivePrefix = "arXiv",
    primaryClass = "hep-th",
    month = "8",
    year = "2024"
}

@article{Nguyen:2025zhg,
    author = "Nguyen, Kevin",
    title = "{Lectures on Carrollian Holography}",
    eprint = "2511.10162",
    archivePrefix = "arXiv",
    primaryClass = "hep-th",
    month = "11",
    year = "2025"
}

@article{Briceno:2025cdu,
    author = "Brice{\~n}o, Mat{\'\i}as and Gonz{\'a}lez, Hern{\'a}n A. and Henneaux, Marc and P{\'e}rez, Alfredo",
    title = "{Matching conditions at null infinity in the presence of logarithms: the role of advanced and retarded radiation}",
    eprint = "2510.21072",
    archivePrefix = "arXiv",
    primaryClass = "hep-th",
    doi = "10.1007/JHEP02(2026)103",
    journal = "JHEP",
    volume = "02",
    pages = "103",
    year = "2026"
}

@article{Nguyen:2025sqk,
    author = "Nguyen, Kevin and Salzer, Jakob",
    title = "{Operator product expansion in Carrollian CFT}",
    eprint = "2503.15607",
    archivePrefix = "arXiv",
    primaryClass = "hep-th",
    doi = "10.1007/JHEP07(2025)193",
    journal = "JHEP",
    volume = "07",
    pages = "193",
    year = "2025"
}

@article{Poulias:2025eck,
    author = "Poulias, Georgios and Vandoren, Stefan",
    title = "{On Carroll partition functions and flat space holography}",
    eprint = "2503.20615",
    archivePrefix = "arXiv",
    primaryClass = "hep-th",
    doi = "10.1007/JHEP06(2025)232",
    journal = "JHEP",
    volume = "06",
    pages = "232",
    year = "2025"
}

@article{Surubaru:2025fmg,
    author = "Surubaru, Iustin and Zhu, Bin",
    title = "{Carrollian amplitudes and holographic correlators in AdS3/CFT2}",
    eprint = "2504.07650",
    archivePrefix = "arXiv",
    primaryClass = "hep-th",
    doi = "10.1103/7t4n-7c6j",
    journal = "Phys. Rev. D",
    volume = "112",
    number = "2",
    pages = "026023",
    year = "2025"
}

@article{Lipstein:2025jfj,
    author = "Lipstein, Arthur and Ruzziconi, Romain and Yelleshpur Srikant, Akshay",
    title = "{Towards a flat space Carrollian hologram from AdS$_{4}$/CFT$_{3}$}",
    eprint = "2504.10291",
    archivePrefix = "arXiv",
    primaryClass = "hep-th",
    doi = "10.1007/JHEP06(2025)073",
    journal = "JHEP",
    volume = "06",
    pages = "073",
    year = "2025"
}

@article{Fiorucci:2025twa,
    author = "Fiorucci, Adrien and Pekar, Simon and Marios Petropoulos, P. and Vilatte, Matthieu",
    title = "{Carrollian-Holographic Derivation of Gravitational Flux-Balance Laws}",
    eprint = "2505.00077",
    archivePrefix = "arXiv",
    primaryClass = "hep-th",
    reportNumber = "CPHT-RR035.042025",
    doi = "10.1103/qv17-ks32",
    journal = "Phys. Rev. Lett.",
    volume = "135",
    number = "26",
    pages = "261602",
    year = "2025"
}

@article{Agrawal:2025bsy,
    author = "Agrawal, Shreyansh and Nguyen, Kevin",
    title = "{Soft theorems and spontaneous symmetry breaking}",
    eprint = "2504.10577",
    archivePrefix = "arXiv",
    primaryClass = "hep-th",
    doi = "10.1103/nbpg-mscq",
    journal = "Phys. Rev. D",
    volume = "112",
    number = "2",
    pages = "L021903",
    year = "2025"
}

@article{Have:2024dff,
    author = "Have, Emil and Nguyen, Kevin and Prohazka, Stefan and Salzer, Jakob",
    title = "{Massive carrollian fields at timelike infinity}",
    eprint = "2402.05190",
    archivePrefix = "arXiv",
    primaryClass = "hep-th",
    reportNumber = "UWThPh 2024-6",
    doi = "10.1007/JHEP07(2024)054",
    journal = "JHEP",
    volume = "07",
    pages = "054",
    year = "2024"
}

@article{Barnich:2011mi,
    author = "Barnich, Glenn and Troessaert, Cedric",
    title = "{BMS charge algebra}",
    eprint = "1106.0213",
    archivePrefix = "arXiv",
    primaryClass = "hep-th",
    reportNumber = "ULB-TH-11-10",
    doi = "10.1007/JHEP12(2011)105",
    journal = "JHEP",
    volume = "12",
    pages = "105",
    year = "2011"
}

@article{Penedones:2010ue,
    author = "Penedones, Joao",
    title = "{Writing CFT correlation functions as AdS scattering amplitudes}",
    eprint = "1011.1485",
    archivePrefix = "arXiv",
    primaryClass = "hep-th",
    doi = "10.1007/JHEP03(2011)025",
    journal = "JHEP",
    volume = "03",
    pages = "025",
    year = "2011"
}

@article{Choi:2024ygx,
    author = "Choi, Sangmin and Laddha, Alok and Puhm, Andrea",
    title = "{Asymptotic Symmetries for Logarithmic Soft Theorems in Gauge Theory and Gravity}",
    eprint = "2403.13053",
    archivePrefix = "arXiv",
    primaryClass = "hep-th",
    month = "3",
    year = "2024"
}

@article{Choi:2024ajz,
    author = "Choi, Sangmin and Laddha, Alok and Puhm, Andrea",
    title = "{The classical super-rotation infrared triangle. Classical logarithmic soft theorem as conservation law in gravity}",
    eprint = "2412.16142",
    archivePrefix = "arXiv",
    primaryClass = "hep-th",
    doi = "10.1007/JHEP04(2025)138",
    journal = "JHEP",
    volume = "04",
    pages = "138",
    year = "2025"
}

@article{Campiglia:2019wxe,
    author = "Campiglia, Miguel and Laddha, Alok",
    title = "{Loop Corrected Soft Photon Theorem as a Ward Identity}",
    eprint = "1903.09133",
    archivePrefix = "arXiv",
    primaryClass = "hep-th",
    doi = "10.1007/JHEP10(2019)287",
    journal = "JHEP",
    volume = "10",
    pages = "287",
    year = "2019"
}

@article{Fiorucci:2024ndw,
    author = "Fiorucci, Adrien and Matulich, Javier and Ruzziconi, Romain",
    title = "{Superrotations at spacelike infinity}",
    eprint = "2404.02197",
    archivePrefix = "arXiv",
    primaryClass = "hep-th",
    doi = "10.1103/PhysRevD.110.L061502",
    journal = "Phys. Rev. D",
    volume = "110",
    number = "6",
    pages = "L061502",
    year = "2024"
}

@article{Compere:2023qoa,
    author = "Comp{\`e}re, Geoffrey and Gralla, Samuel E. and Wei, Hongji",
    title = "{An asymptotic framework for gravitational scattering}",
    eprint = "2303.17124",
    archivePrefix = "arXiv",
    primaryClass = "gr-qc",
    doi = "10.1088/1361-6382/acf5c1",
    journal = "Class. Quant. Grav.",
    volume = "40",
    number = "20",
    pages = "205018",
    year = "2023"
}

@article{Hartong:2025jpp,
    author = "Hartong, Jelle and Have, Emil and Nenmeli, Vijay and Oling, Gerben",
    title = "{Boundary Energy-Momentum Tensors for Asymptotically Flat Spacetimes}",
    eprint = "2505.05432",
    archivePrefix = "arXiv",
    primaryClass = "hep-th",
    month = "5",
    year = "2025"
}

\end{document}